\documentclass[dvipdfmx]{aastex63}

\usepackage{ulem}
\usepackage{url}

\received{2022 June 3}
\revised{2022 September 15}
\accepted{2022 September 20}
\submitjournal{ApJ}

\shorttitle{IMF in the outer Galaxy. Sh 2-209} 

\shortauthors{Yasui et al.}

\begin{document}

\title{Mass function of a young cluster in a low-metallicity
environment. Sh 2-209}

\correspondingauthor{Chikako Yasui}
\email{ck.yasui@gmail.com}

\author[0000-0003-3579-7454]{Chikako Yasui}
\affil{National Astronomical Observatory of Japan 2-21-1
Osawa, Mitaka, Tokyo, 181-8588, Japan}
 
\author[0000-0003-4578-2619]{Naoto Kobayashi}
\affil{Institute of Astronomy, School of Science, University of Tokyo,
2-21-1 Osawa, Mitaka, Tokyo 181-0015, Japan}
\affil{Kiso Observatory, Institute of Astronomy, School of Science,
University of Tokyo, 10762-30 Mitake, Kiso-machi, Kiso-gun, Nagano
397-0101, Japan}

\author[0000-0003-0769-8627]{Masao Saito}
\affil{National Astronomical Observatory of Japan 2-21-1
Osawa, Mitaka, Tokyo, 181-8588, Japan}
\affil{Department of Astronomical Science, School of Physical Science,
SOKENDAI (The Graduate University for Advanced Studies), Osawa, Mitaka,
Tokyo 181-8588, Japan}

\author[0000-0003-1604-9127]{Natsuko Izumi}
\affil{Institute of Astronomy and Astrophysics, Academia Sinica, No. 1,
Section 4, Roosevelt Road, Taipei 10617, Taiwan}

\author[0000-0003-2380-8582]{Yuji Ikeda}
\affil{Photocoding, 460-102 Iwakura-Nakamachi, Sakyo-ku, Kyoto,
606-0025, Japan}




\begin{abstract}

We present deep near-infrared (NIR) imaging of Sh 2-209 (S209), a
low-metallicity (${\rm [O/H]} = - 0.5$ dex) \ion{H}{2} region in the
Galaxy.
From the NIR images, combined with astrometric data from Gaia
EDR3, we estimate the distance to S209 to be 2.5 kpc.
This is close enough to enable us to resolve cluster members clearly
($\simeq$1000 AU separation) down to a mass-detection limit of
$\simeq$0.1 $M_\odot$, and we have identified two star-forming clusters
in S209, with individual cluster scales $\sim$1 pc.
We employ a set of model luminosity functions to derive the underlying
initial mass functions (IMFs) and ages for both clusters. The IMFs we
obtained for both clusters exhibit slightly flat high-mass slopes
($\Gamma \simeq -1.0$) compared to the Salpeter IMF ($\Gamma = -1.35$),
and their break mass of $\simeq$0.1 $M_\odot$ is lower than those
generally seen in the solar neighborhood ($\sim$0.3 $M_\odot$).
In particular, because the S209 main cluster is a star-forming cluster
with a larger number of members ($\sim$1500) than the number ($\sim$100)
in regions previously studied in such environments, it is possible for
the first time to derive the IMF in a low-metallicity environment with
high accuracy over the wide mass range 0.1--20 $M_\odot$.

\end{abstract}


\keywords{Stellar mass functions (1612) ---
Luminosity function (942) ---
Galaxy abundances (574) ---
Open star clusters (1160) ---
Infrared sources (793) ---
Low mass stars (2050) ---
Pre-main sequence stars (1290) ---
Star formation (1569) ---
H II regions (694)}



 \section{Introduction} \label{sec:intro}

Stars are fundamental components of galaxies, and because the stellar
mass determines a star's subsequent evolutionary path, the initial mass
function (IMF)---i.e., the initial distribution of masses for a
population of stars---is one of the fundamental parameters that
determines the physical and chemical evolution of a stellar system. In
addition to informing our understanding of stellar origins and
evolution, the IMF provides important input to many astrophysical
studies \citep{Bastian2010}.
The IMF was first introduced by \citet{Salpeter1955}, and detailed
derivations have been carried out for various regions, including the
field, young clusters and associations, and old globular clusters.
The results indicate that the vast majority of stellar systems
originated from a universal IMF that follows a power law with the
Salpeter index $\Gamma = 1.35$ for masses $\gtrsim$1 $M_\odot$ and which
flattens ($\Gamma \sim 0$) around 0.5 $M_\odot$.
The IMF is also known to exhibit a turnover at $M_C \sim 0.3$ $M_\odot$,
which is called the ``characteristic mass'' \citep{Elmegreen2008}.
However, because there are many different physical and chemical
environments in the universe, whether the IMF is universal or whether it
is sensitive to environmental conditions remains a crucial question.

Metallicity---the abundance of elements heavier than hydrogen and
helium---is known to increase during cosmic evolution due to the
synthesis of elements in stars and supernovae.
At present, only 2\% (by mass) of the baryons in our solar system are
heavy elements.
Nevertheless, metallicity is believed to be one of the most critical
factors for star formation because it sensitively affects the heating
and cooling processes during star formation as well as radiative
transfer. 
Because of these effects, theoretical studies have suggested that the
IMF will move away from the Salpeter form in low-metallicity
environments \citep[e.g.,][]{Omukai2005}.
Some observations also indicate that high-mass IMFs tend to become
top-heavy as the metallicity decreases.
However, there is currently no conclusive evidence for this, so at
present it should be considered just as suggestive evidence
\citep{Kroupa2013}.
Observations of globular clusters show that---with decreasing
metallicity coupled with increasing {density}---the stellar IMF shifts
toward a top-heavy form \citep{Marks2012}.
However, because globular clusters are very old ($\sim$$10^{10}$ yr),
they have already experienced mass segregation and self-regulation, so
the present-day mass function (PDMF) may not reflect the
IMF. Conversely, in young clusters that have not undergone
{much $N$-body relaxation or mass segregation,} it is
reasonable to assume that the PDMF is similar to the IMF
\citep{Lee2020}.
In addition, in young clusters ($\sim$$10^6$ yr), massive stars
($\gtrsim$20 $M_\odot$) have not yet ended their lives, which makes it
possible to derive IMFs over a wide mass range, while in contrast the
mass range of stars in globular clusters is only $\lesssim$ 1$M_\odot$
\citep{Paresce2000}.

The most typical examples of such studies in low-metallicity
environments involve the Large and Small Magellanic Clouds (LMC/SMC).
They are nearby dwarf galaxies at a distance $D \simeq 50$ kpc.
Based on observations of star-forming regions in the solar neighborhood,
the scale of an individual star-forming cluster---the minimum unit of
star formation where many stars are born almost simultaneously---has
been estimated to be $\sim$1 pc \citep{Adams2006}.
This corresponds to the very small scale of
{$\sim$4 arcsec} at the distance of the LMC/SMC. However, star-forming
regions in the LMC/SMC generally extend to $\sim$1 arcmin, which
corresponds to $\sim$15 pc, a much larger spatial scale than the scale
of an individual cluster. For this reason, a single star-forming region
can have a complex star-formation history, and the existence of multiple
generations of stars has been pointed out \citep{De Marchi2017}.
In addition, based on observations in the solar neighborhood, the
separation of individual stars in a star-forming cluster has been
estimated to be $\sim$0.1 pc on average \citep{Adams2006}, which
corresponds to $0\farcs4$ in the LMC/SMC. However, this is only an
average value; in reality, the high stellar density in the center of a
cluster makes it challenging to observe individual stars in the cluster
with sufficient resolution. Although it is not clear whether this is the
direct reason, in some regions investigators have derived the standard
IMF \citep{Kerber2006}, while in others the IMF is flatter (e.g., R136:
\citealt{Sirianni2000}).
For a large variety of the physical environments found in the Galaxy and
the LMC/SMC, observations based on spectroscopic classifications of
stars in OB associations and in star clusters have obtained IMFs that
are comparable to the Salpeter IMF for massive stars
($M \gtrsim 10 M_\odot$) \citep{Massey2003}.
However, a recent result that took into account the star-formation
history found indications of a top-heavy IMF in the mass range 15--200
$M_\odot$ in the 30 Doradus starburst star-forming region in the LMC
\citep{Schneider2018}.
In any case, it is difficult to obtain IMFs extending down to low-mass
stars using the mass coverage characteristic of LMC/SMC observations,
and some effort to bridge the gap between detailed observations in the
solar neighborhood and LMC/SMC observations is needed.

In order to extend detailed derivations of IMFs in low-metallicity
environments to lower stellar masses, we have been focusing on the outer
Galaxy to Galactocentric distances $R_G \gtrsim$15 kpc.
Despite the relatively small distances to these regions ($D\sim 5$--10
kpc), their metallicities have been found to be as low as $\sim$$-$1
dex, which is similar to the metallicity at $z
\thickapprox 2$ in the cosmic chemical evolution
\citep[e.g.,][]{Pei1995}.
Due to these relatively small distances, the spatial resolution in these
regions is about 10 times higher than that in the LMC/SMC, which are
commonly observed low-metallicity environments, making it much easier to
resolve individual stars. Furthermore, the relatively high sensitivity
of these observations makes it possible to achieve a low mass limit.
Therefore, the outer Galaxy provides good alternative targets to nearby
dwarf galaxies.
In our previous systematic studies of low-metallicity regions in the
outer Galaxy, we have explored star-forming regions traced by molecular
clouds and H$\alpha$ and have selected regions for which low
metallicities (${\rm [O/H]} \lesssim -0.5$ dex) have actually been
determined: Cloud 2 \citep{{Yasui2006}, {Yasui2008}}, S207
\citep{Yasui2016a}, S208 \citep{Yasui2016b}, and S127 \citep{Yasui2021}.
From Subaru near-infrared (NIR) imaging of these star-forming regions,
we have confirmed that we can detect stars down to mass $\sim$0.1
$M_\odot$, which allows us to cover the peak of the IMF sufficiently.
The observations also allow us to observe individual star-forming
clusters on scales $\sim$1 pc, which are comparable to those observed in
the solar neighborhood.
{This enables us get closer} to the genuine IMF because
the effects of complex star-formation histories are eliminated.
Based on the fitting of K-band luminosity functions, we found that the
IMF in the outer Galaxy is consistent with that in the solar
neighborhood with regard to both the high-mass slope and the
characteristic mass \citep{{Yasui2017}, {Yasui2008ASPC}}, suggesting
that---down to $\sim$$-$1 dex---the IMF does not depend on metallicity
down to the substellar mass regime ($\sim$0.1 $M_\odot$).
However, previously targeted star-forming clusters have only about 100
members \citep{{Yasui2010},{Yasui2021}}, so their IMFs could not be
determined with high accuracy.

From our Subaru NIR imaging survey of star-forming clusters in the outer
Galaxy, we found active star-formation activity in Sh 2-209 (S209),
which---from previous studies---is thought to be located in a
low-metallicity environment ([O/H$] = -0.5$ dex) at $R_{\rm G} = 18$
kpc.
There are two star-forming clusters in S209, one of which has been
confirmed to have approximately 1500 cluster members \citep{Yasui2010}.
The scale of S209 is comparable to that of one of the largest
star-forming regions in the solar neighborhood, the Orion Nebula Cluster
\citep[ONC;][]{Hillenbrand1997}, and S209 can thus be regarded as a
prototype star-forming region in the outer Galaxy.
Both mid-infrared (MIR) and radio observations \citep{{Klein2005},
{Richards2012}} also suggest that high-mass star formation is occurring
in this region. Therefore, S209 is the best (and currently the only)
star-forming region that can be targeted for deriving a detailed IMF in
a low-metallicity environment.

In this paper, we report the first deep NIR imaging of S209 and the
derivation of IMFs for the two young clusters in this region.
One of the advantages of deriving the IMF based on imaging rather than
spectroscopy is that it allows detections down to low-mass stars with
higher sensitivities and thus enables the derivation of IMFs down to
such masses.
Our IMF derivation mainly follows the method used by \citet{Muench2002},
who derived the IMF for the Trapezium cluster, which has an age $\sim$1
Myr and which is a star-forming region of the same scale as S209.
However, because there are no previous studies of the age of the cluster
in S209, we have developed our own method, which includes age as a
parameter in fitting the K-band luminosity function (KLF).
In Section~\ref{sec:S209}, we summarize the properties of this region
from previous studies and archival data for S209: distance, metallicity,
and star-formation activity.
In addition to the previous distance estimate of 10 kpc to S209, the
possibility has recently been raised that the distance may actually be
2.5 kpc. 
In Section~\ref{sec:obs_data}, we describe NIR imaging observations and
data analysis obtained with the multi-object near-infrared camera and
spectrograph (MOIRCS) at the Subaru Telescope.
In Section~\ref{sec:result}, we present NIR imaging results for the two
young clusters in S209.
In Section~\ref{sec:discussion}, we discuss the properties of the two
S209 clusters using the results from Section~\ref{sec:result}: the
high-mass stars in S209, the distance, the environments around the
star-forming clusters, and the cluster scales.
In Section~\ref{sec:S209_IMF}, we construct model KLFs, which we use to
derive the IMF (Section~\ref{sec:Model}), and we present the best-fit
IMFs we obtained for both S209 clusters (Section~\ref{sec:bestfit_imf}).
In Section~\ref{sec:imf_lowmeta}, we discuss the IMFs obtained for the
S209 clusters (Section~\ref{sec:S209_derived_imf}), compare them with
the IMF derived for the solar neighborhood
(Section~\ref{sec:IMF_solar}), compare them with the IMFs derived for
low-metallicity environments, and finally discuss a possible
metallicity-dependence of the IMFs (Section~\ref{sec:metal_imf}).
In Section~\ref{sec:conclusion}, we conclude and discuss future
prospects.


\section{S\lowercase{h} 2-209} \label{sec:S209}

The nebulosity S209 is an extended \ion{H}{2} region listed in a
catalogue obtained from H$\alpha$ surveys \citep{Sharpless1959}.
It is located at $l = 151.6062^\circ$ and $b = -0.2400^\circ$ in the
Galactic plane, with coordinates $(\alpha_{\rm 2000.0}, \delta_{\rm
2000.0}) = (04^{\rm h} 11^{\rm m} 06.7^{\rm s}, +51^\circ 09' 44'')$
from SIMBAD\footnote{This research has made use of the SIMBAD database,
operated at Centre de Donn\'ees Astronomiques de Strasbourg, France.}
\citep{Wenger2000}.
In this section, we summarize the properties of the targeted
star-forming region S209 obtained from previous works, which we list in
Table~\ref{tab:targets}.
The S209 nebulosity also produces strong MIR emission. We show a
large-scale NIR and MIR pseudocolor image of S209 in
Figure~\ref{fig:3col_2MASS_WISE} and an H$\alpha$ and radio-continuum
image in Figure~\ref{fig:S209_Ha}.

\subsection{Distance} \label{sec:properties}

From spectroscopic and photometric observations of three possibly
dominant exciting stars---with spectral types B1III, O9III, and B1
(hereafter CW1, CW2, and CW3, respectively)---\citet{Chini1984}
estimated the photometric distance to S209 to be 10.9 kpc.
Assuming the Galactocentric distance of the Sun to be $R_\odot = 8.0$
kpc, then the Galactocentric distance to S209 is $R_G \simeq 18$ kpc.
The kinematic distance has also been estimated to be $\simeq$10 kpc,
taking the radial velocity of the local standard of rest to be $V_{\rm
LSR} \sim -50$ km s$^{-1}$, as derived from various observations:
$V_{\rm LSR} =-52.2$ km s$^{-1}$ from CO observations by
\citet{Blitz1982}; $V_{\rm LSR} =-50.0$ km s$^{-1}$ from H$\alpha$
Fabry--Perot observations by \citet{Fich1990}; $V_{\rm LSR}=-52.2$ km
s$^{-1}$ from Fabry--Perot observations of \ion{H}{2} regions by
\citet{Caplan2000}; $V_{\rm LSR}=-48.8$ km s$^{-1}$, $V_{\rm
LSR}=-50.11$ km s$^{-1}$, and $V_{\rm LSR}=-48.3$ km s$^{-1}$ from CO,
\ion{H}{1}, and H$\alpha$, respectively, by \citet{Foster2015}.
According to \citet{Foster2015}., who obtained the most recent
determination, the estimated distance is 10.58$\pm$0.57 kpc, which is
very consistent with the photometric distance.

Astrometric distances have also been obtained from Gaia Early Data
Release 3 (Gaia EDR3; \citealt{Gaia2021}).
For the probable dominant exciting stars CW1, CW2, and CW3, the
parallaxes {were derived} with relatively high accuracy.
The stars CW1, CW2, and CW3 were identified with the Gaia source IDs
271701112917796096, 271701009838634752, and 271701112917794176,
respectively; the parallaxes of these sources divided by the
corresponding parallax errors are 14.7, 5.2, and 3.8, respectively.
Following \citet{Lindegren2021}, we corrected the parallax zero-point
and obtained the distance to each source and its error. In this way, we
found the distances to CW1, CW2, and CW3 to be $2.6^{+0.2}_{-0.1}$,
$3.0^{+0.6}_{-0.4}$, and $3.1^{+1.0}_{-0.6}$ kpc, respectively, all of
which lie in the range $\simeq$2.5--3.0 kpc.
The properties obtained from Gaia EDR3 are summarized in
Table~\ref{tab:gaia}.

\subsection{Oxygen Abundance and Metallicity}\label{sec:meatlicity}

For eight \ion{H}{2} regions, \citet{Vilchez1996} measured several
optical emission-line fluxes---e.g., [\ion{O}{2}] $\lambda\lambda$3727,
H$\beta$, [\ion{O}{3}] $\lambda$4959 and 5007, \ion{He}{1}
$\lambda$5876, H$\alpha$, [\ion{O}{2}] $\lambda\lambda$7320 and 7330,
and P8--P15---and they found the oxygen abundance of S209 to be $12 +
\log ({\rm O/H}) = 8.3$.
Based on Fabry--Perot observations, \citet{Caplan2000} measured several
optical emission-line fluxes in 36 \ion{H}{2} regions:
[\ion{O}{2}] $\lambda\lambda$3726 and 3729, H$\beta$, [\ion{O}{3}]
$\lambda$ 5007, [\ion{He}{1}] $\lambda$5876, and H$\alpha$.
Subsequently, \citet{Deharveng2000} determined the oxygen abundances for
several H II regions, as well as the extinctions, electron densities and
temperatures, and ionic abundances
(O$^+$/H$^+$, O$^{++}$/H$^+$, and He$^+$/H$^+$).
They found the oxygen abundance of S209 to be $12 + \log ({\rm O/H}) =
8.18$.
\citet{Rudolph2006} reanalyzed the elemental abundances of 117
\ion{H}{2} regions with updated physical parameters. They determined the
oxygen abundance of S209 using the data of \citet{Vilchez1996} and
\citet{Caplan2000} to be $12 + \log ({\rm O/H}) = 8.15^{+0.16}_{-0.26}$
and $8.44^{+0.15}_{-0.22}$, respectively.
These abundances correspond to $\simeq$$-0.5$ dex, assuming a solar
abundance of $12 + \log ({\rm O/H}) = 8.73$ \citep{Asplund2009}.

The electron temperatures ($T_{\rm e}$) are also sensitive indicators of
heavy-element abundances, with higher temperatures corresponding to
lower abundances \citep{Shaver1983}. 
The estimated temperatures are very high for S209: $\sim$11000 K
(10855$\pm$3670 K by \citealt{Omar2002}; 10510$\pm$90 K for the central
region and 12570$\pm$360 K for the northern region from
\citealt{Quireza2006ApJ}; and 10795$\pm$985 K from
\citealt{Balser2011}).
These are among the highest temperatures of any \ion{H}{2} regions in
the Galaxy (see Figure~3 in \citealt{Quireza2006ApJ}), which suggests
that S209 is in very-low-metallicity region.
According to the relationship between electron temperature and oxygen
abundance obtained by \citet{Shaver1983}---$12 + \log {\rm (O/H)} = 9.82
- 1.49 T_{\rm e} / 10^4$---the temperature of S209 ($\sim$11000 K)
suggests an oxygen abundance of 8.2, which is consistent with the
abundance determinations discussed above.
\citet{Fernandez-Martin2017} also estimated the oxygen abundance from
the derived electron temperature for S209, obtaining a value that is in
very good agreement with the above estimate.

\subsection{Star-forming Activity} \label{sec:SFinS209}

\begin{figure}[!h]
\centering
 \vspace{12em}
 \hspace{-3em}
\includegraphics[width=18cm]{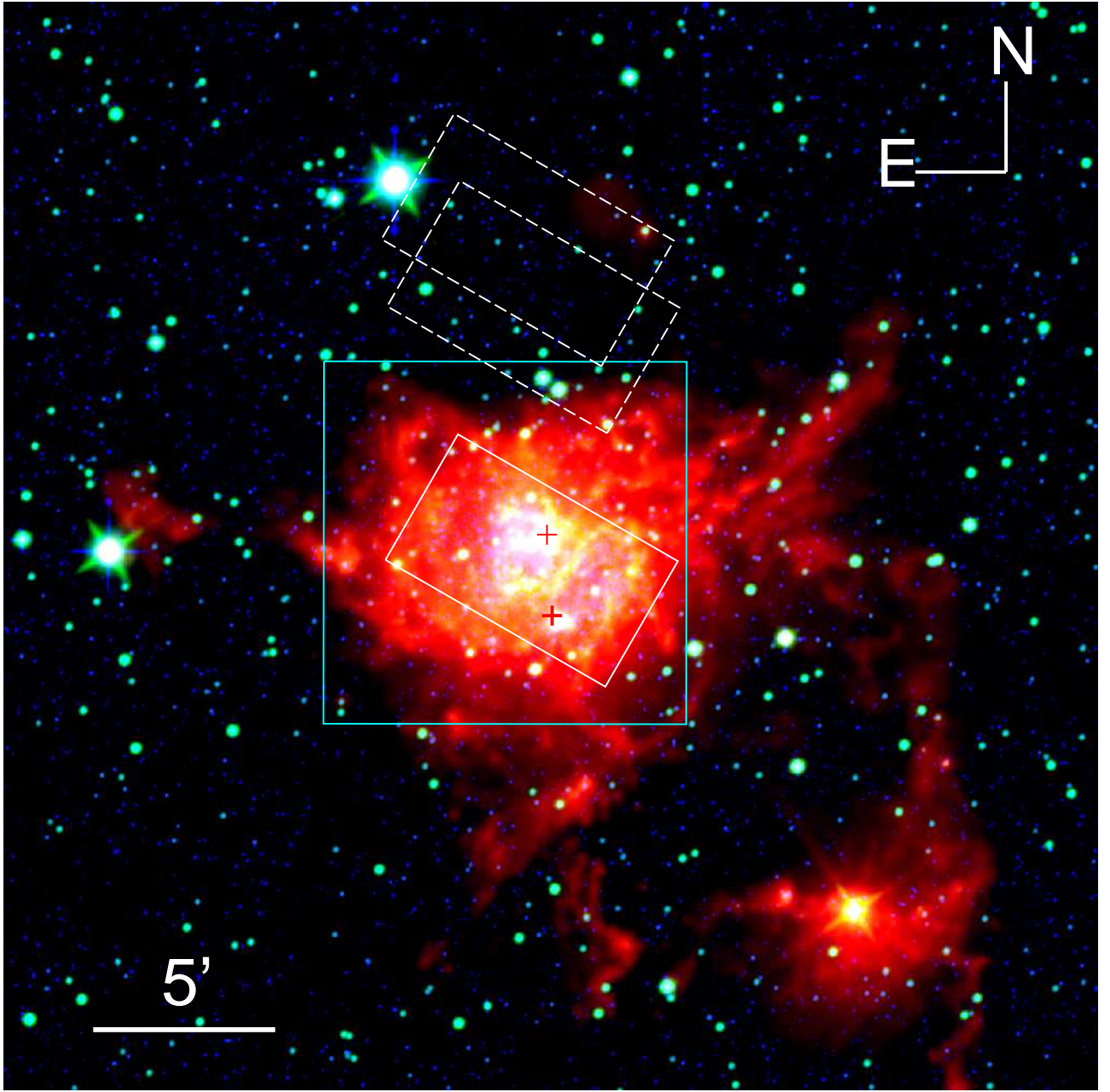}
\vspace{-12em}
\caption{Pseudocolor image of S209 with a wide field of view
 ($\sim$30$'\times$30$'$) centered at {$(\alpha_{\rm 2000.0},
 \delta_{\rm 2000.0}) = (04^{\rm h} 11^{\rm m} 04\fs9, +51^\circ
 09\arcmin 58\farcs9)$} in Equatorial coordinates and $(l, b) =
 (151.60^\circ, -0.24^\circ$) in Galactic coordinates.
Here, 1 arcmin corresponds to 0.75 pc (3.0 pc) for a distance to S209 of
2.5 kpc (10 kpc).
We produced this image by combining the 2MASS $K_S$-band (2.16 $\mu$m;
blue), WISE band 1 (3.4 $\mu$m; green), and WISE band 3 (12 $\mu$m;
red).
The small white boxes show the locations and sizes of the MOIRCS fields
of view (solid and dashed lines for the S209 and sky frames,
respectively), while the cyan box shows the location and size of
Figure~\ref{fig:S209_Ha}.
The red plus signs show the locations of IRAS point sources.}
\label{fig:3col_2MASS_WISE} 
\end{figure}

Figure~\ref{fig:3col_2MASS_WISE} shows a NIR and MIR pseudocolor image
of S209 with a wide field of view ($\sim$30$'\times$30$'$) that is
centered at $(l, b) = (151.60^\circ, -0.24^\circ$) in Galactic
coordinates.
We produced this figure by combining 2MASS \citep{Skrutskie2006}
{\it K}$_S$-band (2.16 $\mu$m, blue), WISE
\citep{Wright2010} band 1 (3.4 $\mu$m; green), and WISE band 3
(12 $\mu$m; red) images.
The positions of two sources in the IRAS Point Source Catalog
\citep{Beichman1988}---IRAS 04073+5102 and IRAS 04073+5100---are shown
with red plus signs.
The 12 $\mu$m emission is mainly from polycyclic aromatic hydrocarbons,
and it traces the photodissociation regions around \ion{H}{2} regions.
This figure shows that the star-forming region extends over 10 arcmin
square. The star-forming cluster [BDS2003]65 has been identified by
\citet{Bica2003} from 2MASS images; it is centered at
$(\alpha_{\rm 2000.0}, \delta_{\rm 2000.0}) = (04^{\rm h} 11^{\rm m}
10^{\rm s}, +51^\circ 09' 58'')$ and has an angular dimension of
$\sim$2.5 arcmin.
\citet{Richards2012} presented MIR data taken with the {\it Spitzer}
Infrared Array Camera (IRAC) in the 3.6, 4.5, 5.8, and 8.0 $\mu$m
bands. From the spatial distribution of objects with infrared excesses,
they suggested that young stellar object candidates are distributed
along ridges of the bright extended 8.0 $\mu$m emission surrounding the
\ion{H}{2} region.

\begin{figure}[!h]
\begin{center}
\vspace{12em}
 \hspace{-5em}
 \includegraphics[width=13cm]{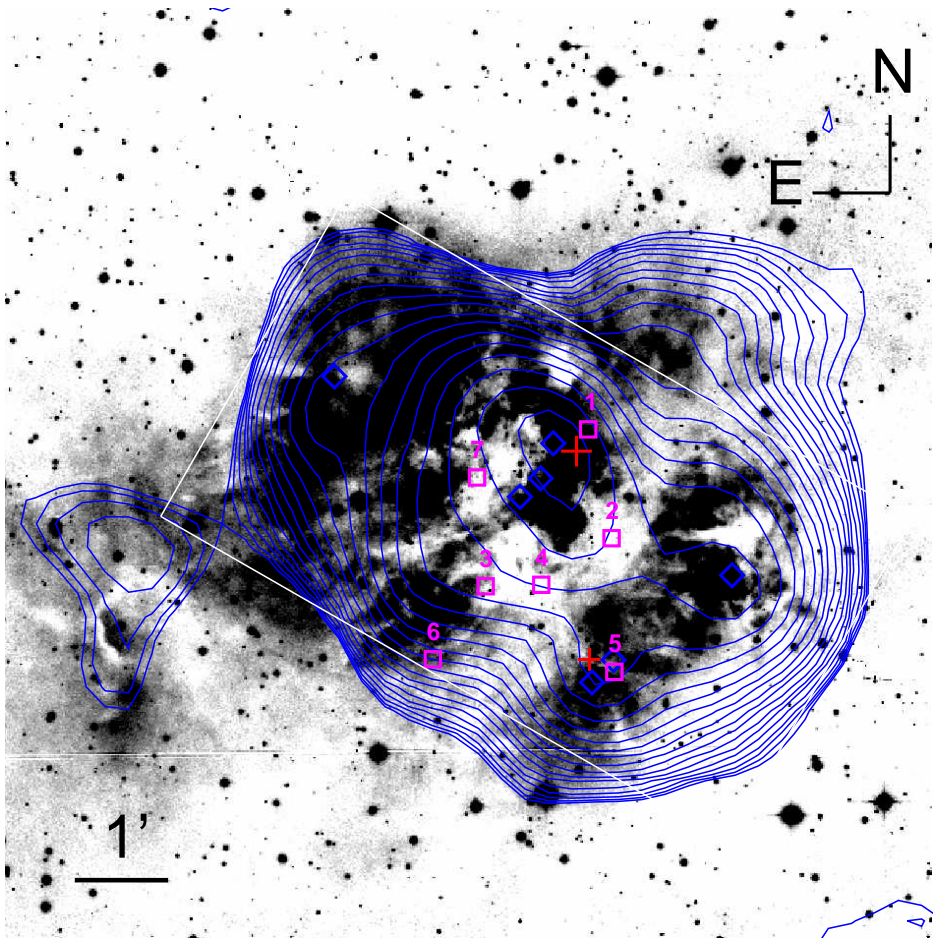}
\vspace{-10em}
\caption{IPHAS H$\alpha$ image of S209 with a field of view of
$13\arcmin \times 17\arcmin$, corresponding to the cyan box in
Figure~\ref{fig:3col_2MASS_WISE}.
The 1.4 GHz radio-continuum emission from the NVSS is also shown, using
blue contours. The contours are plotted at 1 mJy beam$^{-1} \times 2^0$,
$2^{1/2}$, $2^1$,...
The blue diamonds show the NVSS radio point sources, while {the magenta
boxes} show the peak positions of the molecular cores presented by
\citet{Klein2005}.
The red plus sings show the locations of IRAS point sources. The white
box shows the location and size of the MOIRCS field of view.}
\label{fig:S209_Ha}
\end{center}
\end{figure}

The photoionized \ion{H}{2} region is traced by H$\alpha$ emission and
radio-continuum emission.
Figure~\ref{fig:S209_Ha} shows an H$\alpha$ image obtained by the Isaac
Newton Telescope Photometric H-Alpha Survey (IPHAS; \citealt{Drew2005}).
The 1.4 GHz radio continuum from the NRAO VLA Sky Survey (NVSS;
\citealt{Condon1998}) is shown with blue contours superimposed on the
H$\alpha$ image, while the locations of NVSS radio sources are shown
with blue diamonds.
This figure also shows that the star-forming region extends over a
radius of $\sim$3 arcmin, which is consistent with
Figure~\ref{fig:3col_2MASS_WISE}.
From the radio-continuum properties of \ion{H}{2} regions,
\citet{Richards2012} estimated that S209 contains $\sim$2800 $M_\odot$
of ionized gas, which is comparable to the masses of the W49a and
Quintuplet \ion{H}{2} regions, and that the Lyman-continuum photon flux
is $\sim$$10^{49}$ s$^{-1}$. 
Based on these physical parameters, they suggested that the star-forming
cluster is very massive ($>$$10^3$ $M_\odot$).
This is consistent with our previous results \citep{Yasui2010}, in which
we detected more than 2000 sources in NIR observations. However, note
that \citet{Richards2012} assumed a distance of 9.8 kpc (based on the
kinematic distance), and the estimates of these values can change
significantly if the distances are significantly different, as discussed
in Section~\ref{sec:properties}.

Molecular-cloud cores are traced by millimeter-wavelength continuum emission.
\citet{Klein2005} obtained a millimeter-continuum map of the S209
\ion{H}{2} region with the Heinrich Hertz Submillimeter Telescope.
They identified seven molecular cores (the magenta boxes in
Figure~\ref{fig:S209_Ha}) with a total mass of $\sim$$2\times
10^4$$M_\odot$.
From the coincidence of the millimeter-continuum emission with a dark
region in the faint optical-emission nebula, they suggested that the
molecular cores are located in front of the \ion{H}{2} region.
They also found indications of triggered star formation from a ring of
cloud cores around the star cluster, and they suggested the existence of
another cluster around IRAS 04073+5100.

\section{OBSERVATIONS AND DATA REDUCTION} \label{sec:obs_data} 

\subsection{Subaru MOIRCS Imaging} \label{sec:obs_MOIRCS}

We obtained deep $JHK_S$-band images for each band with the 8.2 m Subaru
telescope using the wide-field NIR camera and spectrograph, MOIRCS
\citep{{Ichikawa2006},{Suzuki2008}}.
The MOIRCS instrument uses two 2K ``HAWAII-2'' imaging arrays, which
yield a $4' \times 7'$ field of view ($3.5' \times 4'$ for each
channel), with a pixel scale of $0\farcs117$ pixel$^{-1}$.
The instrument uses the Mauna Kea Observatory (MKO) NIR photometric
filters \citep{{Simons2002},{Tokunaga2002}}.

We performed these observations on 2006 September 3 UT. The observing
conditions were photometric, and the seeing was excellent
($\sim$$0\farcs35$--$0\farcs45$) throughout the night.
Because linearity of the detector output is not guaranteed for counts
over $\sim$20,000 analog-to-digital units (ADUs), we also obtained
short-exposure images on 2006 November 8 UT, when the conditions were
very humid ($\sim$45\%--75\%) and the seeing was $\sim$1\farcs$2$.
For the long-exposure images, the exposure times were 120, 20, and 30s
for the {\it J}, {\it H}, and $K_S$ bands, respectively, whereas the
exposure time for the short-exposure images was 13s for all bands.
The total integration times for the long-exposure images were 480, 480,
and 720s for the {\it J}, {\it H}, and $K_S$ bands, respectively,
whereas the total integration time for the short-exposure images was 52s
for all bands.
We centered the images of S209 at
$\alpha_{\rm 2000} = 04^{\rm h} 11^{\rm m} 08\fs3$,
$\delta_{\rm 2000} = +51^\circ 09' 29\farcs3$,
so they covered the central region of the \ion{H}{2} region described in
Section~\ref{sec:SFinS209} (see the white boxes in
Figures~\ref{fig:3col_2MASS_WISE} and \ref{fig:S209_Ha}
for the MOIRCS field; hereafter, the ``S209 frame''). 
For background subtraction, we also obtained images of the sky
(hereafter, the ``sky frame'') away from the S209 nebulosity.
For the long-exposure observations, the images were centered $\sim$9
arcmin north of S209, at $\alpha_{\rm 2000} = 04^{\rm h} 11^{\rm m}
09^{\rm s}$, $\delta_{\rm 2000} = +51^\circ 18\arcmin 24\arcsec$
with a $4\arcmin \times 7\arcsec$ field of view. 
For the short-exposure observations, they were centered 7 arcmin north
of S209, at
$\alpha_{\rm 2000} = 04^{\rm h} 11^{\rm m} 08^{\rm s}$, $\delta_{\rm
2000} = +51^\circ 16' 33''$, which is slightly south of the sky frame
used for the long-exposure images.
The sky frames are shown as dashed white boxes in
Figure~\ref{fig:3col_2MASS_WISE}). We summarize the details of the
observations in Table~\ref{tab:LOG}.

\subsection{Data Reduction and Photometry}

\begin{figure}[!h]
\begin{center}
\vspace{5em}
\hspace{-15em}
 \includegraphics[width=10cm]{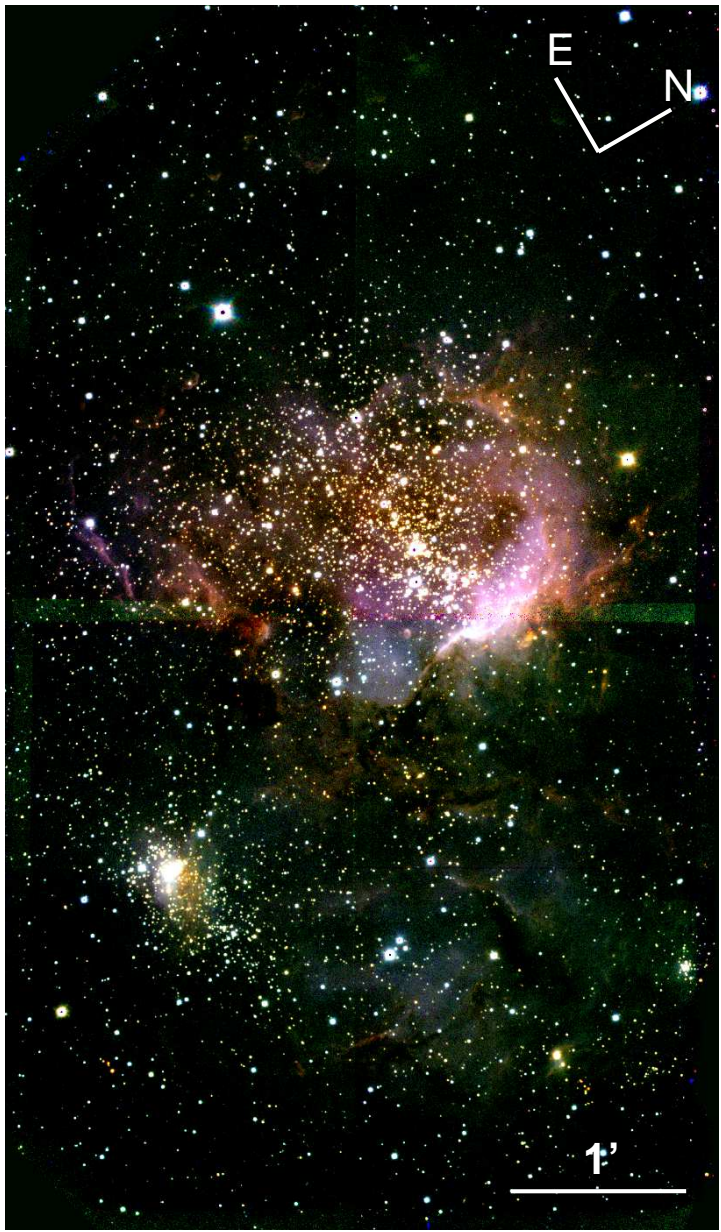}
\vspace{-3em}

\caption{We produced this color image by combining the $J$
(1.26\,$\mu$m), $H$ (1.64\,$\mu$m), and $K_S$ (2.14\,$\mu$m)-band images
obtained with MOIRCS at the Subaru telescope in September 2008. The
field of view of the image is $\sim$$4\arcmin \times 7\arcmin$.}
\label{fig:3col_S209}
\end{center}
\end{figure}

We reduced all the data in each band using standard IRAF\footnote{IRAF
is distributed by the National Optical Astronomy Observatories, which
are operated by the Association of Universities for Research in
Astronomy, Inc., under cooperative agreement with the National Science
Foundation.} procedures, including flat fielding, bad-pixel correction,
median-sky subtraction, image shifts with dithering offsets, and image
combination.
We used sky flats kindly provided by the MOIRCS support astronomer,
Dr. Ichi Tanaka, who obtained them using data from the closest run.
In addition to the above standard procedures, we applied distortion
corrections before combining the images using the
``MCSRED''\footnote{\url{http://www.naoj.org/staff/ichi/MCSRED/mcsred\_e.html}}
reduction package for the MOIRCS imaging data.
We constructed a pseudocolor image of S209 by combining the
long-exposure images for the {\it J} (1.26 $\mu$m, blue), {\it H} (1.64
$\mu$m, green), and $K_S$ (2.15 $\mu$m, red) bands
(Figure~\ref{fig:3col_S209}).

For the long-exposure images, we obtained photometry by fitting the
point-spread function (PSF) using IRAF/DAOPHOT.
To derive the PSF, we selected stars that were bright but not saturated,
with their highest pixel counts below the non-linear sensitivity regime
(20,000 ADU), that were not close to the edge of the frame, and that did
not have any nearby stars with magnitude differences
$<$4 mag.
We performed the PSF photometry in two iterations using the ALLSTAR
routine: the first used the original images, and the second used the
images remaining after the sources from the first iteration had been
subtracted.
We obtained PSF-fit radii of 3.2, 3.5, and 3.5 pixels for the full
widths at half maximum in the {\it J}, {\it H}, and $K_S$ bands, and we
set the inner radii and the widths of the sky annuli to be,
respectively, four and three times as large as the PSF-fit radii.
We used the MKO standard GSPC P330-E \citep{Leggett2006} to calibrate
the photometry.
Based on the pixel-to-pixel noise in the long-exposure images for the
S209 frame, the 10$\sigma$ limiting magnitudes are $J \simeq 22.0$ mag,
$H \simeq 21.0$ mag, and $K_S \simeq 20.6$ mag.
The limiting magnitudes are slightly different for MOIRCS channel 1 and
channel 2, while those for the sky frames are $J\simeq 22.4$ mag, $H
\simeq 21.2$ mag, and $K_S \simeq 21.0$ mag.
The limiting magnitudes in the all {\it JH}$K_S$
bands for the S209 frames are slightly brighter than those for the sky
frames, despite the longer exposure times, probably due to the S209
nebulosity.
The limiting magnitudes are summarized in Table~\ref{tab:limit}.

Bright stars with magnitudes $J \lesssim 17$ mag, $H \lesssim 16$ mag,
and $K_S \lesssim 15$ mag are saturated in the long-exposure images for
both the S209 frames and the sky frames.
We therefore used the short-exposure images for the photometry of such
bright stars, employing the same procedure as for long-exposure images
but using PSF-fit radii of 13, 15, and 10 pixels for the {\it J}-, {\it
H}-, and {\it K}-bands, respectively.
For the photometric calibration, we used stars for which magnitudes can
be determined with small uncertainties in both the short- and
long-exposure images (magnitudes $J < 18$ mag, $H < 17$ mag, and $K_S <
16$ mag, with magnitude uncertainties $<$0.05 mag).
However, in the sky frames, note that the overlap between the
long-exposure image and the short-exposure image was only about half of
the field of view (Figure~\ref{fig:3col_2MASS_WISE}).
Although it is possible to use only the overlapped area as the sky
frame, we actually used the photometric data for all the stars in the
frames of the long- and short-exposure images, aiming to obtain a higher
signal-to-noise ratio (S/N) by including the wider area, because we use
the sky frames as control fields in later sections.
The photometric results show that the short-exposure images are properly
photometric in all JHK bands for all sources with $K<16.8$ mag---except
for the saturation of very bright sources---and that the long-exposure
images are properly photometric except for very faint sources with $K
\ge 16.8$ mag, which are below the sensitivity limit.
Therefore, for the stars in the sky frame, we decided to use the
long-exposure images for stars with $K \ge 16.8$ mag and the
short-exposure images for stars with $K < 16.8$ mag.
For the stars in the S209 frame, we preferentially adopted infrared
luminosities from the photometry obtained from the long-exposure images
because they have higher sensitivities and higher angular resolution.
Finally, in both the S209 and the sky frames, we found that very bright
stars---with magnitudes 
$J \lesssim 12$ mag, $H \lesssim 11$ mag, and $K_S \lesssim 11.5$
mag---were saturated even in the short-exposure images.

\section{The Two Young Clusters in S209} \label{sec:result}

In this section, we identify the two star-forming clusters found in the
MOIRCS NIR images and study their reddening properties.
To reduce contamination by foreground stars and enable the
mass-detection limit to {extend below $\sim$0.1 $M_\odot$,} we next
extract a sample from each cluster for delimited ranges of extinction
and mass, which we term {``Mass-$A_V$--limited'' samples} (see
Section~\ref{sec:sample_S209}).
We use these samples to derive the cluster KLFs, which we employ in
later sections to obtain reliable IMFs by fitting them to the model
KLFs.

\subsection{Identification of the Young Clusters}
\label{sec:ident_cluster}

Using the pseudocolor image (Figure~\ref{fig:3col_S209}), we identified
enhancements in the stellar density, compared to the surrounding area,
on the northeast and southwest sides of the field observed with MOIRCS.
Hereafter, we refer to these enhancements as the ``main cluster'' and
the ``sub-cluster,'' respectively.
Both clusters seem to be associated with, or at least are located close
to, IRAS objects: the main cluster with IRAS 04073+5102 and the
sub-cluster with IRAS 04073+5100.

\begin{figure}[!h]
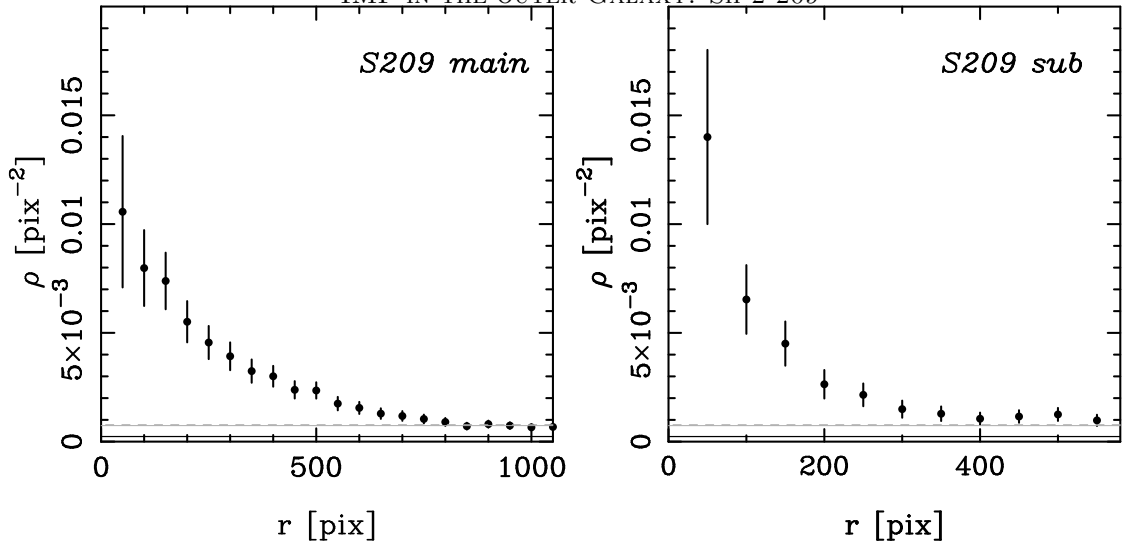

\begin{center}
\hspace{-5em}
 \includegraphics[scale=0.43]{S209_fig4a.eps}
 \includegraphics[scale=0.43]{S209_fig4b.eps}
\caption{Radial variation of the projected stellar density of stars
({\it filled circles}) in the Sh 2-209 main-cluster region ({\it left})
and the sub-cluster region ({\it right}).
The error bars represent Poisson errors (3$\sigma$). 
The gray horizontal solid lines and dashed lines represent the density
of stars in the background region and their Poisson errors (3$\sigma$),
respectively. The black horizontal solid lines represent the density of
stars in the control field.}
\label{fig:profile_S209}
\end{center}
\end{figure}

\begin{figure}[!h]
 \begin{center}
  \vspace{2em}
  \hspace{-15em}
  \includegraphics[width=10cm]{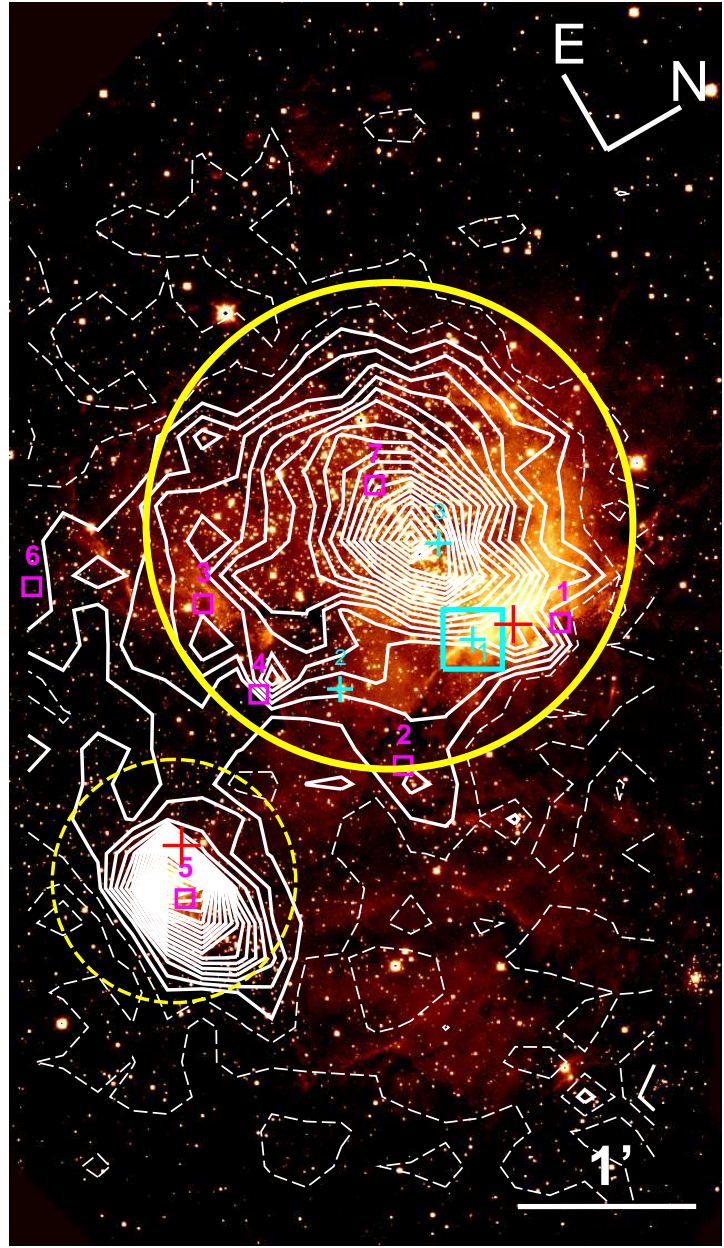}
  \vspace{-2em}
\caption{In the left panel, the stellar density of detected sources in
the MOIRCS Ks-band image is shown with white contours, superposed on the
MOIRCS $K_S$-band image, whose field of view is the same as
Figure~\ref{fig:3col_S209}.  The contour levels are as follows:
1$\sigma$ and 2$\sigma$ higher stellar densities than the average
stellar density in the control field are indicated by dotted lines,
3$\sigma$, 4$\sigma$, 5$\sigma$,..., and 23$\sigma$ higher than the
average stellar density in the control field are shown as solid lines.
The defined regions of the S209 main and sub-cluster are shown as solid
yellow and dotted circles, respectively.
The red plus signs show the IRAS point sources, IRAS 04073+5102 and IRAS
04073+5100.
The cyan plus signs show the three brightest stars identified in the
optical bands by \citet{Chini1984} (CW1, CW2, and CW3; see details in
main text).
The magenta squares show the molecular-cloud cores identified from
millimeter-continuum emission \citep{Klein2005}. 
An enlarged view of the area around CW1 ($20\arcsec \times 20\arcsec$;
shown as a cyan box in the left panel) is shown in the small panel on
the lower right. The image is from MOIRCS $K_S$-band images, and the
orientation is the same as in the left panel.}
\label{fig:3col_S209_sd2}
 \end{center}
\end{figure}

First, we defined the cluster regions. We positioned many circles, each
with a radius of 50 pixels ($\sim$6$\arcsec$), in the S209 frame, moved
the centers in 1 pixel steps, and counted the numbers of stars included
in all the circles (4.1 $\pm$ 3.4 stars on average).
From these circles, we found two regions of localized high density that
contain maxima in the numbers of stars (84 for the northern and 110 for
the southern regions), and this enabled us to define the centers of the
clusters with an accuracy of $\sim$5$\arcsec$:
$(\alpha_{2000},\delta_{2000}) = (04^{\rm h}11^{\rm m}10.9^{\rm s},
+51\arcdeg09\arcmin56.5\arcsec)$ and $(\alpha_{2000}, \delta_{2000}) =
(04^{\rm h}11^{\rm m}03.7^{\rm s}, +51\arcdeg07\arcmin53.8\arcsec)$.
Figure~\ref{fig:profile_S209} shows the radial variation of the
projected stellar density using stars detected at 10$\sigma$ and above
($K_S \lesssim$20.5 mag).
The horizontal dashed gray lines represent the stellar density of a
region located more than 800 pixels from the main cluster and more than
400 pixels from the sub-cluster (hereafter, the ``background region'').
We defined each cluster as a circular region with a radius defined such
that the stellar density exceeds that of the entire sky frame by
3$\sigma$: {700 pixels (82$\arcsec$) for the main cluster and 350 pixels
(41$\arcsec$) for the sub-cluster.}
The main and sub-cluster regions defined here are shown in
Figure~\ref{fig:3col_S209_sd2} as solid and dotted yellow circles,
respectively. The estimated cluster radii of 82$\arcsec$ and 41$\arcsec$
correspond to 1.0 pc and 0.5 pc, respectively, at $D=2.5$ kpc and to 4.1
pc and 2.1 pc at $D=10$ kpc.

The main cluster corresponds to that identified as [BDS2003]65 in
\citet{Bica2003}, and the existence of the sub-cluster was suggested by
\citet{Klein2005} from an MSX source and faint NIR emission.
Both clusters are located near the region where the WISE band 3 (12
$\mu$m) emission is very strong (Figure~\ref{fig:3col_2MASS_WISE}); this
combination is often seen in clusters \citep{Koenig2012}.
Among the seven cores identified from millimeter-continuum emission by
\citet{Klein2005} (see Section~\ref{sec:SFinS209})---which are shown as
magenta squares in Figure~\ref{fig:3col_S209_sd2}---core 7 is closest to
the center of the cluster, with an offset of $\sim$20$\arcsec$, and
cores 1, 2, 3, and 4 surround the cluster.
In addition, the center of the sub-cluster is almost coincident with
core 5.

The cluster region probably should be identified by comparison with the
stellar density in the control field, rather than with that of the
background region in the S209 frame.
In Figure~\ref{fig:profile_S209}, the black horizontal solid lines
represent the density of stars in the control field.
They show that the stellar density in the control field is overall lower
than that in the background region. This may be due to the occurrence of
star-forming activities throughout the S209 frame, as suggested in
Figures~\ref{fig:3col_2MASS_WISE} and \ref{fig:S209_Ha}.
For the S209 frame and control frame, we counted the number of stars in
a circle of $r=40$ pixels in 20-pixel steps.
In the control field, the number is estimated to be $N = 2.8 \pm 2.1$. 
Based on the counts in the control field, the distribution of the counts
in the S209 frame is shown as dotted contours for 1$\sigma$ and
2$\sigma$, and solid contours for 3$\sigma$, 4$\sigma$,..., and
23$\sigma$ (Figure~\ref{fig:3col_S209_sd2}).
This figure also shows the high counts compared to the control field
over the entire S209 frame, except for a part of area in the north,
indicating the presence of star-forming regions over the entire frame.
The identified clusters generally cover a region of 3$\sigma$ or larger,
while there are regions with higher density than the control field over
a wider area, and the star-forming cluster may actually be a bit larger
than defined here.

A similar situation is seen in the well-known nearby star-forming region
IC 348, as suggested by, e.g., \citet{Muench2003}.
They defined the central region as the ``core'' and the surrounding
region as the ``halo.''
Using this nomenclature, we are discussing the star-forming activities
in the core region of S209 in the present paper.
As the main purpose of this paper is to derive the IMF more precisely in
a low-metallicity environment, we therefore limit the cluster regions
strictly as a first step.
This avoids defining cluster regions that are too large, which would
lead to a large age spread and the inclusion of many generations of
star-formation activity.
We have been working on multi-wavelength observations in a larger area
of S209 (cf., Izumi et al., in preparation, for radio observations using
the NRO 45 m radio telescope and the Very Large Array) to elucidate the
star-formation history in this region.
We will discuss the differences in the IMF at different locations in the
clusters along with these results in future work.

\subsection{Reddening Properties} \label{sec:reddening}

\begin{figure}
\begin{center}
  \vspace{10em}
  \hspace{-5em}
 \includegraphics[scale=0.45]{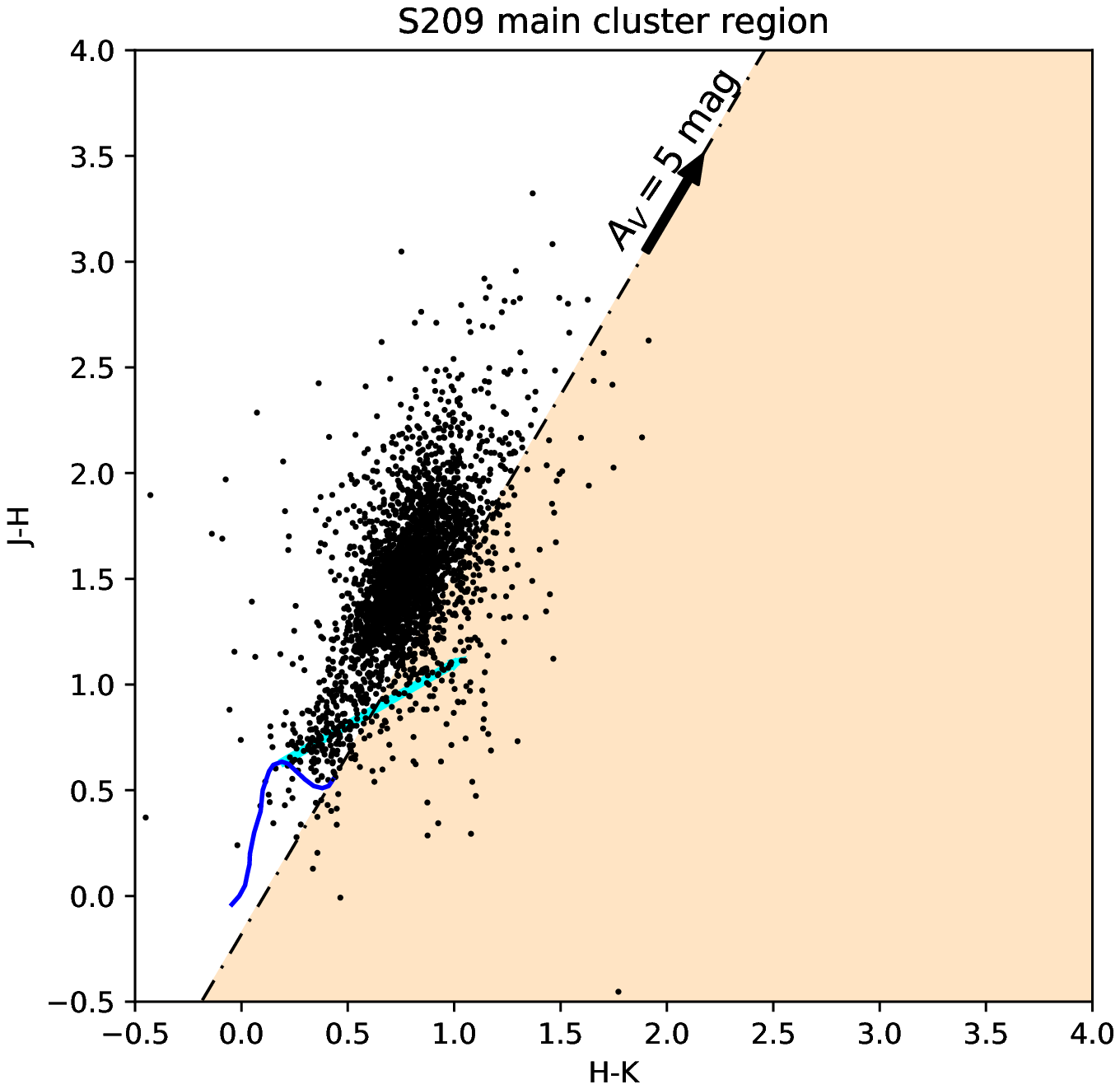}
 \includegraphics[scale=0.45]{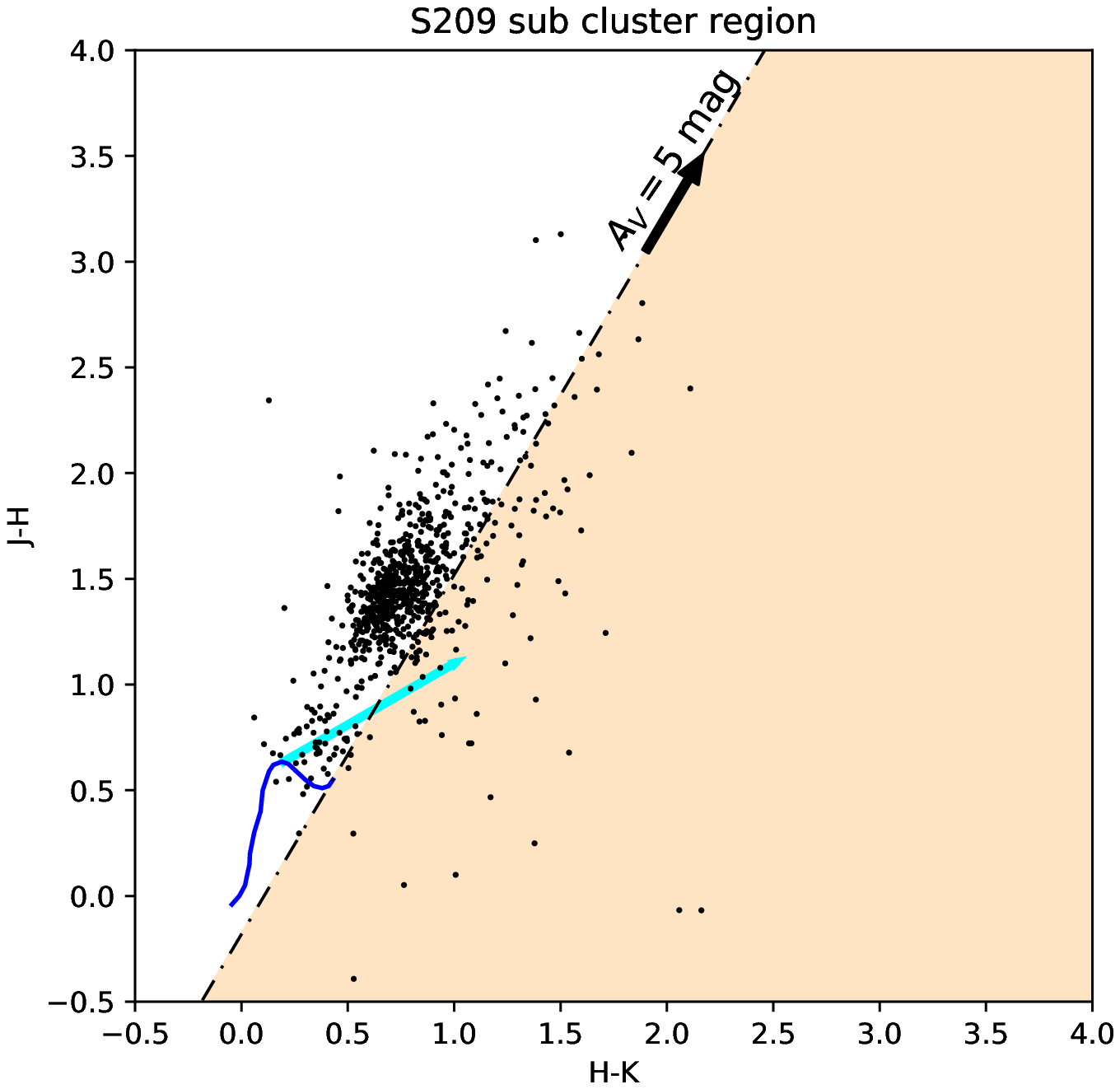}
  \vspace{-10em}
\caption{$(J-H)$ vs. $(H-K)$ color--color diagrams for the Sh 2-209 main
cluster region ({\it left}) and the sub-cluster region ({\it right}).
The blue curve in the lower left portion of each diagram is the locus of
points corresponding to unreddened main-sequence stars.
The dot--dashed line, which intersects the main-sequence curve at the
maximum {\it H}$-$$K_S$ values (the point on the curve corresponding to
spectral type M6) and which is parallel to the reddening vector,
represents the border between stars with and without circumstellar
disks.
The loci of classical T Tauri star (CTTS) are shown with cyan lines.}
\label{fig:colcol_S209}
\end{center}
\end{figure}

Stars in star-forming regions often exhibit both extinction due to the
interstellar medium and to the intra-cluster medium, as well as an
infrared color excess due to circumstellar media such as dust disks
\citep[e.g.,][]{Lada1992}.
Using the {\it JHK} color--color diagram, we estimate the extinction and
infrared excess of each source, and derive the distributions of
extinction and infrared excess for the S209
clusters.
We constructed $J-H$ versus $H-K_S$ color--color diagrams for stars in
the S209 main and sub-cluster regions in the left and right panels of
Figure~\ref{fig:colcol_S209}, respectively.
All sources that are detected at more than 10$\sigma$ in all {\it JHK}
bands are plotted.
The dwarf-star track from \citet{Yasui2008} for spectral types from late
B to M6 in the MKO system is shown as a blue curve in each panel.
The locus of classical T Tauri stars (CTTS), originally derived by
\citet{Meyer1997} in the CIT system, is shown as a cyan line in the MKO
system \citep{Yasui2008}.
The arrow shows the reddening vector for $A_V = 5$ mag.

\begin{figure}
\begin{center}
  \vspace{10em}
  \hspace{-5em}
 \includegraphics[scale=0.45]{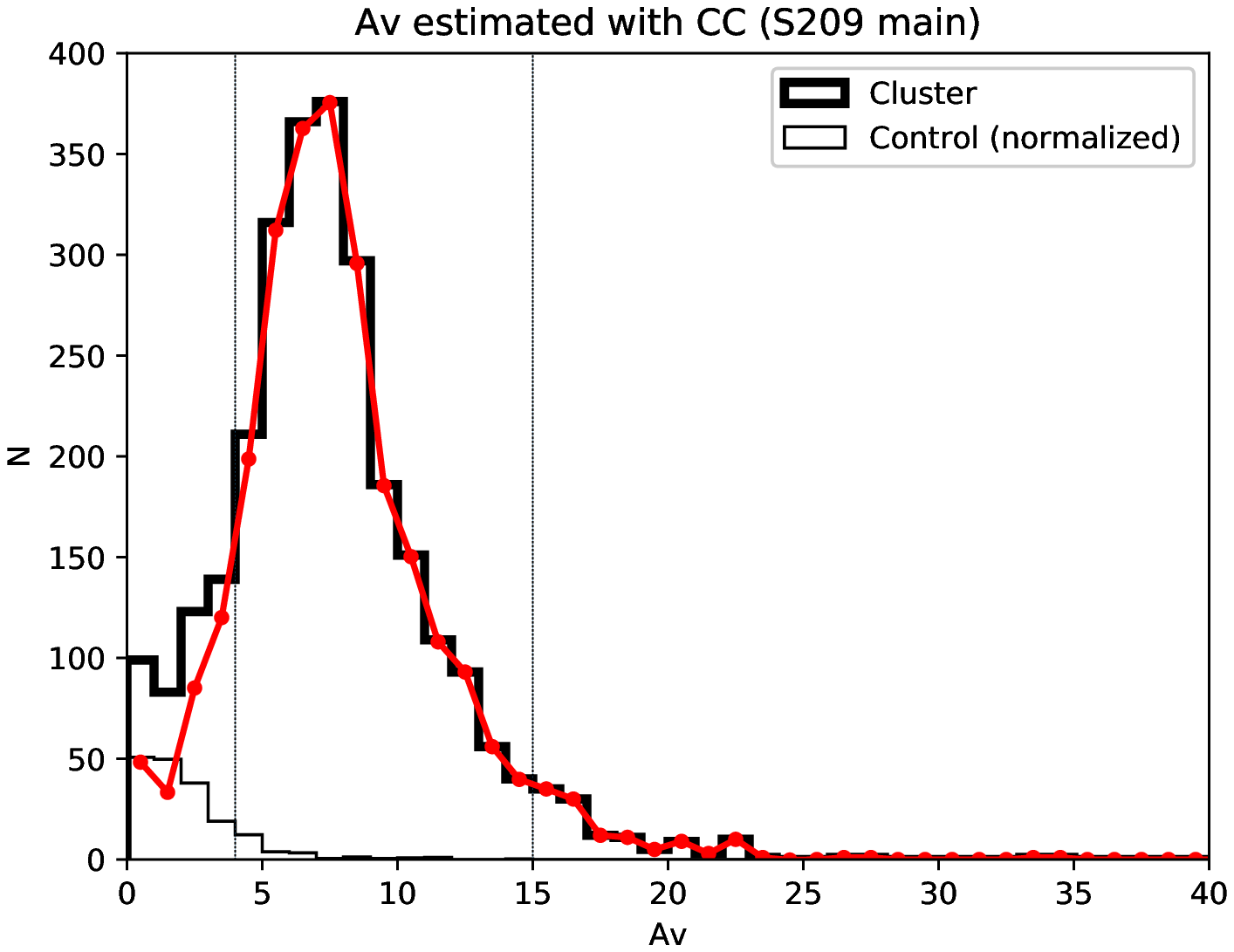}
 \includegraphics[scale=0.45]{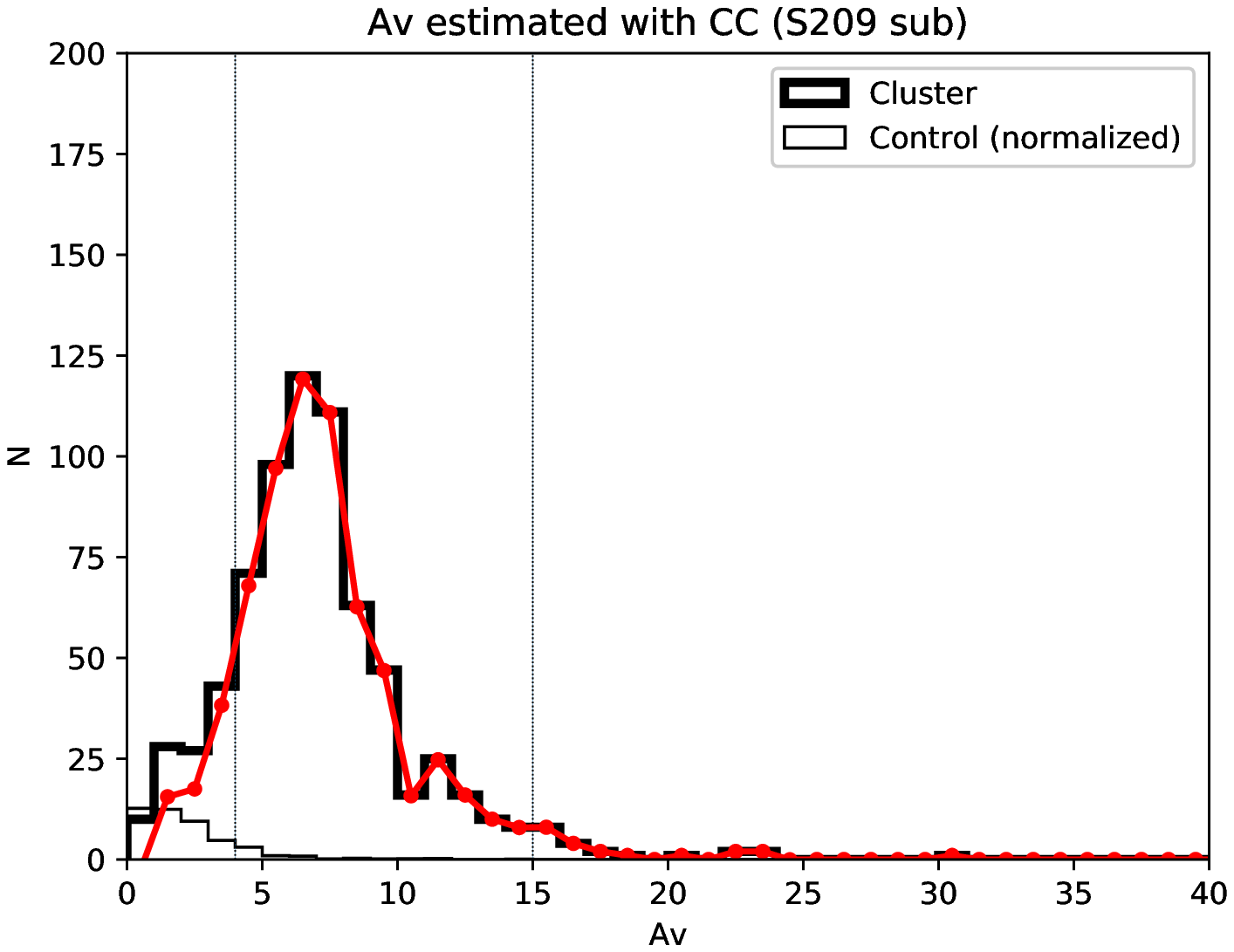}
  \vspace{-10em}
\caption{Distributions of $A_V$ for stars in the regions of the S209
main cluster (left panel) and of the sub-cluster (right panel).
The distributions for stars in the cluster regions are shown as thick
lines, while those in the control field are shown as thin lines. The
distributions for stars in the control field are normalized to match the
total area of each cluster region. The distributions obtained by
subtracting the normalized distribution for stars in the control field
from the distribution for stars in the cluster regions are shown as red
lines.}
\label{fig:Av_CC}
\end{center}
\end{figure}

\begin{figure}[!h]
 \begin{center}
  \vspace{10em}
  \hspace{-5em}
  \includegraphics[scale=0.45]{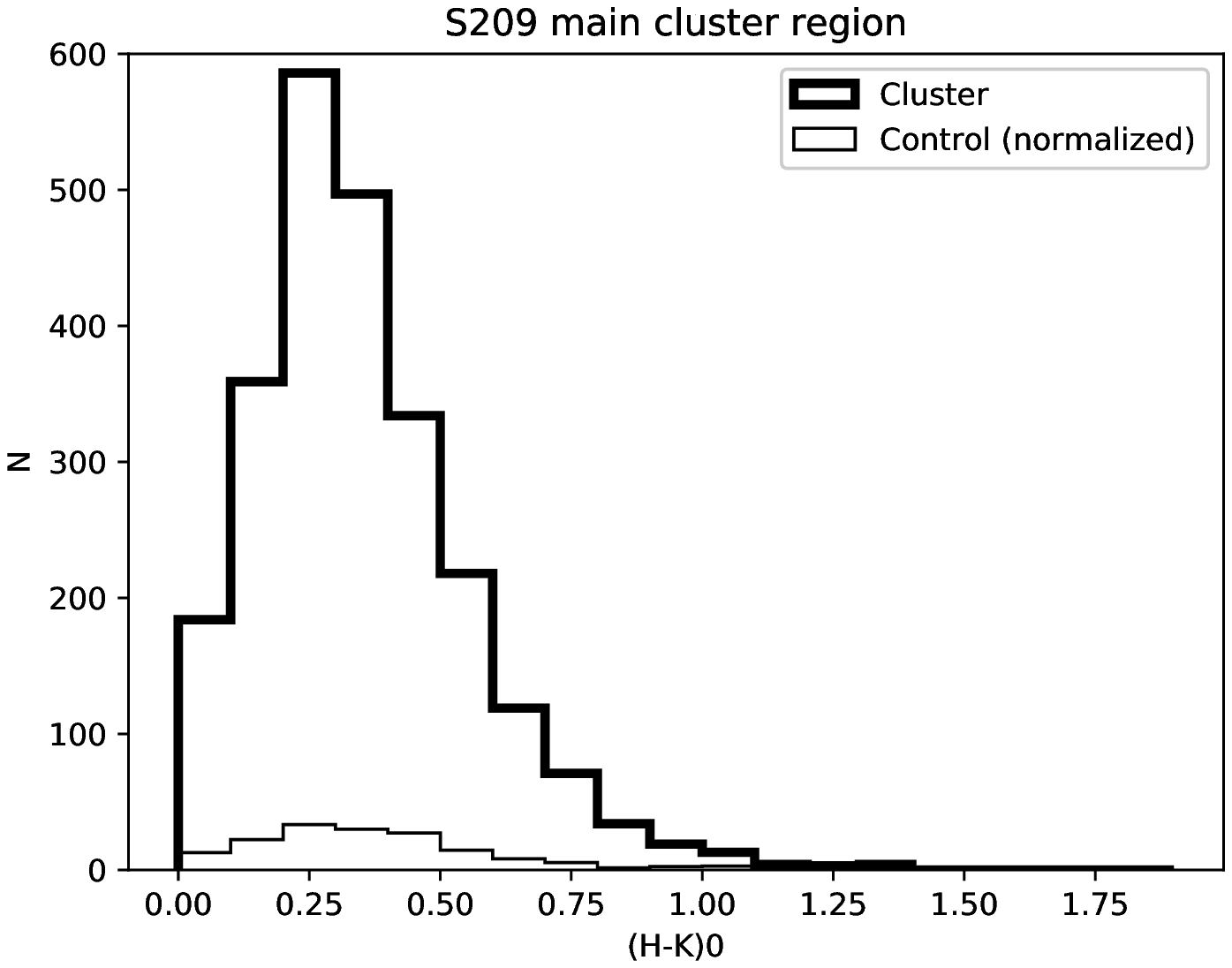}
  \includegraphics[scale=0.45]{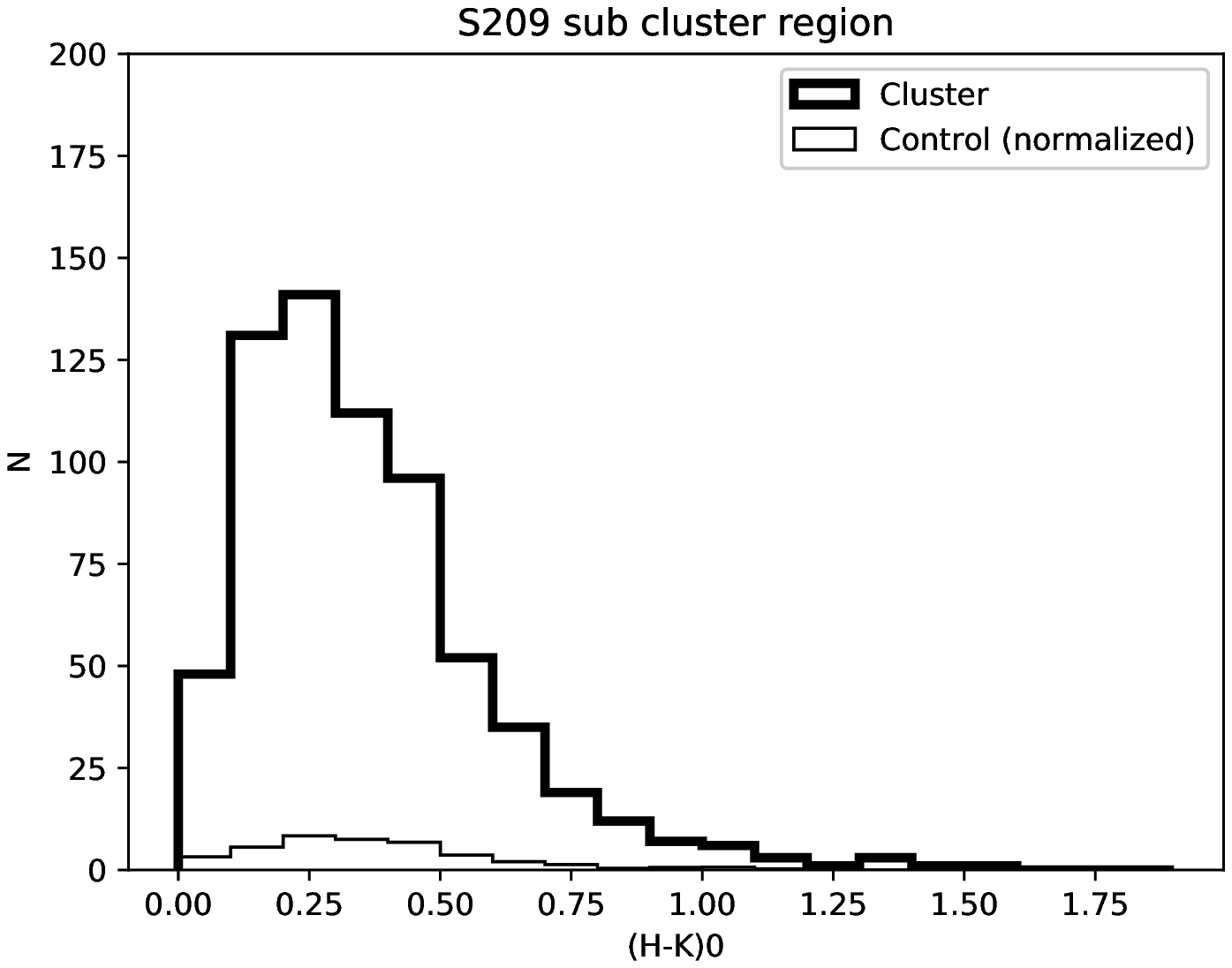}
  \vspace{-10em}
\caption{Distributions of $(H-K)_0$ for stars in the regions of the S209
main cluster (left panel) and of the sub-cluster (right panel). The
distributions for stars in the cluster regions are shown as thick lines,
while those in the control field are shown as thin lines. The
distributions for stars in the control field are normalized to match the
total area of each cluster region.}
\label{fig:HK0_CC}
\end{center}
\end{figure}

We derived the distributions of extinction and infrared excess for the
S209 clusters, which we use in Section~\ref{sec:Model} as probability
distribution functions for modeling.
First, we determined the extinction and the infrared excess for each
source in the S209 cluster regions using the JHK color--color diagram
following the procedure of \citet{Muench2002}.
We estimated the extinction ($A_V$) and the intrinsic $(H-K)$ color
[$(H - K)_0$] for each star by dereddening it along the reddening vector
to the young-star locus in the color--color diagram.
For convenience, we approximated the young-star locus by extending the
CTTS locus, and we used only those stars that are above the CTTS locus.
We derived the $A_V$ values based on the distance required to achieve
the obtained amount of dereddening with the reddening law of
\citet{Rieke1985}.
When each star is dereddened to the young-star locus on the color--color
diagram, we regard the $H-K$ value on the young-star locus as $(H-K)_0$.
The obtained distributions of $A_V$ and $(H - K)_0$ are shown as thick
lines in Figures~\ref{fig:Av_CC} and \ref{fig:HK0_CC}, respectively.

\begin{figure}
\begin{center}
  \vspace{10em}
  \hspace{-5em}
  \includegraphics[scale=0.45]{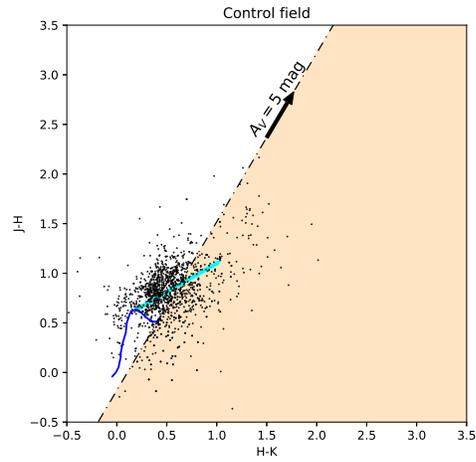}
  \vspace{-10em}
\caption{Same figure as Figure~\ref{fig:colcol_S209}, but for the
control field.}
\label{fig:colcol_CF}
\end{center}
\end{figure}

The obtained distributions include all sources in the cluster regions,
some of which are expected to be foreground or background stars.
To remove these effects, we also derived the distributions in
the control field. 
Using the JHK color--color diagram for stars in the control field, shown
in Figure~\ref{fig:colcol_CF}, we derived distributions of $A_V$ and $(H
- K)_0$.
The obtained distributions were normalized to match the total area of
each cluster region and are shown as thin lines in
Figures~\ref{fig:Av_CC} and \ref{fig:HK0_CC}.
Figure~\ref{fig:Av_CC} shows the distribution obtained by subtracting
the distribution for stars in the control field from the distribution
for stars in the cluster regions--- the red lines---which we consider to
be the distribution for cluster members.

The $A_V$ distribution for stars in the control field decreases
monotonically as the $A_V$ value increases, and it becomes almost zero
at $A_V = 5$ mag.
In contrast, the distributions for stars in both cluster regions
increase from $A_V = 0$ mag to 7-8 mag and then decrease, and they
become almost zero at $A_V = 20$ mag.
We therefore consider that most of the stars with large $A_V$ values
belong to the star-forming region.
The reason why stars in the cluster region have such large $A_V$ values
may be that they are subject to large extinction by the interstellar
medium due to the relatively large distance to S209.
In addition, the large dispersion of the $A_V$ values suggests that
star-forming molecular clouds still exist in this star-forming region.

The $(H-K)_0$ distributions for stars in both cluster regions and in the
control field increase from $(H-K)_0 = 0$ mag to 0.2--0.3 mag and then
decrease.
The distributions for the cluster stars extend up to $(H-K)_0 \simeq
1.5$ mag, whereas the distribution for field stars reaches zero around
$(H-K)_0 = 0.8$ mag.
The larger $(H-K)_0$ values for stars in the cluster region is due to
the color excess produced by circumstellar material around the young
stars.
Here, we assume that this excess is only a $K$-band excess ($K$-excess).
We estimate the average value of $(H-K)_0$ for stars in the control
field to be 0.3 mag, and---for simplicity---we use this as a typical
$(H-K)_0$ value for stars without circumstellar material
[$(H-K)_{0, {\rm crit}} = 0.3$].

\begin{figure}[!h]
 \begin{center}
  \vspace{10em}
  \hspace{-5em}
 \includegraphics[scale=0.45]{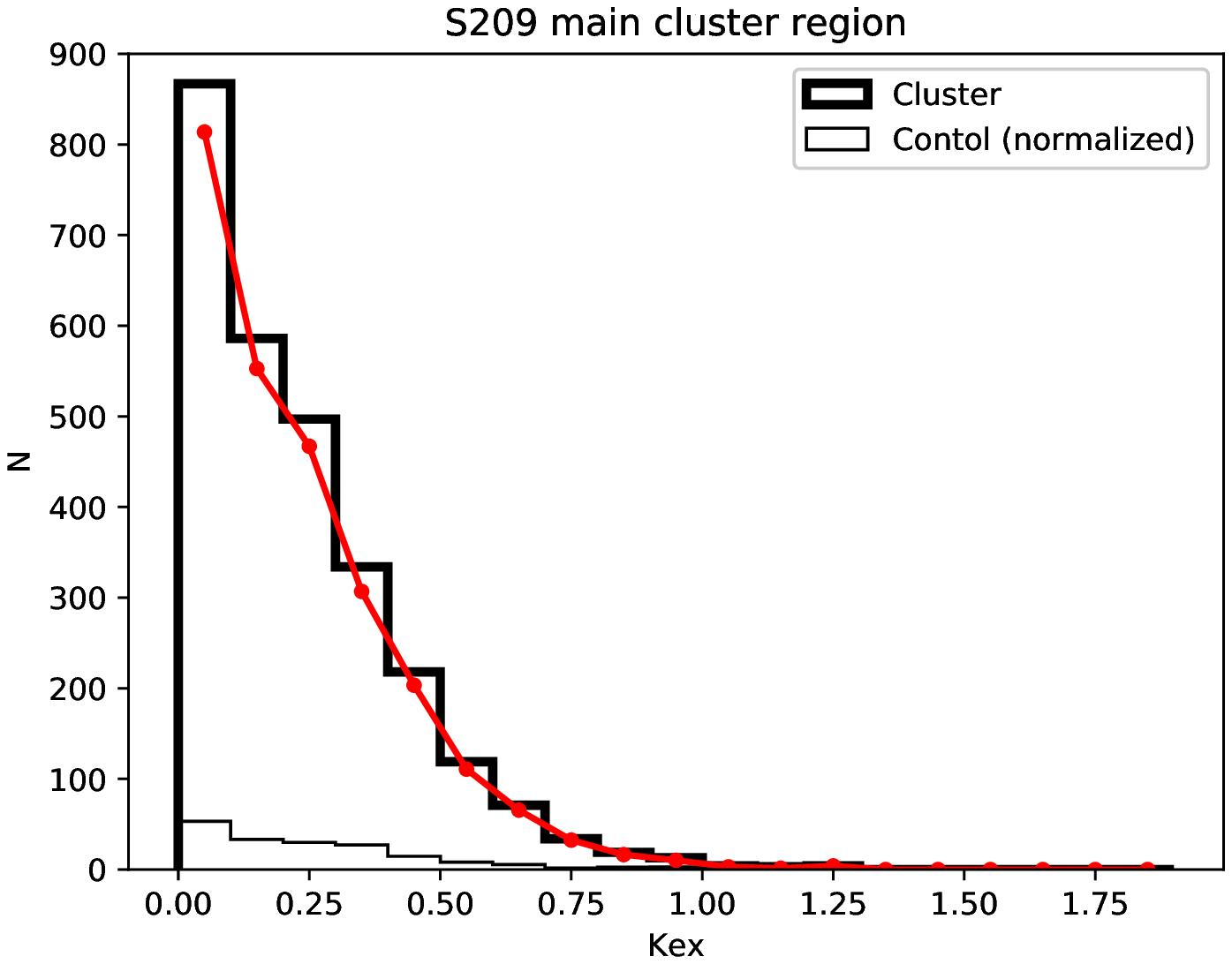}
 \includegraphics[scale=0.45]{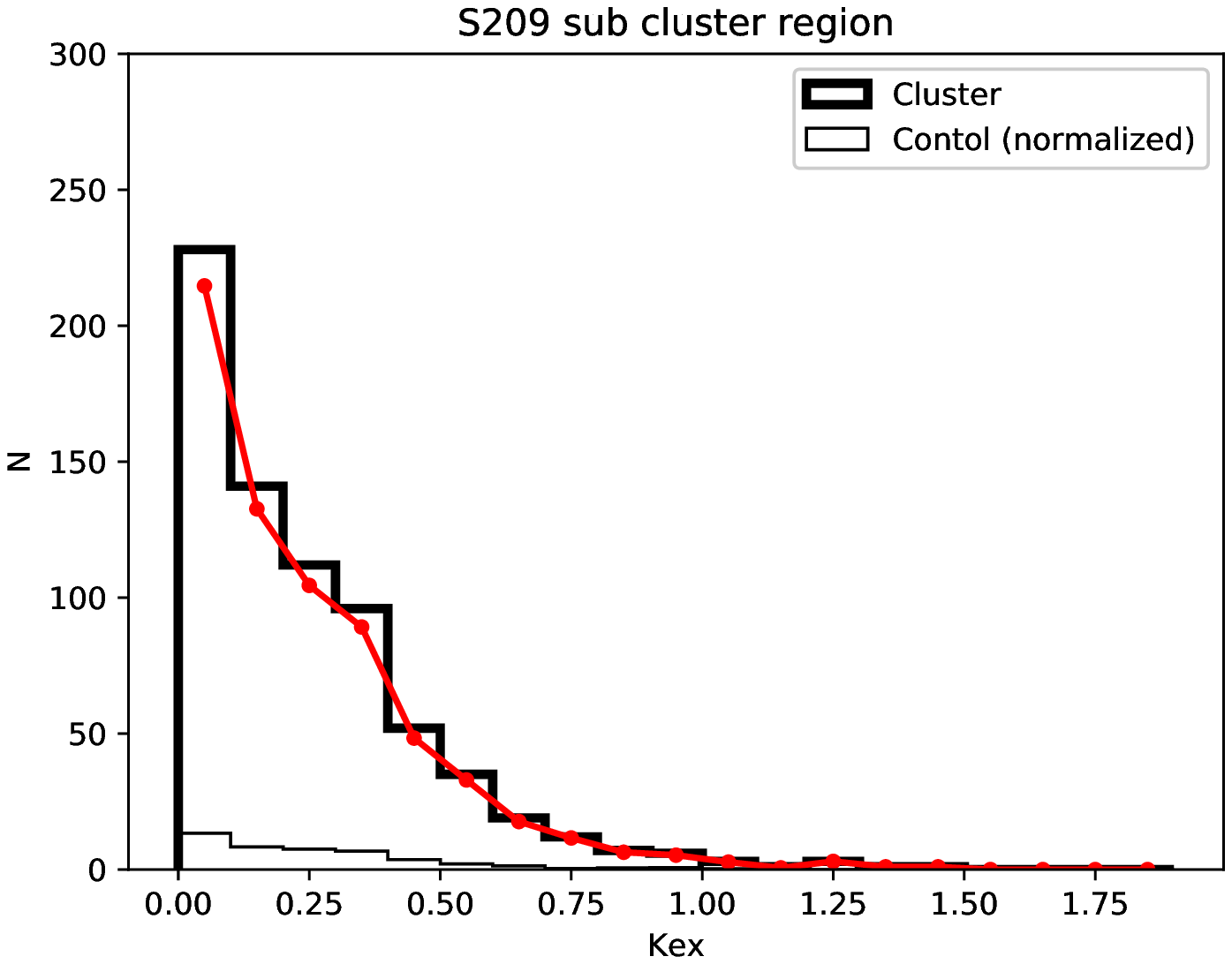}
  \vspace{-10em}
\caption{$K$-excess distributions for stars in the regions of the S209
 main cluster (left panel) and of the sub-cluster (right panel).
The distributions for stars in the cluster regions are shown as thick
lines, while those in the control field are shown as thin lines.
The distributions for stars in the control field are normalized to match
the total area of each cluster region. The distributions obtained by
subtracting the normalized distribution for stars in the control field
from the distribution for stars in each cluster region are shown as red
lines.}
\label{fig:Kex_S209}
\end{center}
\end{figure}

We consider stars with $(H-K)_0 \ge (H-K)_{0, {\rm crit}}$ to be stars
with a $K$-excess, which we define as $(H-K)_0$ minus $(H-K)_{0, {\rm
crit}}$, while stars with $(H-K)_0 \le (H-K)_{0, {\rm crit}}$ are
considered to be stars without a $K$-excess.
The resulting distributions of $K$-excess are shown in
Figure~\ref{fig:Kex_S209}.
The distributions for stars in the S209 main and sub-cluster regions are
shown in the left and right panels, respectively.
{As in Figures~\ref{fig:Av_CC} and \ref{fig:HK0_CC},} distributions for
stars in the cluster regions are shown with thick lines, while those for
stars in the control field are shown with thin lines.
The subtracted distributions are shown with red lines.
In a later section, we use the subtracted distributions of stars for the
S209 main and sub-clusters to construct model KLFs when evaluating the
IMF.

\subsection{Mass-$A_V$--limited Samples} \label{sec:sample_S209}
In this section, we extract Mass-$A_V$--limited samples from the
color--magnitude diagram for each cluster.
In our previous papers about young star-forming clusters in
low-metallicity environments in the Galaxy, we used the $J-K$ vs. $K$
color--magnitude diagram to select members of star-forming clusters
\citep[e.g.,][]{Yasui2021}.
The main goal of that paper was to obtain rough estimates of the cluster
IMF and age and to identify stars with protoplanetary disks.
In contrast, the aim of the present paper is to derive a precise IMF
down to the regime of substellar masses.
Because stars in star-forming regions usually have large
extinctions---especially for targets at large distances---the longer NIR
wavelengths are more effective for deep detections \citep{Muench2002}.
Here, we therefore used the longer-wavelength $H - K$ vs. $K$
color--magnitude diagram to select star-forming cluster members.

\begin{figure}[!h]
\begin{center}
  \vspace{10em}
  \hspace{-5em}
 \includegraphics[scale=0.45]{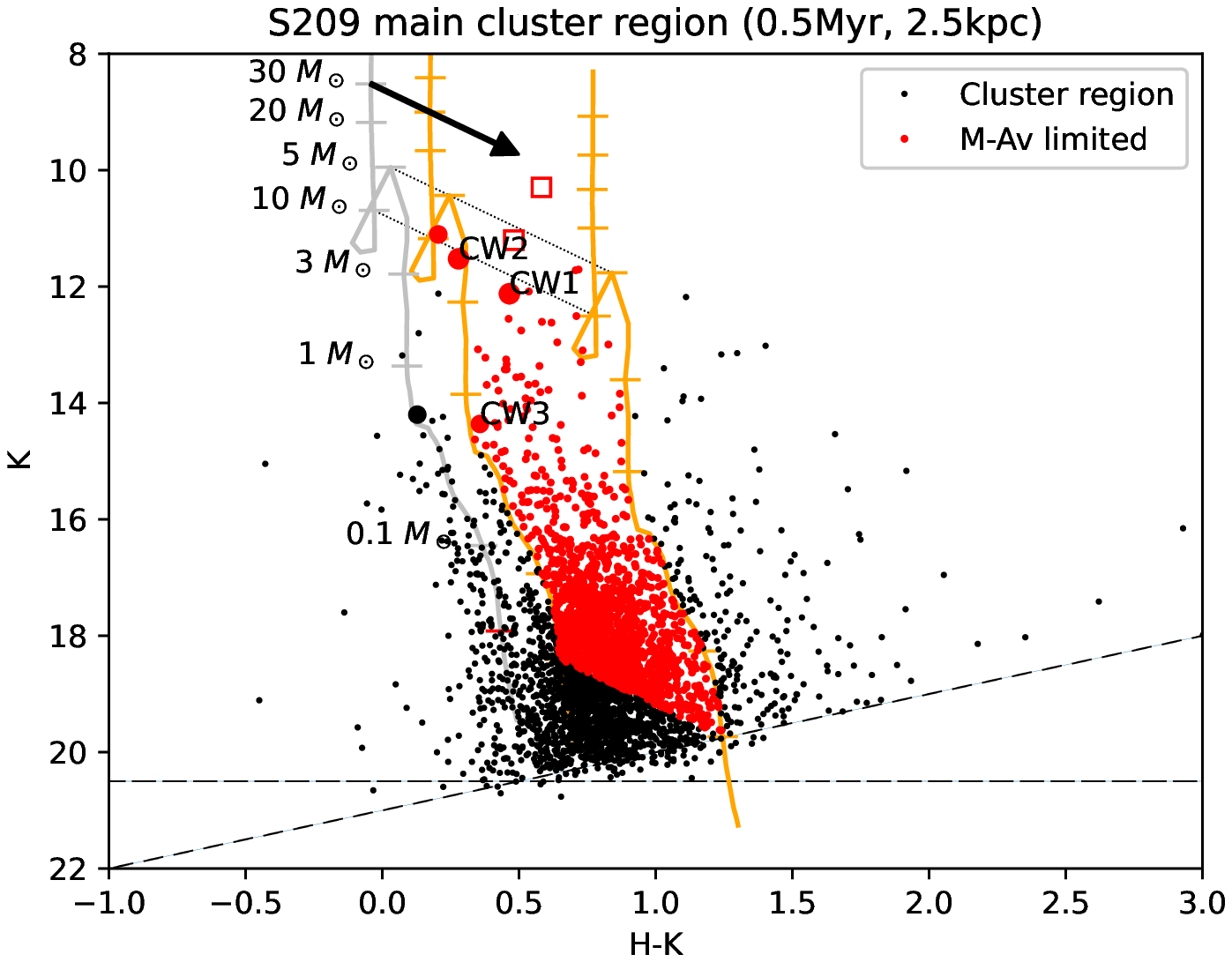}
 \includegraphics[scale=0.45]{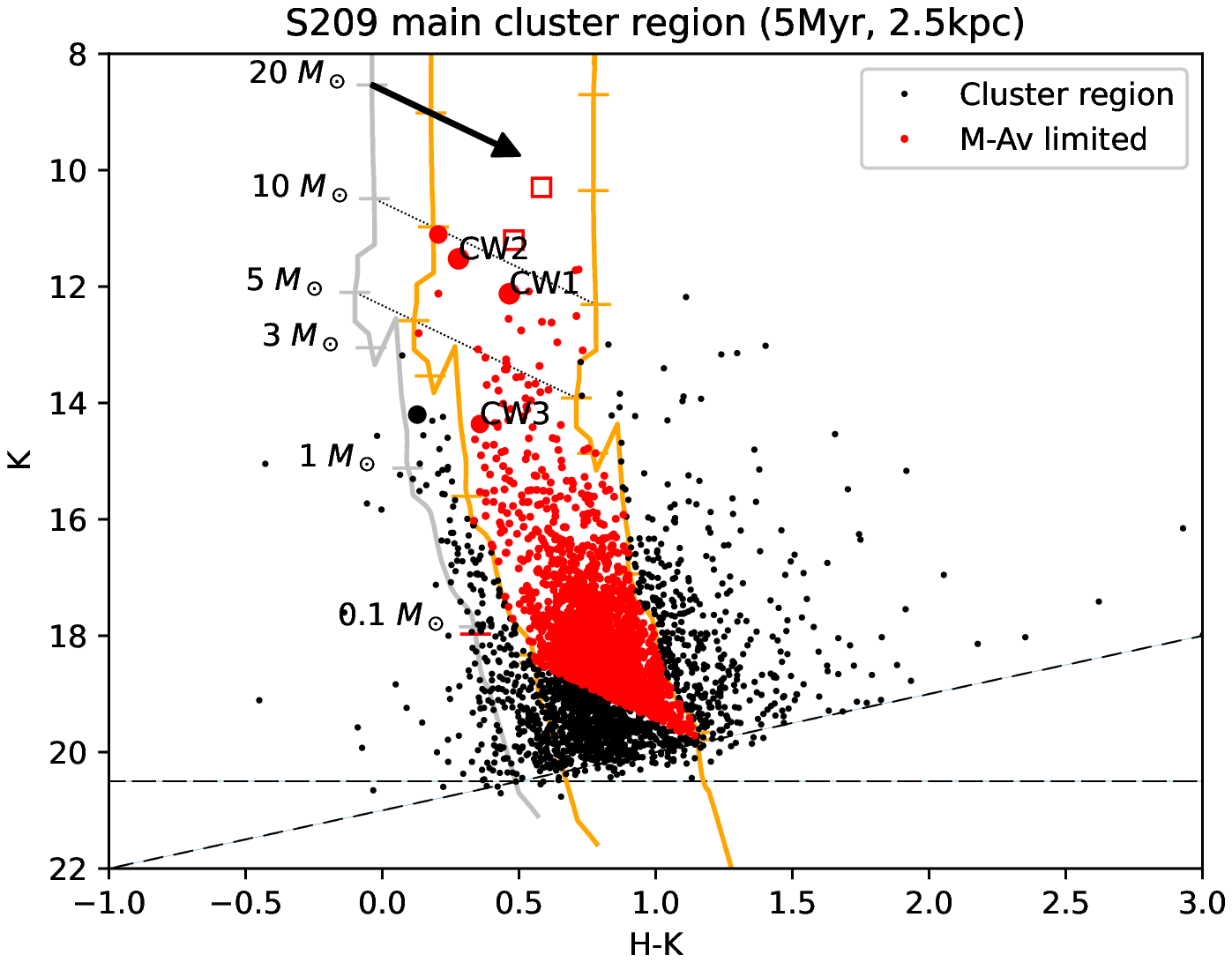}

 \hspace{-5em}
 \includegraphics[scale=0.45]{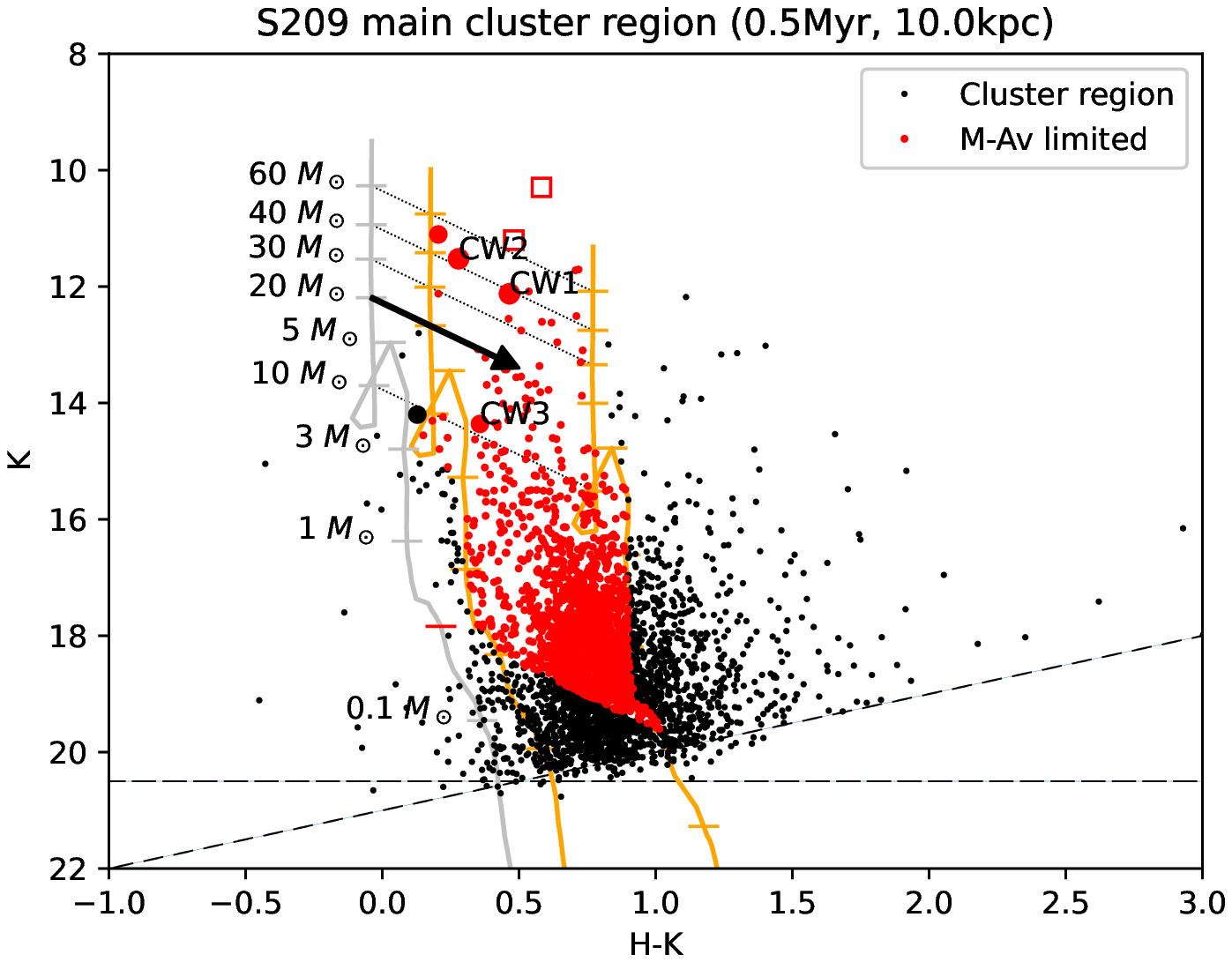}
 \includegraphics[scale=0.45]{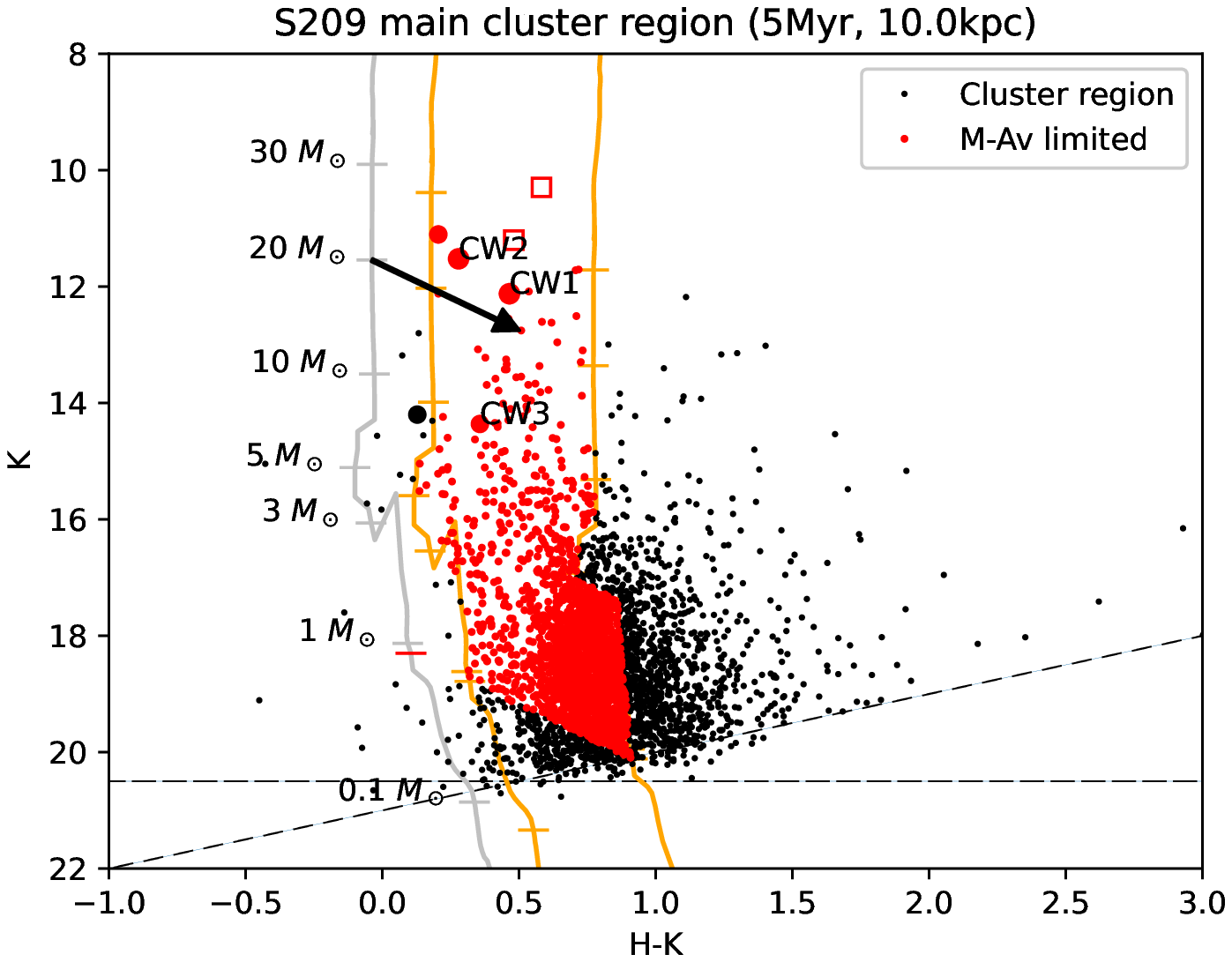}
  \vspace{-10em}
\caption{($H-K_S$) vs. $K_S$ color--magnitude diagram for the S209 main
cluster.
Only point sources that are in the cluster region and are detected with
more than 10 $\sigma$ in both the H and $K_S$ bands are plotted.
The gray lines show isochrone models from \citet{Lejeune2001} for masses
$M/M_\odot > 7$; from \citet{Siess2000} for the mass range $3 <
 M/M_\odot \le 7$; and from \citet{D'Antona1997} for the mass range
$0.017 \le M/M_\odot \le 3$.
A distance of 2.5 kpc (10.0 kpc) is assumed in the top (bottom) panels,
while an age of 0.5 Myr (5.0 Myr) is assumed in the left (right) panels.
The black arrows show the reddening vectors for $A_V=10$ mag from the
isochrone models.
Stars that are in the cluster region on the sky and are located between
the orange lines on the color--magnitude diagram ($4 \le A_V \le 15$
mag), are identified as cluster members.
Identified cluster members are shown as red dots, while sources that are
located in the cluster region but which are not considered to be cluster
members are plotted with black dots.
The dashed lines show the 10$\sigma$ limiting magnitudes.
The stars in Table~\ref{tab:gaia}, for which astrometric distances were
derived using Gaia EDR3, are shown as large circles, while the stars for
which saturation was observed in the MOIRCS images are shown as open
squares using 2MASS magnitudes.
For these sources as well, those within the $A_V$ range containing
identified cluster members are shown in red, and those outside this
range are shown in black.}
\label{fig:CM_S209}
\end{center}
\end{figure}

\begin{figure}[!h]
\begin{center}
  \vspace{10em}
  \hspace{-5em}
  \includegraphics[scale=0.45]{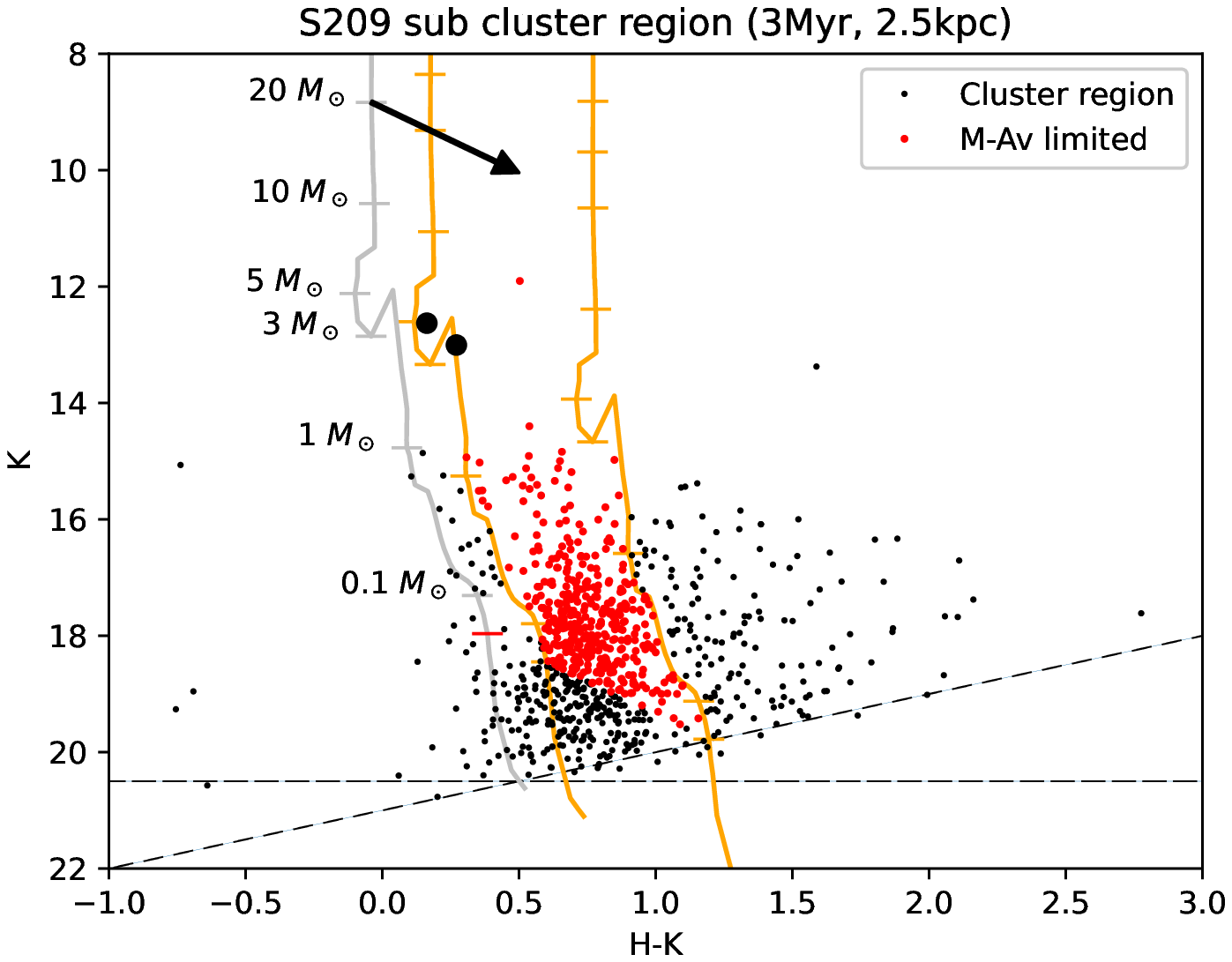}
  \includegraphics[scale=0.45]{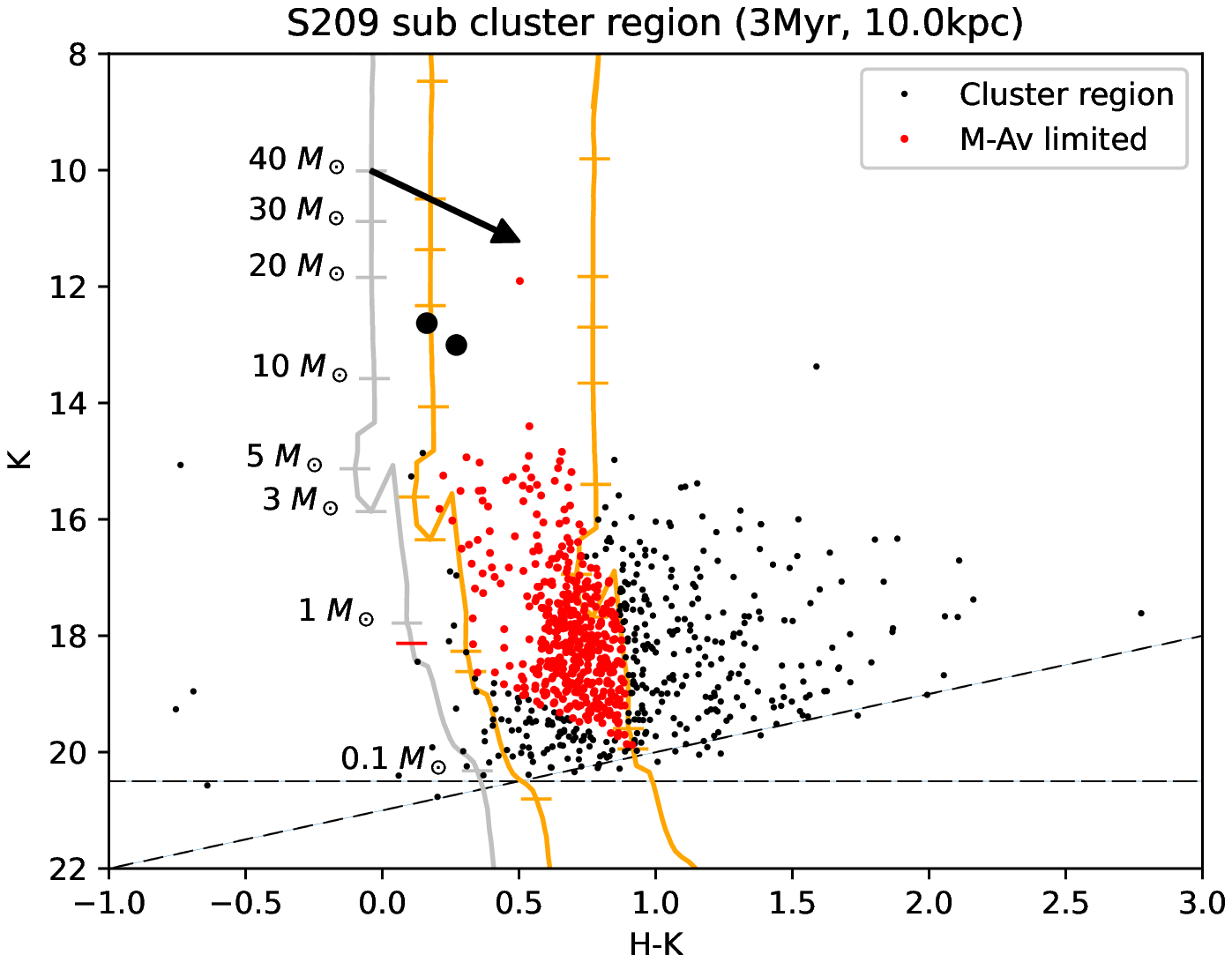}
  \vspace{-10em}
\caption{Same figure as Figure~\ref{fig:CM_S209}, but for the S209
sub-cluster. Both the left and right panels assume an age of 3 Myr,
while the distance is assumed to be 2.5 kpc (10 kpc) in the left (right)
panel.}
\label{fig:CM_S209sub}
\end{center}
\end{figure}

We constructed the $H - K_S$ versus $K_S$ color--magnitude diagrams for
the point sources detected in the S209 cluster regions
(Figures~\ref{fig:CM_S209} and \ref{fig:CM_S209sub} for the main and
sub-cluster regions, respectively).
All point sources that are detected with more than 10$\sigma$ in both
the {\it H} and $K_S$ bands are shown.
The dashed lines mark the 10$\sigma$ limiting magnitudes.
The CW sources (CW1, CW2, and CW3) are shown as filled circles labeled
with their names.
Stars that are very bright and that saturated in the MOIRCS images are
also plotted in this figure, shown with open squares.

The gray lines in these figures show isochrone tracks: from
\citet{Lejeune2001} for masses $M/M_\odot > 7$; from \citet{Siess2000}
for the mass range $3 < M/M_\odot \le 7$; and from
\citet{{D'Antona1997},{D'Antona1998}} for the mass range $0.017 \le
M/M_\odot \le 3$.
Because the distances and ages remain to be determined for the two S209
clusters, we assumed the distance to be either 2.5 kpc, which is the
astrometric distance from Gaia EDR3, or 10 kpc, which is the
kinematic/photometric distance (see Section~\ref{sec:S209}).
Also, because the ages of these clusters are not yet known, for the S209
main cluster we used here isochrone tracks for the ages 0.5 Myr and 5
Myr as examples.
These ages correspond to the youngest and oldest extreme cases for which
young stars still have disks traced with $K$-band excess
\citep{{Lada1999}, {Yasui2010}}.
For the sub-cluster, we used the intermediate age of 3 Myr.
The tickmarks on the isochrone models, which are shown in the same
colors as the isochrone tracks, correspond to the positions of stellar
masses 0.1, 1, 3, 5, 10, 20, 30, 40, and 60 $M_\odot$.
The arrows show the reddening vector for $A_V = 10$ mag.

Here, we define {the Mass-$A_V$--limited sample} for each assumed
distance and age.
To limit the extinction range, we note that Figure~\ref{fig:Av_CC} shows
that the number of stars in the control field is large when $A_V$ is
smaller, so more contamination is expected to exist in both cluster
regions in this $A_V$ range.
Therefore, we set the minimum value of $A_V$ to be 4 mag in selecting
the $A_V$--limited samples, so that the number of stars in the cluster
region is significantly larger than that in the control field.
On the large-$A_V$ side, for $A_V \gtrsim 10$ mag there are almost no
stars in the control field.
Therefore, the larger the value of $A_V$, the more stars can be
extracted in the cluster field.
However, if the extinction range is set to a very large value, the
mass-detection limit becomes large, even though the number of stars in
the cluster is not very large.
To take smaller masses into account in deriving the IMF, the upper limit
of the $A_V$ range must therefore be a reasonable value.
In this case, we set $A_V = 15$ mag, which allows the mass-detection
limit to extend down to $\lesssim$0.1 $M_\odot$ for ages {$\le$5} Myr at
the distance of 2.5 kpc (0.04 $M_\odot$ for 0.5 Myr and {0.09} $M_\odot$
for 5 Myr),
while it only reaches $\lesssim$1.0 $M_\odot$ for the distance of 10 kpc
(0.25 $M_\odot$ for 0.5 Myr and 1.0 $M_\odot$ for 5 Myr).

Similarly to the lower-mass stars, the estimated masses of higher-mass
stars vary greatly with the assumed age and distance.
At a distance of 2.5 kpc and an age of 0.5 Myr, the masses of CW1 and
CW2 are both estimated to be $\sim$10 $M_\odot$ and that of CW3 is
$\lesssim$1 $M_\odot$.
In contrast, at the same distance but an age of 5 Myr, the masses of CW1
and CW2 are again estimated to be $\sim$10 $M_\odot$ while that of CW3
becomes $\sim$3 $M_\odot$.
Conversely, at the distance of 10 kpc and an age of 0.5 Myr, the masses
of CW1 and CW2 are estimated to be $\sim$40 $M_\odot$ while that of CW3
is $\lesssim$10 $M_\odot$.
Finally, for the same 10 kpc distance but an age of 5 Myr, the masses of
CW1 and CW2 are estimated to be $\sim$20--30 $M_\odot$ while that of CW3
becomes $\sim$10 $M_\odot$. We discuss this issue further in
Section~\ref{sec:gaia_moircs}.

\begin{figure}[!h]
\begin{center}
  \vspace{10em}
  \hspace{-5em}
  \includegraphics[scale=0.45]{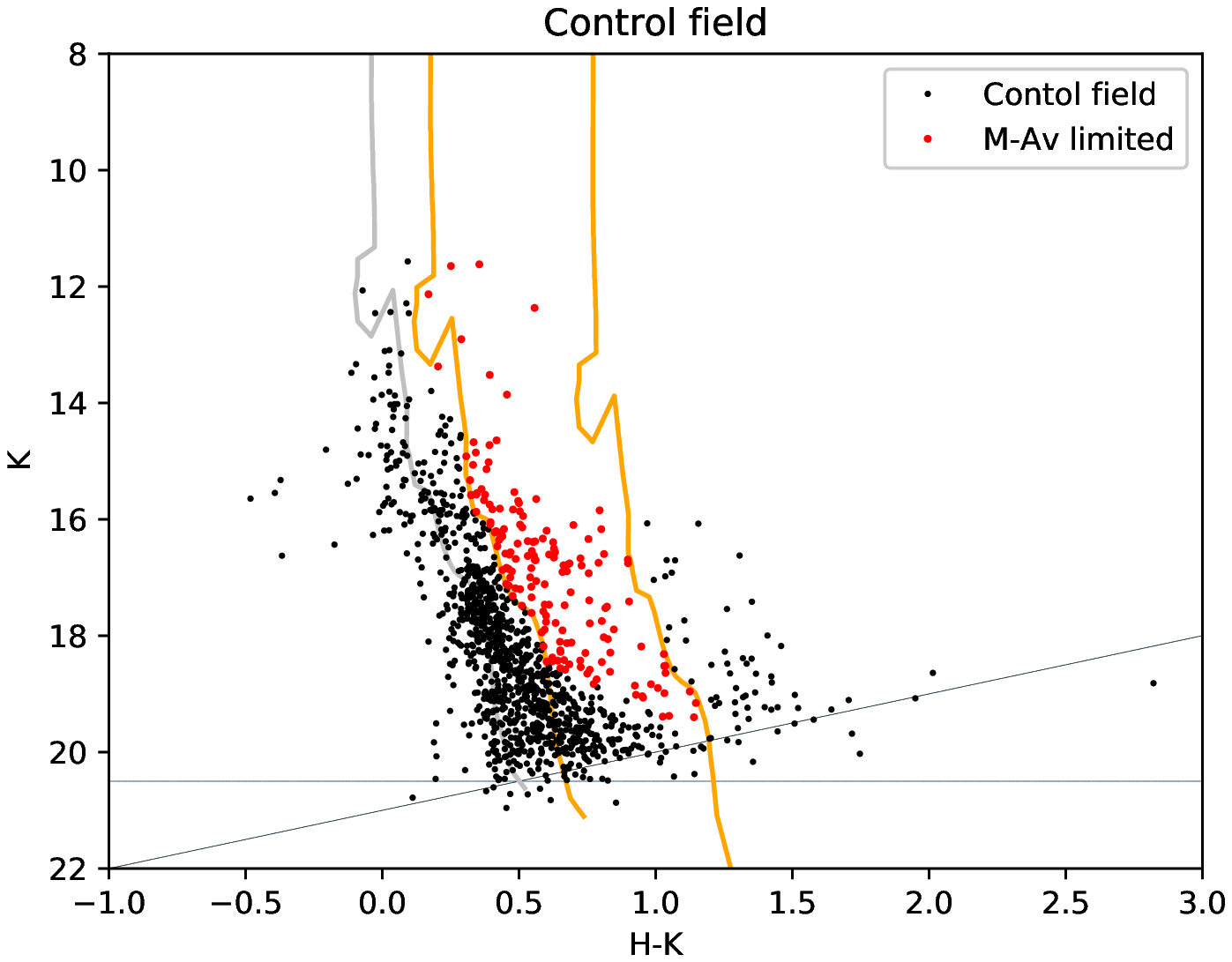}
  \includegraphics[scale=0.45]{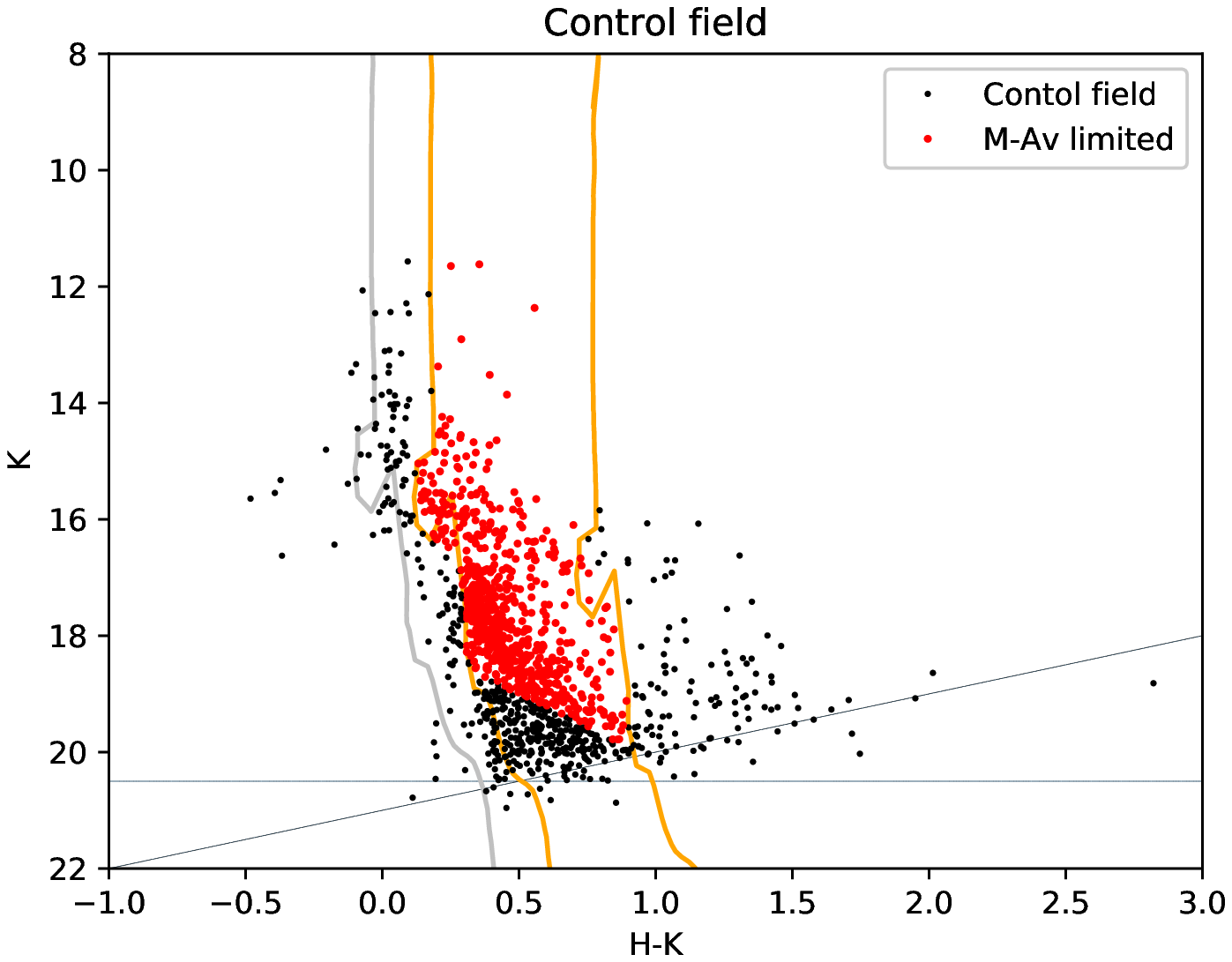}
  \vspace{-10em}
\caption{Same figure as Figure~\ref{fig:CM_S209}, but for the control
field. For reference, isochrone tracks for an age of 3 Myr and a
distance of 2.5 kpc (10 kpc) are shown in the left (right) panel.}
\label{fig:CM_S209_CF}
\end{center}
\end{figure}

In Figures~\ref{fig:CM_S209} and \ref{fig:CM_S209sub}, {the
Mass-$A_V$--limited samples} selected here are shown in red, while
others are shown in black.
Because the samples obtained here do not completely exclude foreground
(and sometimes background) sources, we obtained a
pseudo-Mass-$A_V$--limited sample for the control field using the same
method as for the cluster regions (Figure~\ref{fig:CM_S209_CF}).
By subtracting the number of pseudo-sources obtained in the control
field---which we normalized to match the total area of each cluster
region---from the number of sources in each cluster region, we estimate
the final numbers of sources for the S209 main and sub-clusters to be
$\simeq$1500 and 350, respectively.

\subsection{Cluster KLFs} \label{sec:KLF_S209} 

\begin{figure}[!h]
\begin{center}
  \vspace{10em}
  \hspace{-5em}
 \includegraphics[scale=0.45]{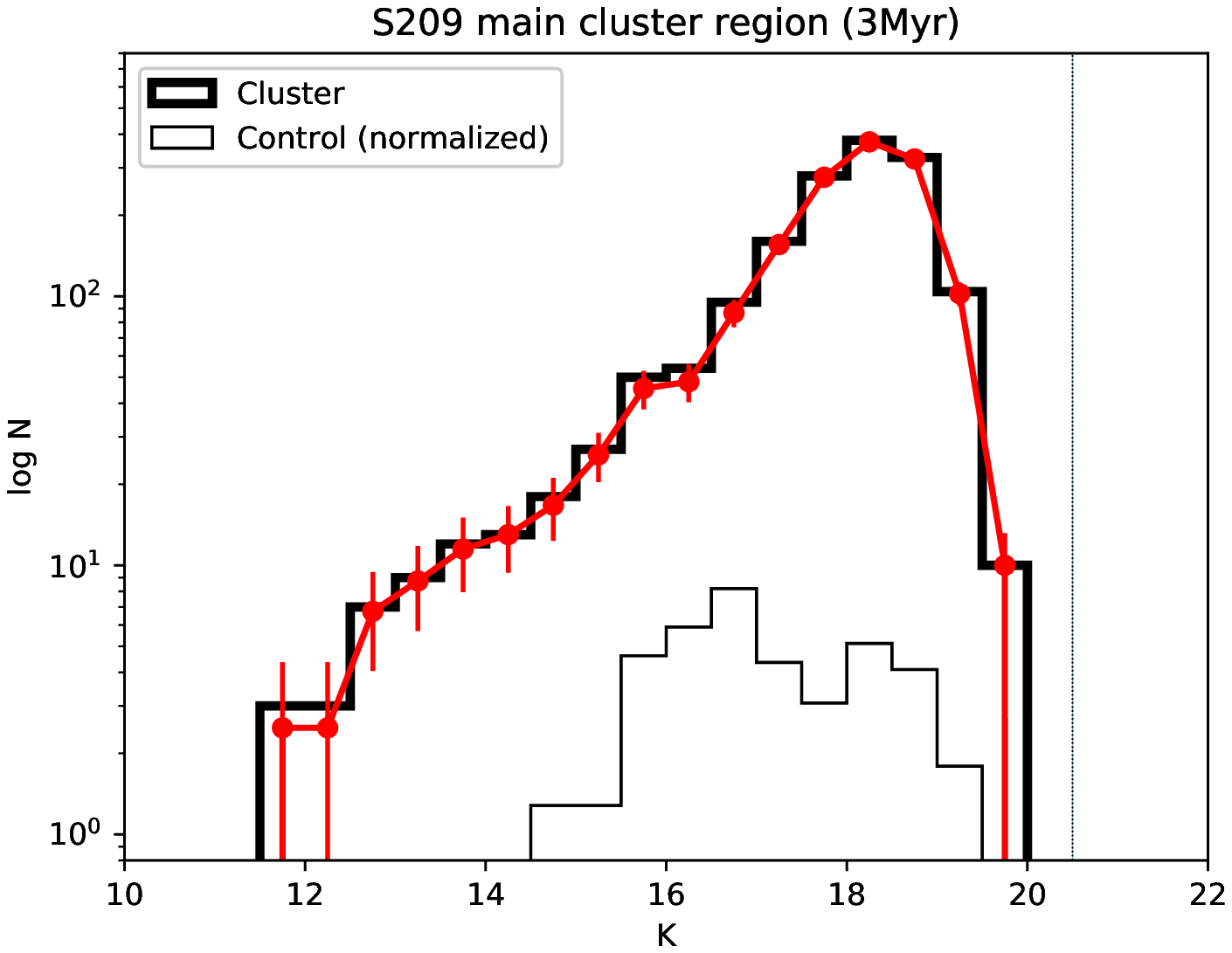}
 \includegraphics[scale=0.45]{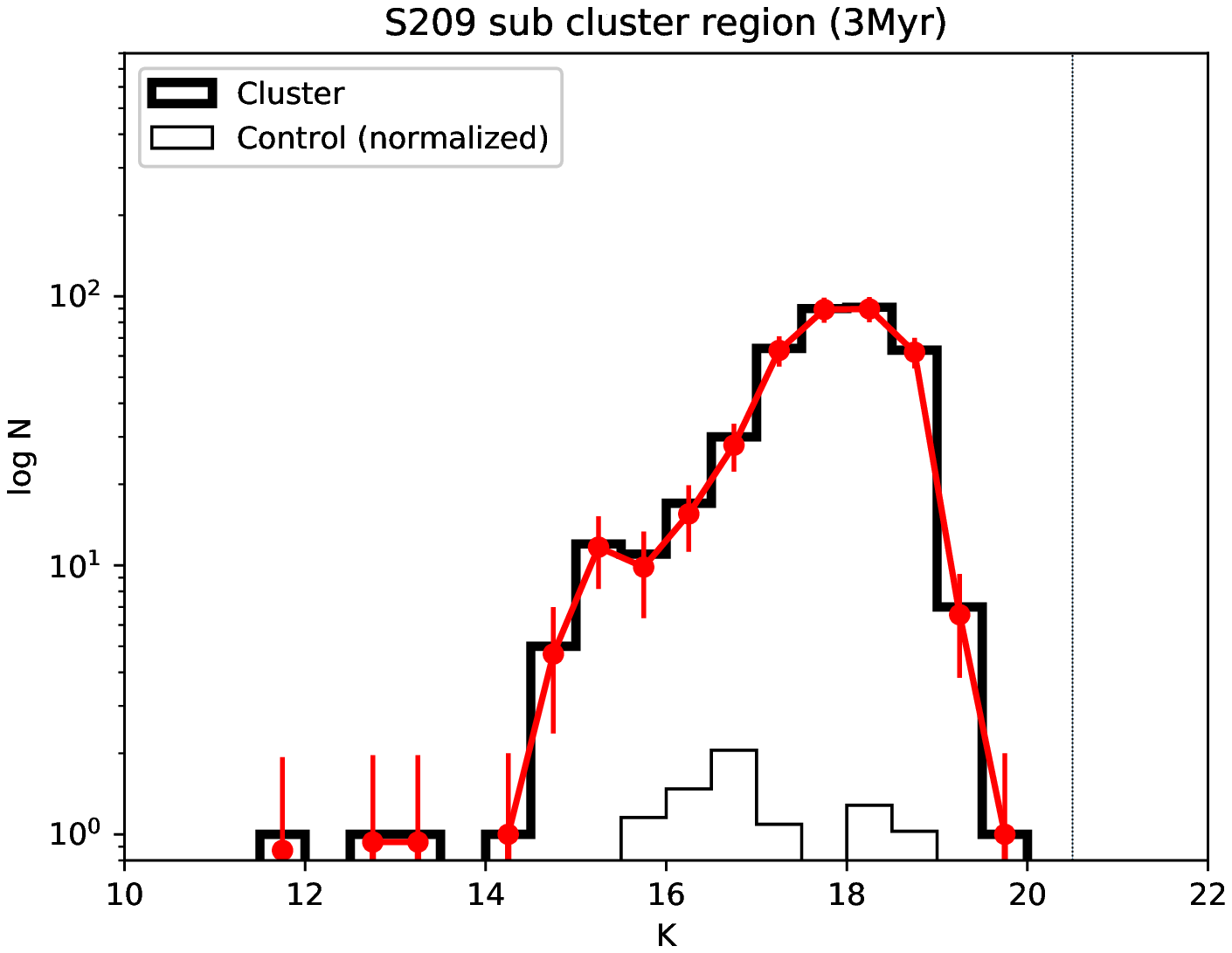}
  \vspace{-10em}
\caption{Examples of cluster KLFs for the S209 main cluster (left) and
sub-cluster (right). Both examples are for an age of 3 Myr and a
distance of 2.5 kpc.}
\label{fig:obsKLF}
\end{center}
\end{figure}

We constructed the KLF for each cluster using the Mass-$A_V$--limited
samples extracted in Section~\ref{sec:sample_S209}, assuming various
ages (0.5--10 Myr) and distances (2.5 and 10 kpc).
Figure~\ref{fig:obsKLF} shows an example for an age of 3 Myr and a
distance of 2.5 kpc.
In this figure, the KLFs for sources in the cluster regions (the cluster
region KLFs) are shown with thick black lines and those for sources in
the control field (the control field KLF) with thin lines (the main and
sub-clusters are shown in the left and right panels, respectively).
The control field KLF is normalized to match the total area of each
cluster region.
We subtracted the normalized counts from the control-field KLF from the
counts for each cluster region KLF to obtain the cluster KLFs, which are
shown as thick red lines in Figure~\ref{fig:obsKLF}.
For the distances and ages assumed here, the KLFs for the S209 clusters
increase monotonically up to {$K \sim 18$ mag} for both clusters, peak
in the {18.0--18.5 mag} bins for the main cluster and in the {17.5--18.5
mag} bins for the sub-cluster, and then decrease at still larger {\it K}
magnitudes.

\begin{figure}[h]
\begin{center}
  \vspace{10em}
  \hspace{-5em}
  \includegraphics[scale=0.45]{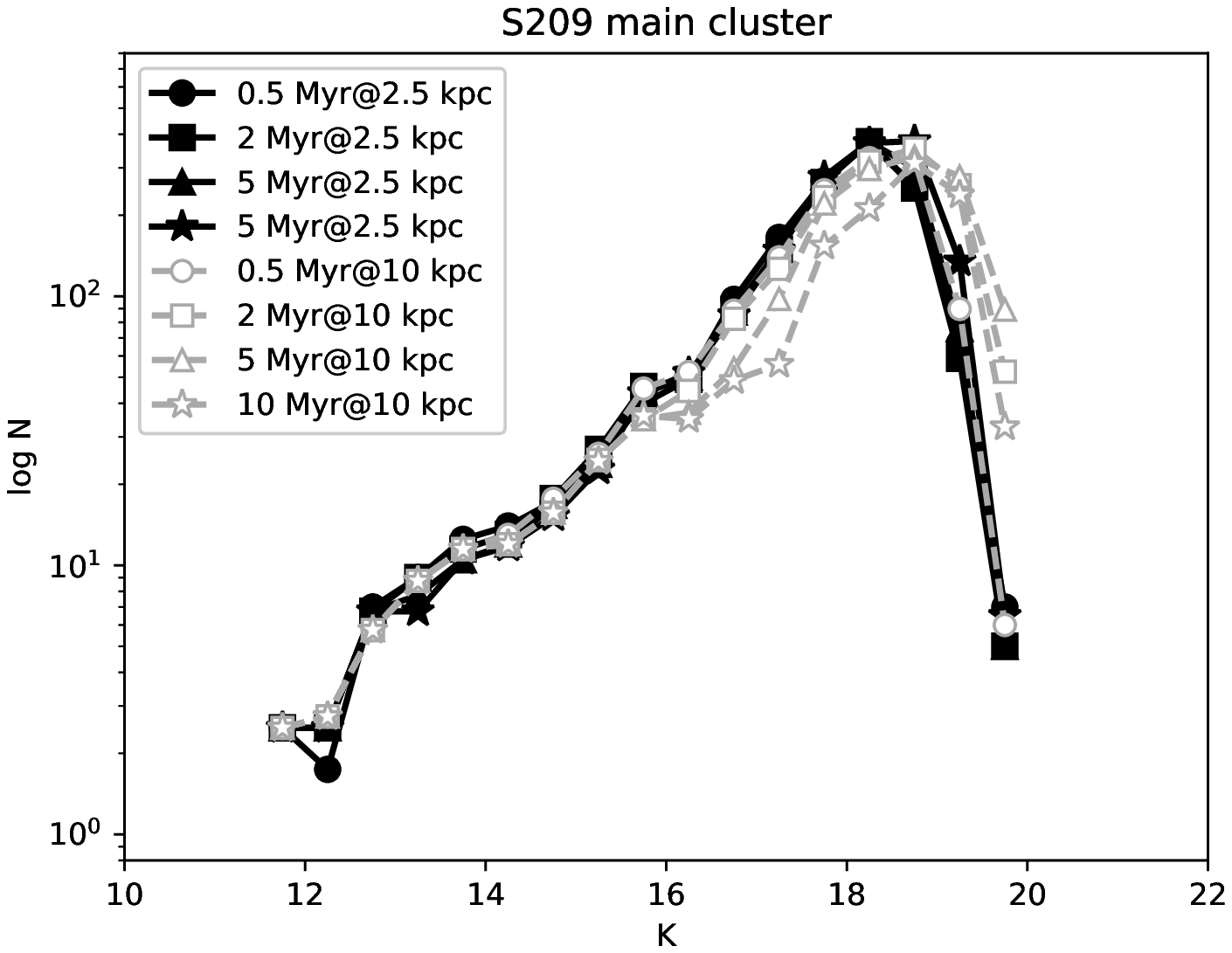}
  \includegraphics[scale=0.45]{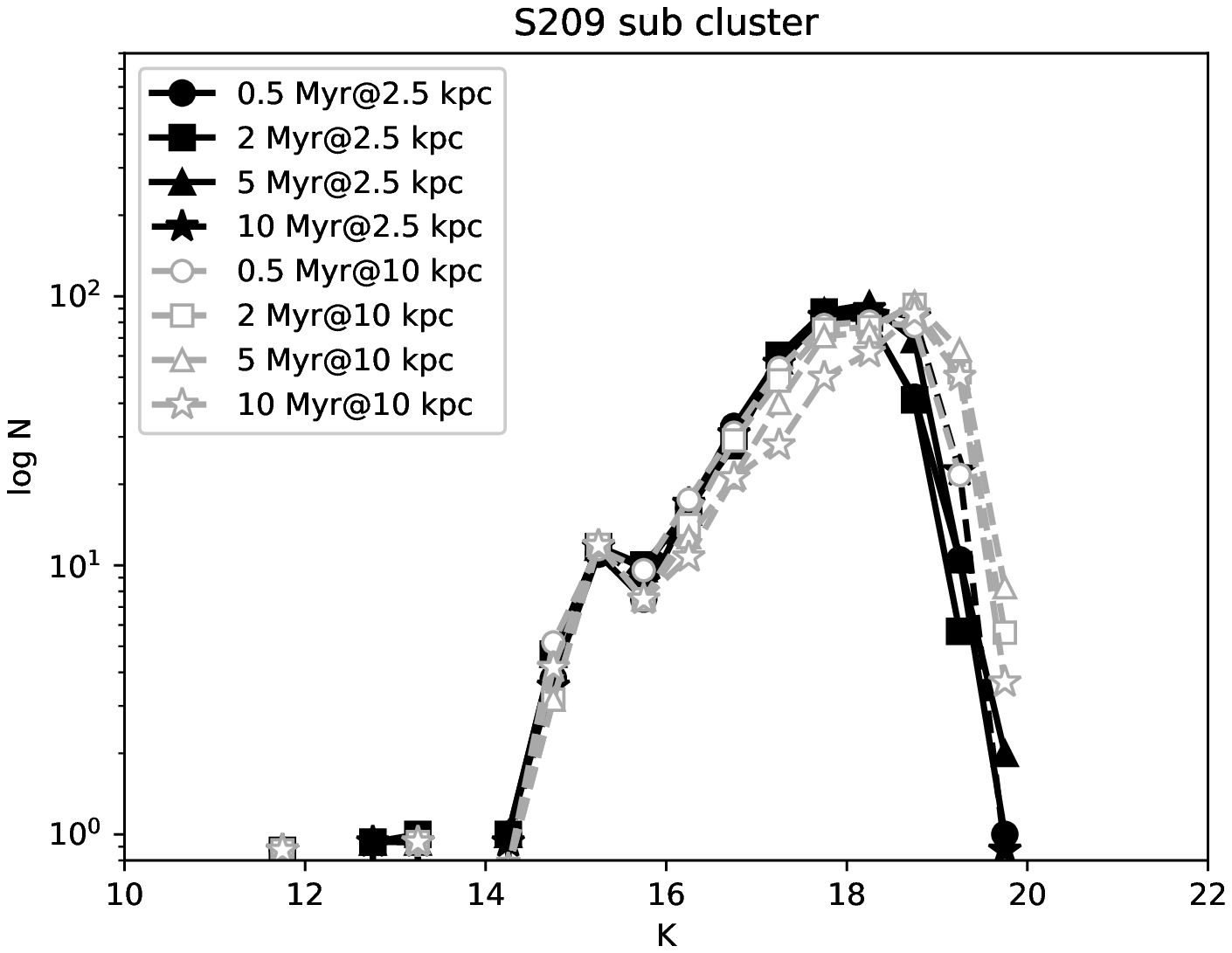}
  \vspace{-10em}
\caption{S209 cluster KLFs for different ages and distances.
The left panel shows the S209 main cluster KLFs and the right panel
shows the S209 sub-cluster KLFs.
The KLFs obtained by assuming a distance of 2.5 kpc are shown as black
filled symbols and solid lines, while those obtained by assuming a
distance of 10 kpc are shown as gray open symbols and dashed
lines. Different symbols represent different ages: circles for 0.5 Myr;
squares for 2 Myr; triangles for 5 Myr; and stars for 10 Myr.}
\label{fig:clKLFs}
\end{center}
\end{figure}

Because the ages and distances of the S209 clusters are not yet known,
we derived the cluster KLFs using the same method for different
distances and ages.
In Figure~\ref{fig:clKLFs}, the black filled symbols and gray open
symbols show the results for distances of 2.5 kpc and 10 kpc,
respectively (the left and right panels for the main and sub-clusters,
respectively).
For both clusters, the KLF increases almost monotonically up to the peak
magnitude, and then decreases monotonically.
For the same distance, the peak magnitude of the KLF is independent of
age over the entire range {0.5--10 Myr}, but the peak value is different
at different distances ({18.0--18.5 mag} for the 2.5 kpc distance and
{18.5--19.0 mag} for the 10 kpc distance for the main cluster;
{17.5--18.5 mag} for the 2.5 kpc distance and {18.5--19.0 mag} for the
10 kpc distance for the sub-cluster).

In general, the KLF peak reflects the IMF peak. However, the counts may
just appear to decrease after $K\sim 18$ mag due to the way in which we
extracted the Mass-$A_V$--limited sample here.
In fact, the color--magnitude diagrams (Figures~\ref{fig:CM_S209} and
\ref{fig:CM_S209sub}) show that for $K \lesssim18$ mag there are stars
in all $A_V$ ranges, whereas for $K \gtrsim 18$ mag there are stars only
in the larger $A_V$ ranges.
In particular, the counts in the faintest K-magnitude bin are extremely
low for the same reason that the range of $A_V$ sampled for the faintest
stars in K magnitudes is very narrow in the diagram.
We discuss this point further in Section~\ref{sec:S209_derived_imf}.

In later sections, we compare the cluster KLFs obtained here with model
KLFs to derive the IMFs.
In creating the model KLFs, we consider the same Mass-$A_V$--limited
sample as for the observations.
A similar method was used for the Trapezium cluster by 
\citet{Muench2002}. 
However, for nearby {clusters} such as Trapezium, it was necessary to
consider background sources as contamination in addition to foreground
sources.
The background sources are complicated to consider because they can be
subject to reddening by molecular clouds in star-forming regions.
In contrast, for a region in the outer Galaxy, previous studies have
shown that most of the contamination can be considered as due to
foreground objects \citep{{Yasui2008}, {Yasui2016b}, {Yasui2021}}, so it
is not necessary to take into account the complex reddening of
background sources; instead, we can simply subtract the number of stars.


\section{Discussion of results from MOIRCS NIR imaging} \label{sec:discussion}

Based on the results obtained from NIR imaging in
Section~\ref{sec:result}, we here discuss further the two young S209
clusters.
We discuss the masses of high-mass stars from the color--magnitude
diagram in Section~\ref{sec:high-mass}; 
which distance is most reliable, 10 kpc (kinematic/photometric distance)
or 2.5 kpc (astrometric distance) in Section~\ref{sec:gaia_moircs};
and summarize the scales of the clusters in Section~\ref{sec:scale}.

\subsection{High-mass Stars in the S209 Region} \label{sec:high-mass}

From spectroscopic observations {by \citet{Chini1984}}, the spectral
types of CW1 and CW2---two of the three sources that are particularly
bright at optical wavelengths---are estimated to be B1 III and O9 III,
respectively (see Section~\ref{sec:properties}).
These spectral types correspond to masses of about 20 $M_\odot$
\citep{Drilling2000}\footnote{The values of masses are from
\citet{Drilling2000}, but the values in luminosity class V are used for
those without information on luminosity class III below.}.
On the NIR color--magnitude diagram (Figure~\ref{fig:CM_S209}), these
two sources are shown by large circles labeled with their names.
This figure shows that the K-band magnitudes for stars $\ge$10 $M_\odot$
become brighter as the cluster age increases from {0.5 Myr to 5 Myr.}
From this diagram, the masses of CW1 and CW2 for the distance of $D=10$
kpc are estimated to be $\sim$40 $M_\odot$ at the very young age of 0.5
Myr, while they lie between 20 and 30 $M_\odot$ at an age of 5 Myr; all
of these masses are larger than 20 $M_\odot$. 
Conversely, assuming $D=2.5$ kpc, their masses are estimated to be
$\sim$3--10 $M_\odot$ at the very young age of 0.5 Myr and $\lesssim$10
$M_\odot$ at 5 Myr, which are all smaller than 20 $M_\odot$. 
None of the masses estimated here agree very well with the masses
obtained from the spectral types, assuming either distance.
This is probably due to the fact that the spectral type was derived
using photographic plates (not a CCD as in the present observations),
which may lead to large uncertainties in the derivation of the spectral
types.
For example, the mass of a B5 III star---a slightly later type than
B0---is 7 $M_\odot$, which seems to be consistent with the estimate for
$D=2.5$ kpc.
Conversely, the mass of the slightly earlier type O6 V is 37 $M_\odot$,
which seems to be consistent with the estimate for $D=10$ kpc. 
Therefore, it seems that no strong constraints can be imposed from the
CW sources.

Although {the previous study} considered stars that are bright at
optical wavelengths, in the NIR bands we detected two brighter stars in
the S209 cluster region.
Because these star images are saturated in the MOIRCS images, we
obtained their magnitudes from the 2MASS Point Source Catalog 
\citep{Skrutskie2006}. 
They are shown with open squares in Figure~\ref{fig:CM_S209}.
Of these two, 2MASS 04110946+5110005 [K (2MASS) = 10.295, H (2MASS) =
10.877, and J (2MASS) = 11.932] is the brightest source in the K band,
suggesting that it is the highest-mass source.
It is located about 8$''$ south of CW3, which is located near the center
of the S209 main cluster.
The location of this star near the center of the main cluster, together
with the fact that its estimated extinction is large ($>$10 mag, from
the color--magnitude diagram), suggests that this star is located within
S209 and that it is the most massive source, which is responsible for
exciting the \ion{H}{2} region.
From the NIR color--magnitude diagram, the mass of this star is
estimated to be about 20 $M_\odot$ at $D=2.5$ kpc for the cluster age of
0.5--5 Myr,
whereas it is estimated to be $\gg$60 $M_\odot$ at $D=10$ kpc for a
young age of 0.5 Myr.
The KLF fitting in a later section (Section~\ref{sec:bestfit_imf})
suggests that an age of 5 Myr is most plausible at $D=2.5$ kpc,
whereas an age of 0.5 Myr is considered most plausible at $D = 10$ kpc
(although the probability of this was very small and was rejected from
the fitting results).
Because the cluster mass of S209 is estimated to be about $\sim$1000
$M_\odot$ (see Section~\ref{sec:SFinS209} and estimates in later
sections), the relation between the maximum stellar mass and the cluster
mass \citep{Weidner2006} suggests that the maximum stellar mass is
consistent with $\sim$20 $M_\odot$, while 60 $M_\odot$ seems to be too
large.
Therefore, $D=2.5$ kpc seems to be the more plausible distance.
However, note that analyses based on individual high-mass candidates
identified towards the clusters, some of which have previously
determined spectral types, allowed relatively weak constraints on their
individual masses and hence on the clusters distances and ages. Stronger
constraints will be presented in Section~\ref{sec:S209_IMF}, where a
full modelling of the clusters KLF will be discussed.

\subsection{Implications for the Distance of S209 from Gaia Sources}
\label{sec:gaia_moircs}

In Section~\ref{sec:properties}, we checked the astrometric distance
from Gaia EDR3, but only for the CW sources, and we found the distance
to be 2.5--3.0 kpc.
However, we cannot yet reject the possibility that the three CW sources
may not lie within S209.
In the MOIRCS images, a bright rim can be seen in the region of the main
cluster in the direction from west to north (Figures~\ref{fig:3col_S209}
and \ref{fig:3col_S209_sd2}).
This rim has a very similar morphology to the ring-like morphology of
the molecular cores---traced with millimeter-continuum emission in
\citet{Klein2005}---that surrounds the S209 main cluster around Core 1.
The NIR image shows a dark region just outside this rim, as judged by
the very low stellar density (Figure~\ref{fig:3col_S209_sd2}),
suggesting that the bright rim is located at the boundary between the
molecular cloud and \ion{H}{2} region, where a shock is thought to occur
\citep{Elmegreen1977}.
The rim is particularly bright in the west, and it overlaps CW1, or may
even originate from CW1.
The right panel of Figure 4 shows an enlargement of the MOIRCS $K_S$-band
image around CW1 ($20'' \times 20''$).
This suggests that the star is at least in the region where
star-formation activity is occurring.

In addition to the CW sources, we also checked the astrometric distances
from Gaia EDR3 for all sources in the cluster regions for which the
parallaxes divided by parallax errors are $\ge$5 (Table~\ref{tab:gaia}).
We identified two other sources in the {main cluster} region; they have
the Gaia source IDs 271701009838583168 and 271701108615685760.
Following the same procedures that we used for the CW sources
(Section~\ref{sec:properties}), we found the distances to these sources
to be $1.1 \pm 0.1$ and $2.3 \pm 0.1$ kpc, respectively.
The distance to the latter source is relatively close to those of the CW
sources while the distance to the former source is much different,
suggesting that only the latter source lies within the S209 region.
Because the latter source is located near the center of the S209 main
cluster, where some bright (and therefore relatively massive) stars
exist, while the former source is located at the edge of the cluster
region, this is a consistent result.

Two other sources in the sub-cluster region also have parallaxes
obtained from Gaia EDR3 for which the parallaxes divided by parallax
errors are $\ge$5: they have the Gaia source IDs 271697951821873792 and
271701009838585856.
Their distances are estimated to be $1.38 \pm 0.05$ and $0.544 \pm
0.008$ kpc, respectively, which are significantly different from the
estimates for the S209 cluster (2.5--3.0 kpc).
This may suggest that this region is in a different location from the
main cluster to the north.  However, because it seems somewhat unnatural
that completely different star-forming regions should exist
coincidentally in a very close regions on the sky (the separation
between the centers of the two clusters is only 2$\farcm$3), 
it is likely that the two objects for which Gaia parallaxes were
obtained with relatively high accuracy do not lie within S209; instead,
they are likely to be foreground stars.
In fact, our radio observations (Izumi et al. 2022, in prep.) indicate
that the molecular clouds associated with the main and sub-clusters both
exist at $\sim$$-$50 km s$^{-1}$, suggesting that both clusters exist in
the same location.

The four sources identified here from Gaia EDR3 and the CW sources are
shown as large filled circles in the color--magnitude diagrams
(Figures~\ref{fig:CM_S209} and \ref{fig:CM_S209sub} for the S209 main
and sub-cluster regions, respectively).
For the main cluster region, Gaia 271701009838583168 has small
extinction from its position, $(H-K, K) \simeq (0.1, 14)$, while Gaia
271701108615685760 exhibits large extinction from its position, $(H-K,
K) \simeq (0.6, 11)$.
The former source is included in the mass-$A_V$--limited sample in
Section~\ref{sec:sample_S209} (hence shown in red), while the latter
source is not included in the sample (hence shown in black).
This also supports the idea that only the former is a foreground
star. Meanwhile, for the sub-cluster region, both Gaia sources (Gaia
271697951821873792 and 271701009838585856; shown with large black filled
circles) exhibit relatively small extinctions, from their respective
positions: $(H-K, K) \simeq (0.3, 13)$ and $(H-K, K) \simeq (0.2,
12.6)$, respectively. Both show marginal colors, whether or not they are
included in the Mass-$A_V$--limited sample, but there is no major
contradiction in considering them as foreground stars.
Table~\ref{tab:gaia} shows whether each Gaia source is likely to be a
member of S209 (column 7).
The Gaia survey is still an ongoing project, and more precise distance
derivations will be released in the future.
To derive the most reliable distances from the current data, we
calculated the weighted averages of the distance for the objects
considered here to be members of S209,
considering the standard errors of the absolute stellar parallax
($\sigma_P$) for each object.
The weighted average is calculated to be 2.5 kpc, which we adopt as the
astrometric distance in the later sections.

\subsection{Size Scales of the Clusters}
\label{sec:scale}

\begin{figure}
\begin{center}
  \vspace{10em}
  \hspace{-10em}
\includegraphics[scale=0.6]{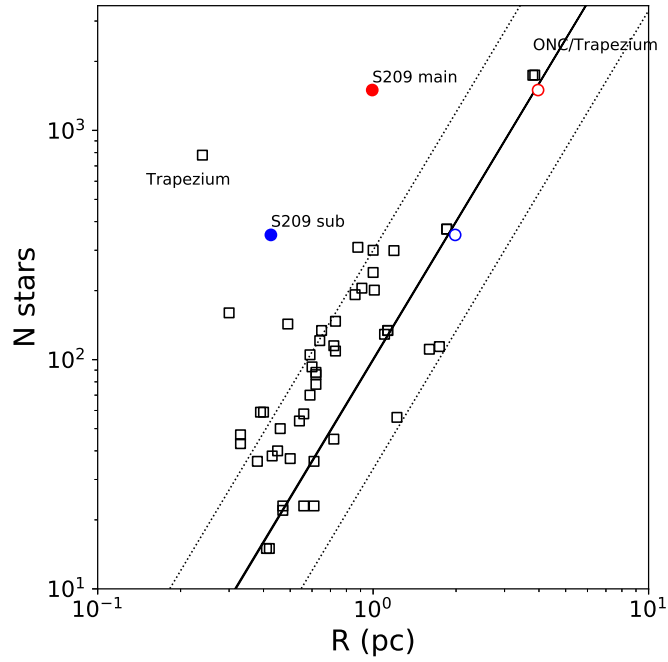}
  \vspace{-10em}
\caption{Correlation between the number of stars in a cluster ($N_{\rm
stars}$) and the radius of the cluster ($R$).
The red and blue circles show the S209 main and sub-clusters,
respectively. The filled circles show the case for the 2.5 kpc distance,
while the open circles show the case for the 10 kpc distance. The open
squares represent clusters in the solar neighborhood using data from
\citet{Adams2006}.
The solid line shows a rough fit to the data for clusters in the solar
neighborhood; most points are scattered within a factor of $\sqrt{3}$ in
$R$, the range shown with dotted lines. The lines correspond to constant
cluster density.}
\label{fig:RvsN_cl}
\end{center}
\end{figure}

Most stars are formed almost simultaneously in star-forming clusters,
which are the smallest units of star formation \citep{LadaLada2003}.
For young clusters in the solar neighborhood, there is a clear
correlation between cluster size and the number of cluster members from
their embedded stage up to ages of $\sim$10 Myr \citep{Adams2006}, as
shown by the open squares in Figure~\ref{fig:RvsN_cl}.
The figure shows that most clusters contain $\sim$10--500 cluster
members ($N_{\rm stars}$) and have radii $R \sim 0.2$--2 pc.
\citet{Adams2006} suggested that the data can be fitted by a relation of
the form $R (N_{\rm stars}) = R_{300} \sqrt{N_{\rm stars} / 300}$ with
$R_{300} = \sqrt{3}$ pc.
This relation is shown in Figure~\ref{fig:RvsN_cl} as a solid line, and
most data points are scattered within a factor of $\sqrt{3}$ of
$R_{300}$, shown with dotted lines.

In Section~\ref{sec:result}, we identified two clusters in S209--the
main cluster and the sub-cluster---and we extracted Mass-$A_V$--limited
samples for both clusters.
We found the main cluster to have $\simeq$1500 cluster
members within a circular region of radius 82$\arcsec$, whereas the
sub-cluster has $\simeq$350 cluster members within a circular region of
radius 41$\arcsec$.
Because 1$\arcmin$ corresponds to 0.75 pc at the 2.5 kpc distance of
S209, these cluster radii for the main and sub-clusters correspond to
1.0 and 0.5 pc, respectively.
We plot the values for the S209 main and sub-clusters in
Figure~\ref{fig:RvsN_cl} as red and blue filled circles, respectively.
The plot shows that both clusters have radii comparable to those of
star-forming clusters in the solar neighborhood ($r \sim 1$ pc),
suggesting that the S209 clusters can be considered to have similar
scales to those of individual star-forming clusters in the solar
neighborhood.
However, they have relatively large values of $N_{\rm star}$ (the main
cluster has among the largest values of $N_{\rm star}$ compared to
nearby clusters; it is comparable to that of ONC/Trapezium).
Thus, although the two S209 clusters have comparable densities, both
densities are higher than the range shown by the dotted line on the
high-density side of the diagram, second only to that of the Trapezium
cluster.
Although there are two possible distances for S209---2.5 kpc and 10 kpc
(see Section~\ref{sec:properties})---we present the 2.5 kpc distance
case here because the discussion in Sections~\ref{sec:high-mass} and
\ref{sec:gaia_moircs} 
suggests that 2.5 kpc is more plausible.
However, for completeness, we also show the 10 kpc case in
Figure~\ref{fig:RvsN_cl}, using red and blue open circles for the S209
main and sub-clusters, respectively.
In either case, the result that the S209 clusters are comparable in size
to the scale of individual clusters is confirmed.

In the following sections, we derive the IMFs for the S209 clusters.
Note that these clusters constitute a very good laboratory for exploring
the metallicity dependence of the IMF because they are individual
star-forming clusters that have the same spatial scale as in the local
region of the Galaxy ($\sim$1 pc), but they exist in a different
metallicity environment.
In addition, the high spatial resolution enables 1000 AU separations to
be resolved---as estimated for the seeing of $\simeq$$0\farcs4$
(Section~\ref{sec:obs_MOIRCS})---at a distance of 2.5 kpc.
In particular, the main cluster is a very suitable target for the first
high-precision derivation of the IMFs in low-metallicity clusters
because it is the first large-scale target ($N_{\rm stars} > 10^3$) in a
low-metallicity environment with detections down to the substellar mass
region ($\lesssim$0.1 $M_\odot$ from Section~\ref{sec:sample_S209}).


\section{Derivation of the S209 Clusters' IMFs}
\label{sec:S209_IMF}

In general, information about the ages of regions and individual stars
is necessary in order to derive the IMF. There are two main methods for
determining the age: one is to combine imaging and spectroscopic
observations, and the other is to use only imaging observations.
In the former method, the temperature of each star is determined from
spectroscopic observations and the luminosity from imaging observations,
and the age and mass of each star is derived by comparing these data
with stellar-evolution tracks in the H--R diagram.
This method has often been used to derive IMFs for nearby star-forming regions
\citep[e.g.,][]{{Hillenbrand1997}, {Luhman2000}}.
However, this method is not very practical when dealing with somewhat
distant regions such as the target here, because it requires too much
observation time to achieve sufficient sensitivity, and it is therefore
difficult to derive the IMF down to sufficiently low stellar masses.

The latter method has often been used to derive ages by comparing
imaging observations with stellar-evolution tracks in a color--magnitude
diagram.
Classical methods derive the age of a star cluster from the property
that stars in the cluster turn to the red as they leave the main
sequence.
However, these methods can only be used for relatively old regions
($>$10 Myr) or for regions with large numbers of massive stars.
The presence of an \ion{H}{2} region in S209
(Section~\ref{sec:SFinS209}) and the fact that it appears very red in
NIR and MIR images suggest that it is very young ($<$10 Myr).
It is therefore difficult to apply this method to S209.
Recently, various attempts have been proposed for determining ages by
using NIR wavelengths as well as optical wavelengths
\citep[cf.,][]{Soderblom2014}, but S209 has large extinction and large
infrared disk excesses ($A_V \simeq 4$--15 mag and a $K$-excess of 0--1.0
mag)---as
well as large dispersions in these values---making it difficult to
derive ages using these methods due to the large color dispersion of the
stars.

In this paper, we therefore derive the IMF and the cluster age
simultaneously as a set, rather than deriving the age independently.
The KLF has often been used to derive the IMF because it has a large
dependence on the IMF \citep{Lada1995}.
Because the KLF is also known to be highly dependent on age
\citep{Muench2000}, we here consider both the cluster age and the IMF as
parameters, and we determine the most reliable age and IMF together by
comparing model KLFs with the KLFs obtained from observations of the
S209 clusters.
Although there are two different possibilities for the distance of
S209---10 kpc and 2.5 kpc (Section~\ref{sec:properties})---because the
discussion in Section~\ref{sec:discussion} (in particular,
Section~\ref{sec:gaia_moircs}) suggests that 2.5 kpc is more plausible,
we primarily consider this distance here.

\subsection{Modeling of the S209 Clusters' KLFs} \label{sec:Model}
In this section, we discuss how to construct model KLFs for the S209
clusters, and we use them to derive the cluster IMF from the
observational cluster KLF obtained in Section~\ref{sec:KLF_S209}.
Our modeling of the cluster KLF basically follows the method of
\citet{{Muench2000},{Muench2002}}.
They derived the IMF for the ONC cluster from the KLF fit using only
imaging data.
Because the age of the ONC cluster is known, they only needed to derive
the IMF.
However, there have been no detailed studies of the star-forming
clusters in S209, and hence there is no information about their ages.
Therefore, we have added age as a parameter to the method described by
\citet{Muench2002}.

Our specific procedure is as follows.
First, we assume that the IMF follows a power law: $dN / d \log m
\propto m^\Gamma$.
The IMFs obtained in the solar neighborhood have negative slopes when
the mass is relatively high ($\gtrsim$1 $M_\odot$),
and the slope becomes constant or positive when the mass exceeds a
certain value \citep[e.g.,][]{{Bastian2010},{Elmegreen2008}}.
Based on this behavior, we define the mass at which the slope changes to
be the ``break mass.''
Actually, in the S209 clusters' observed KLFs, as the magnitudes become
fainter (i.e., as the mass gets smaller), the counts are seen first to
increase and then to decrease. 
Although \citet{Muench2002} employed three break masses---because their
KLF shows a tendency to increase, decrease, and then increase again---we
utilized only two break masses because the S209 clusters' KLFs appear to
show at most only one cycle of an increasing-then-decreasing pattern:
$m_1$, where the IMF transitions from decreasing to roughly flat, and
$m_2$, where the IMF transitions from flat to decreasing.
In our modeling, we first generated stars with masses determined
probabilistically according to the IMF defined here. The IMF mass range
is set to include all the possible ranges that can be obtained from the
Mass-$A_V$---limited samples at each assumed age and distance.

At this stage, it is not clear whether the turnover in the S209
clusters' KLF counts reflects a similar trend in the IMF
(Section~\ref{sec:KLF_S209}),
and the actual IMF can take one of three possible shapes: a one-power
law (a monotonic increase), a two-power law (an increase followed by an
almost flattening), or a three-power law (a decrease followed
by a flattening and an increase). 
Therefore, the two break masses are allowed to overlap for all three
possibilities to be represented.
This allows a one-power-law IMF to be represented by two break masses,
both taking the minimum value of the mass range; a two-power-law IMF to
be represented by two break masses that can take values other than the
minimum or maximum value of the mass range; and a three-power-law IMF
represented by two different break masses that can take values other
than the minimum or maximum of the mass range.

The masses we obtained next need to be converted to magnitudes.
We determined the NIR luminosities using the mass--luminosity (M--L)
relation at each age.
We used the M--L relations employed in Section~\ref{sec:sample_S209} for
the isochrone tracks in the color--magnitude diagram
(Figures~\ref{fig:CM_S209} and \ref{fig:CM_S209sub}). 
After converting to luminosities and then to NIR magnitudes by taking
the distance into account, we added the effects of reddening.
We generated the reddening (extinction and infrared disk excess)
stochastically using the reddening probability distribution for the S209
cluster obtained in Section~\ref{sec:reddening}
(Figure~\ref{fig:Av_CC} for extinction and Figure~\ref{fig:Kex_S209} for
the $K$-excess).
We checked the resulting magnitudes and colors to determine whether or
not they satisfy the condition for a Mass-$A_V$--limited sample given
in Section~\ref{sec:sample_S209}; those that did not satisfy the
condition were not counted in the model KLF.
We repeated this process until the number of stars satisfying this
condition become equal to the number of stars in the
Mass-$A_V$--limited samples for the S209 clusters, which we had
obtained by subtracting the number of samples in the control field from
the number of samples in the S209 field for each age and distance.

For each synthetic cluster, we binned the resulting magnitudes in
half-magnitude bins in order to compare them to the observed cluster
KLFs to derive the best-fit IMF.
For both the S209 main and sub-clusters, we generated 100 independent
KLFs for each age and IMF. We calculated the average of these 100 KLFs
and the corresponding 1$\sigma$ standard deviation.

There are a few things to be noted about this modeling:
\begin{enumerate}
 \item[i)]  Mass range: In constructing a model KLF, we determined the
       mass-detection limit from the color--magnitude diagram
       (Section~\ref{sec:sample_S209}), and we used that mass as the
       lowest mass of the Mass-$A_V$--limited sample.
      However, stars with masses below the mass-detection limit but with
       particularly large {\it K} excess (e.g., $K_{\rm excess} \gtrsim
       1.0$ mag
       may be included in the region where the Mass-$A_V$--limited
       sample is extracted from the color--magnitude diagrams.
      Therefore, we also considered such stars when constructing the
      model KLFs.
      The mass-detection limits obtained from the observations were
      0.04, 0.05, 0.055, 0.09, 0.12, and 0.16 $M_\odot$ for ages 0.5,
      1, 2--3, 5, 7, and 10 Myr, respectively,
      while we considered masses down to 0.02, 0.03, 0.05, and 0.07
      $M_\odot$ ($-$1.7, $-$1.6, $-$1.3, and $-$1.2 on a logarithmic scale)
      for ages 0.5--1, 2--5,  7, and 10 Myr, respectively, in modeling
      the KLFs.

 \item[ii)]  Age spread: We have not taken age spread into account in the
       modeling discussed here because \citet{Muench2000} found that
      variations in the cluster-age spread have only a small effect on
       the form of the KLF and would likely be difficult to distinguish
       observationally.
      In addition, in this paper we defined the cluster region strictly
      as a first step in deriving the IMF for this region
      (Section~\ref{sec:ident_cluster}) in order to avoid a large age
      spread and the inclusion of star-forming activity.
      However, the cluster in reality will have a certain age spread,
      which may vary significantly depending on the metallicity.
      We will therefore revisit this question in future work, in
      combination with the results of ongoing multi-wavelength
      observations of a large area of S209, in order to elucidate the
       star-formation history in this region.

 \item[iii)]  Selection of isochrone models: \citet{{Muench2000}, {Muench2002}}
       found that variations in the pre-main-sequence (PMS) {\it
       M}--{\it L} relation, which result from differences in the
       adopted PMS tracks, produce only small effects in the form of the
       model luminosity functions, and these effects are mostly likely
       not detectable observationally.
       Here, we adopted standard PMS isochrone models:
      \citet{{D'Antona1997},{D'Antona1998}} for the low-mass side and
      \citet{Siess2000} for intermediate masses.
       Both sets of models are included among the models discussed by
       \citet{{Muench2000}, {Muench2002}}.
       Because the targets here---the S209 clusters---are located in a
       low-metallicity environment, models with such metallicities must
       be considered.
       \citet{{D'Antona1997},{D'Antona1998}} and \citet{Siess2000}
       included models for $Z=0.01$ (i.e., $[{\rm M/H}] = -0.3$).
       In addition, \citet{Baraffe1998} provided models with $[{\rm
       M/H}] = -0.5$ (for the mass range 0.079--1.0 $M_\odot$ and the
       age range $>$2 Myr), which is also included in the discussion in
       \citet{Muench2002}.
       \citet{Baraffe1997} also provided models with $[{\rm M/H}] =
       -2.0$ to $-$1.0.
       We compared these models to the solar-metallicity models and
       confirmed that differences in the {\it M}--{\it L} relation due
       to changes in metallicity within the same group of models were
       very small.
       We also found the metallicity differences to be significantly
       smaller than the differences among the models by different
       groups. Therefore, we adopted the models with solar
       metallicity ($[{\rm M/H}] = 0.0$) here.

\end{enumerate}

\subsection{Best-Fit IMFs} \label{sec:bestfit_imf}

By comparing the model KLFs obtained in Section~\ref{sec:Model} with the
S209 cluster KLFs obtained in Section~\ref{sec:KLF_S209}, we derived a
reduced $\chi^2$ and probability for each model KLF.
The parameter ranges are essentially the same as those in \citet{Muench2002}
($\Gamma_1 = -2.0$ to $-$1.0,
$\Gamma_2 = -0.4$ to 0.4,
$\Gamma_3 = -0.4$ to 2.0,
$\log m_1 = -1.1$ to 0.1, and
$\log m_2 = -1.4$ to 0.1). 
However, we have made some changes to these values.
First, we chose the wider range $\Gamma_1 = -2.0$ to $+$0.5, because
this parameter can be either larger or smaller than the Salpeter IMF
index, which is generally found in the solar neighborhood.
We set the minimum values of $\log m_1$ and $\log m_2$ to be the minimum
mass used in the modeling for each age (Section~\ref{sec:Model}),
although we fixed their maximum values to be 0.1, as in
\citet{Muench2002}.
We assume that the parameters $\log m_1$ and $\log m_2$ satisfy $\log
m_1 \ge \log m_2$ within these ranges.

\begin{figure}[!h]
\begin{center}
  \vspace{10em}
  \hspace{-5em}
 \includegraphics[scale=0.45]{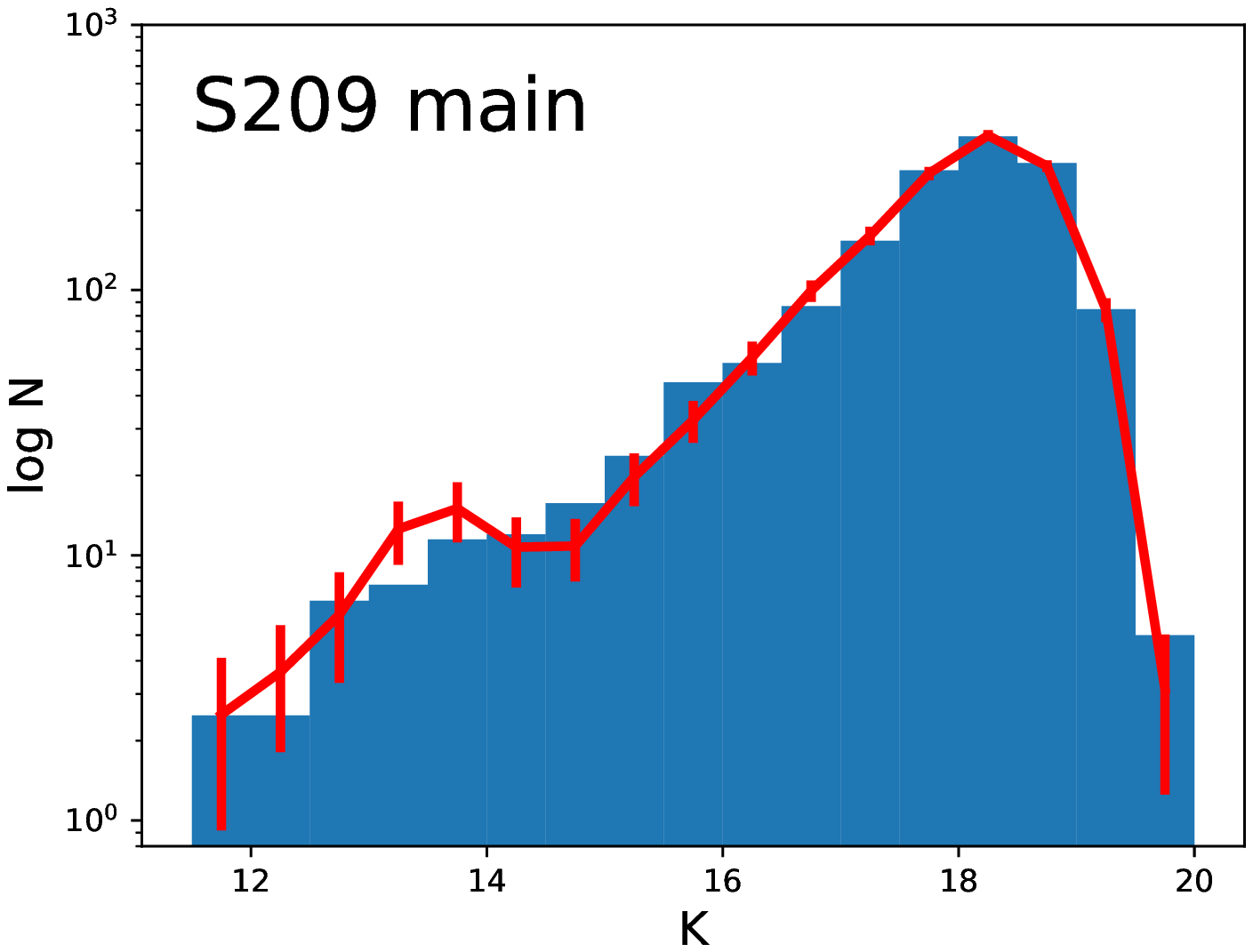}
 \includegraphics[scale=0.45]{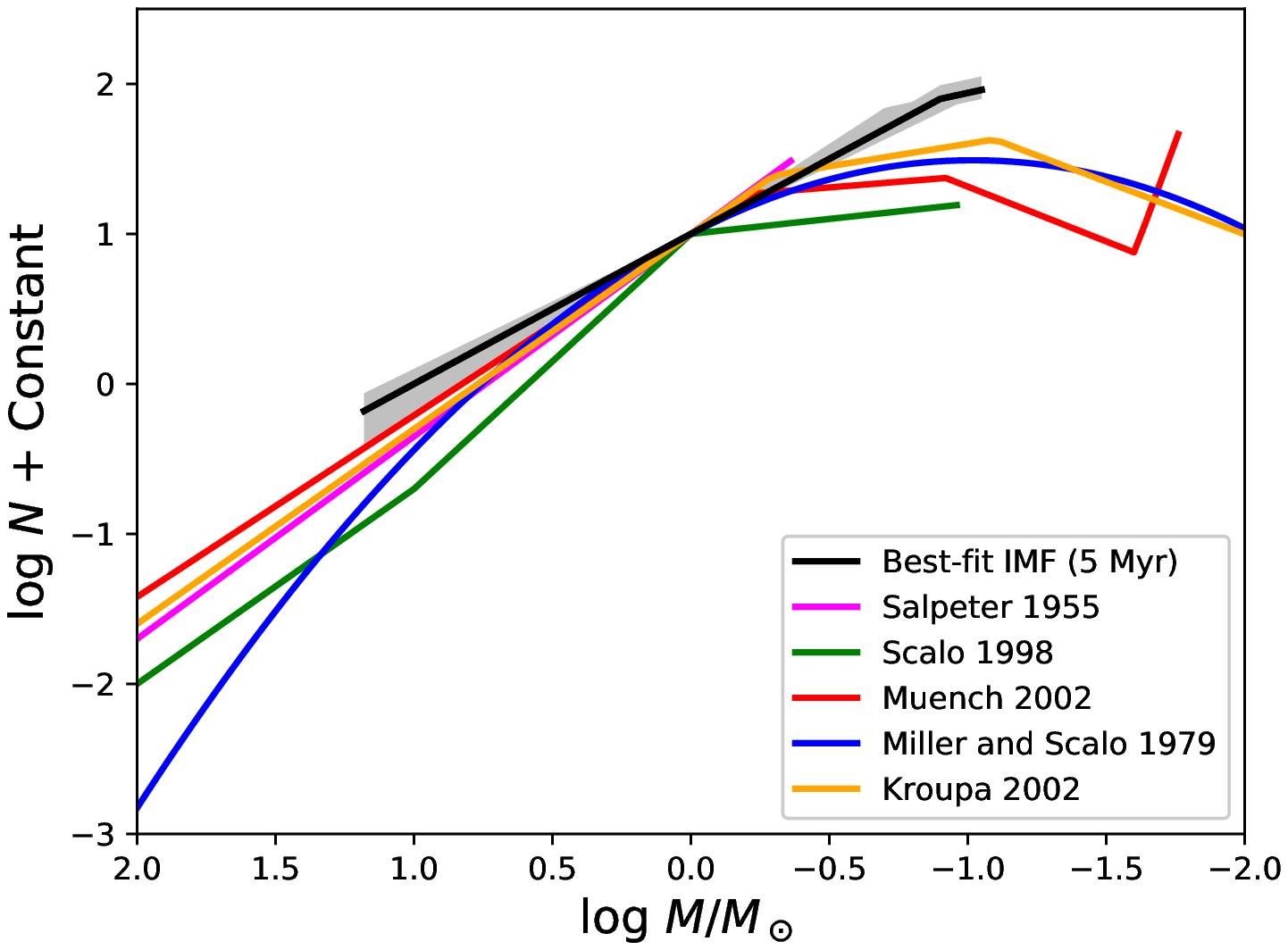}
  \vspace{-10em}
\caption{Model KLF and IMF for the S209 main cluster with the parameters
 of the best-fit IMF.
Left: the best-fit model KLF for the best-fitting age (5 Myr) is shown
as the red line with 1$\sigma$ standard deviation, while the observed
cluster KLF for the age is shown as the blue histogram.
Right: the best-fit IMF is shown as the black line with gray highlighted
regions showing the 90\% confidence level.
The cluster IMF is also compared to IMFs previously obtained in the
field and in the nearby star clusters:
\citet[magenta]{Salpeter1955}, 
\citet[green]{Scalo1998},
\citet[red]{Muench2002}, 
\citet[blue]{Miller1979}, 
and \citet[orange]{Kroupa2002}.}
\label{fig:fit_main_D2.5}
\end{center}
\end{figure}

\begin{figure}[!h]
\begin{center}
  \vspace{10em}
  \hspace{-5em}
 \includegraphics[scale=0.45]{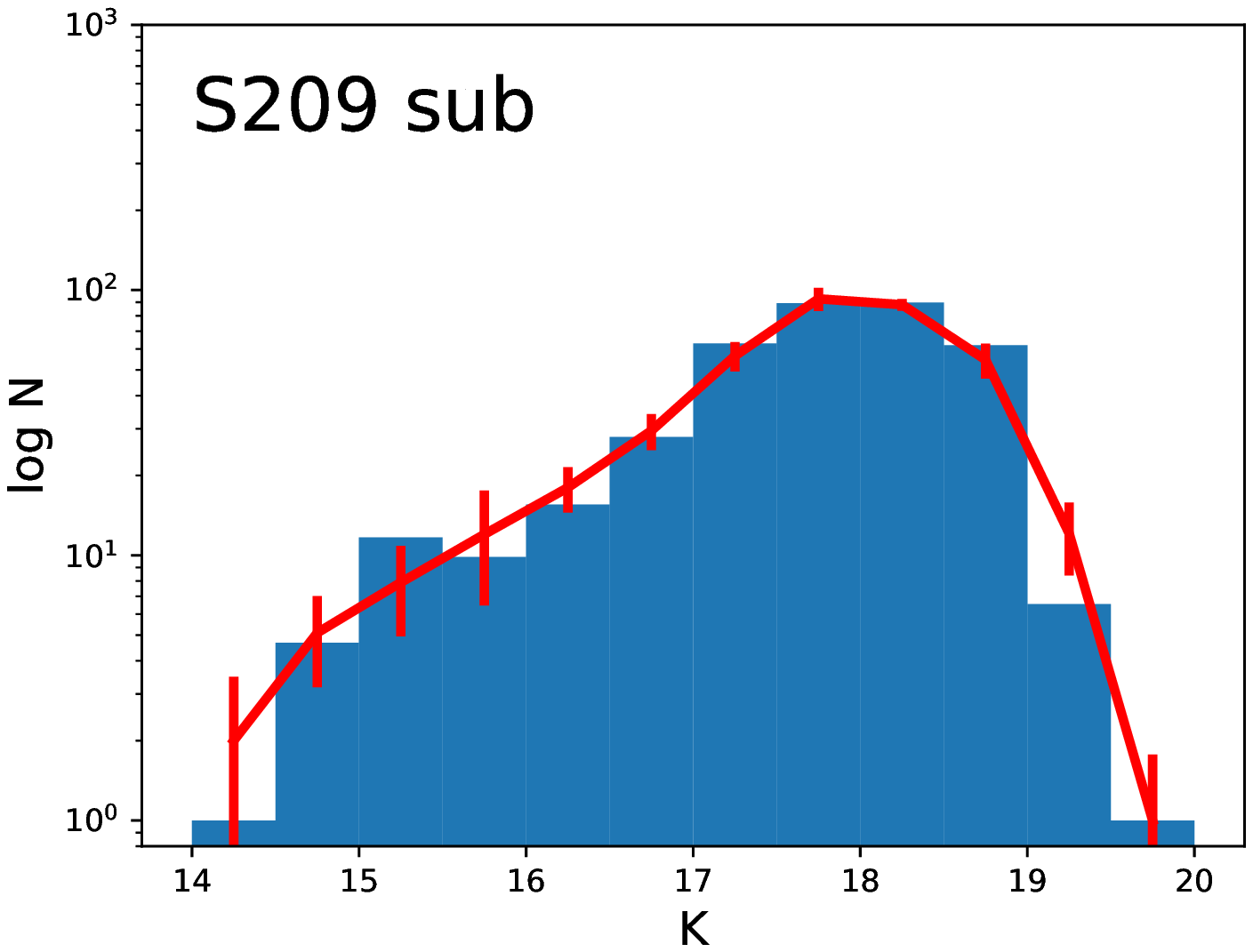}
 \includegraphics[scale=0.45]{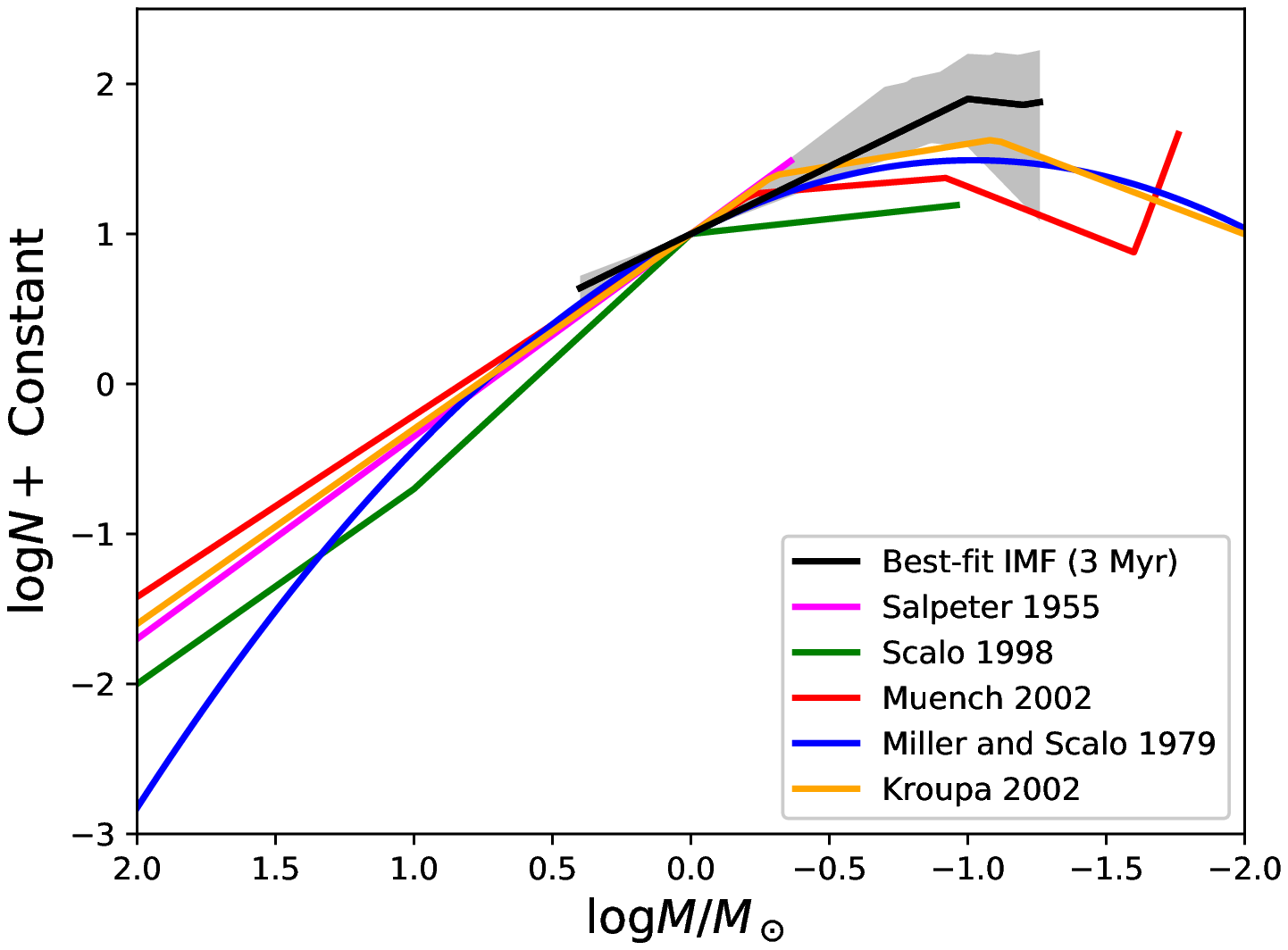}
  \vspace{-10em}
 \caption{Same figure as Figure~\ref{fig:fit_main_D2.5}, but for the
 S209 sub-cluster.}
\label{fig:fit_sub_D2.5}
\end{center}
\end{figure}

The best-fit IMF results we obtained for each age for the S209 main and
sub-clusters are shown in Tables~\ref{tab:fit_main_D2.5} and
\ref{tab:fit_sub_D2.5}, respectively.
For the S209 main cluster, we obtained a reduced $\chi^2$ value of
almost 1 at 5 Myr, corresponding to a probability of 50\%, and we
obtained larger values for both younger and older ages.
For the S209 sub-cluster, the $\chi^2$ value takes its smallest value at
3 Myr (1.01), corresponding to a probability of 72\%, and here, too, we
obtained larger values for both younger and older ages.
In Figures~\ref{fig:fit_main_D2.5} and \ref{fig:fit_sub_D2.5},
respectively, we also show the model KLFs and IMFs for S209 main and
sub-clusters using the parameters for the best fit-IMFs.

For completeness, we also calculated the cases for which the distance is
10 kpc.
We show these results in Tables~\ref{tab:fit_main_D10} and
\ref{tab:fit_sub_D10} for the S209 main and sub-clusters, respectively.
For the main cluster, although we obtained a smaller reduced $\chi^2$
when we assumed a younger age, even for the youngest age assumed here
(0.5 Myr), the value of $\chi^2$ we obtained was significantly greater
than 1 (2.07).
This corresponds to a probability of 2\%, which indicates that this
possibility can be rejected.
However, for the sub-cluster, $\chi^2$ is close to 1 at the young age of
1 and 2 Myr, so this result alone does not allow us to reject this
possibility.
In the next section, we discuss the reasons for the different results
for the two clusters, despite the suggestion that they exist in the same
environment (Section~\ref{sec:gaia_moircs}).


\section{The IMF in a low-metallicity environment}
\label{sec:imf_lowmeta}

\subsection{The IMFs Derived for the S209 Clusters}
\label{sec:S209_derived_imf}

The bottom rows of Tables~\ref{tab:fit_main_D2.5} and
\ref{tab:fit_sub_D2.5} show the parameters of the best-fit KLF for the
S209 main and sub-clusters, with the ranges corresponding to the 90\%
confidence levels shown in brackets.
The right panels of Figures~\ref{fig:fit_main_D2.5} and
\ref{fig:fit_sub_D2.5} show the best-fit IMF (black lines) and the 90\%
confidence level (the gray highlighted regions).
For the model KLFs, we assumed IMFs consisting of three power laws, with
two break masses ($\log m_1$ and $\log m_2$), and we considered masses
extending below the mass-detection limit.
However, the lowest mass that can be derived here for the IMFs for the
S209 clusters is the mass-detection limit obtained from observations,
and we considered masses down to that limit in deriving the final IMF.
Because the age can be $\gtrsim$5 Myr for both the S209 clusters, the
90\% confidence levels indicate that the mass-detection limit should be
considered to be $\simeq$0.1 $M_\odot$; however, the best-fit values of
$\log m_2$ for both clusters are smaller than that.
Therefore, the mass range of the final best-fit IMF does not cover the
second break mass represented by $\log m_2$ but extends only down to the
mass-detection limit.
This suggests that the best-fit IMF is actually a two-power-law IMF,
even though we originally assumed three-power-law IMFs. The parameters
of the best-fit IMF for the S209 main and sub-clusters considering all
the ages are shown in Table~\ref{tab:bestfit_main_sub}, with 1$\sigma$
uncertainties.

The best-fit IMF parameters we obtained for both the IMF slope and the
break mass on the high-mass side are very similar for S209 main and
sub-clusters ($\Gamma_1 = -1.0$, $\log m_1 = -0.9$). 
However, the uncertainties are relatively small for the S209 main
cluster IMF, while they are somewhat larger for the sub-cluster IMF.
The main reason for this may be the large difference in the number of
members in the two clusters ($N_* \simeq 1500$ for the main cluster and
350 for the sub-cluster).
This is directly related to our motivation for using the main
cluster---which has a particularly large number of members---to derive
the IMF with high accuracy in a low-metallicity environment.
The number of cluster members ($N_*$) in the S209 main cluster is about
the same as that in the Trapezium ($N_*\simeq 1500$;
\citealt{Muench2002}), while the number in the sub-cluster is of the
same order as that in IC 348 ($N_*\simeq 150$; \citealt{Muench2003}), but
slightly larger.
We primarily based the present paper on their method of deriving the
IMF, except that we obtained not only the IMFs of the clusters but also
their ages from KLF fitting.
The 1$\sigma$ uncertainties in $\Gamma_1$ and $\log m_1$ are $\Delta
\Gamma_1 \simeq 0.1$ and $\Delta \log m_1 \simeq 0.1$, respectively, for
the Trapezium cluster, and they are $\Delta \Gamma_1 \simeq 0.3$ and
$\Delta \log m_1 \simeq 0.2$, respectively, for the IC 348 cluster.
The uncertainties for the Trapezium and IC 348 clusters are comparable
to those for the S209 main and sub-clusters, respectively, which
suggests that the relationship between $N_*$ and the obtained IMF
uncertainty appears to be quantitatively consistent with these previous
studies of nearby clusters.

As a result of the KLF fitting discussed in
Section~\ref{sec:bestfit_imf}, we found solutions for the S209 main
cluster when we assumed the distance to be $D=2.5$ kpc, but we found no
solutions when the distance was 10 kpc.
However, for the sub-cluster, solutions exist in both cases.
Although this result seems to contradict the suggestion in the
discussion in Section~\ref{sec:discussion} that the two clusters are
located in the same environment at the distance $D=2.5$ kpc, the reason
for this difference seems to be that the uncertainties in the IMF
derivation are due to the differences in $N_*$ between the main and
sub-clusters.
In the main cluster, the possibility $D = 10$ kpc can be rejected due to
the small uncertainties that result from the large value of $N_*$, while
in the sub-cluster KLF fitting alone cannot constrain which possible
distance is the more plausible because the uncertainties in the IMF
derivation are large due to the small value of $N_*$.
In other words, KLF fitting as carried out here indicates that when
$N_*$ is large, fitting that includes distance as a parameter in
addition to IMF and age can be effective.

The other parameter besides the IMF that we obtained in the KLF fitting
is the age of the cluster.
The most plausible ages for the best-fit KLF are 5 Myr for the S209 main
cluster and 3 Myr for the sub-cluster.
The positions of the S209 main cluster and the molecular cores traced
with millimeter-continuum emission in \citet{Klein2005} (cores 1, 2, 3,
and 4 surrounding the cluster) suggest that some high-mass stars in the
main cluster (candidates are discussed in Section~\ref{sec:high-mass})
are dissipating the molecular cloud that was originally the material for
the formation of these stars and that they are also ionizing the
molecular cloud at the position of the bright rim.
Meanwhile, the S209 sub-cluster seems to be completely embedded within
the molecular cloud (core 5).
It is known that stars are born in molecular clouds.
They are initially embedded within those clouds at very young ages
($\lesssim$3 Myr), but the clouds gradually dissipate by ages $\gtrsim$3
Myr \citep{LadaLada2003}.
The relationship between the S209 clusters and their associated
molecular clouds thus suggests that the age of the S209 main cluster is
$\gtrsim$3 Myr while that of the sub-cluster is $\lesssim$3 Myr. 
The ages derived as best-fit values from the KLF fitting roughly match
this discussion.

In summary, for the S209 clusters, we have succeeded in deriving the IMF
for star-forming regions in a low-metallicity environment on the
($\sim$1 pc) scales of individual clusters down to stellar masses as low
as $\simeq$0.1 $M_\odot$.
In particular, because of the large number of cluster members for the
S209 main cluster, we have obtained its IMF with high accuracy for the
first time.
In addition, by targeting such young star-forming clusters, one can
treat a wide range of masses, extending up to 20 $M_\odot$ on the
high-mass side. 
Because these clusters have not yet experienced much N-body relaxation
or mass segregation,
we consider the IMF obtained here to be reasonably similar to the true
IMF \citep{Lee2020}.

\subsection{Comparison with IMFs in the solar neighborhood} 
\label{sec:IMF_solar}

We first compared the high-mass slopes obtained for the IMFs of the S209
clusters to those in the solar neighborhood.
In the solar neighborhood, derivations of the slope have been performed
for various regions, fields, young clusters and associations, and old
globular clusters, and they can generally be explained by Salpeter slope
($\Gamma_1 = -1.35$).
Although examples that differ somewhat from the Salpeter slope have been
reported for some regions, reviews such as \citet{Bastian2010} note that
the differences occur in a very small number of cases and that where
non-Salpeter slopes are reported, the results are often of low
statistical significance or are due to systematic differences in the
approaches.
In contrast, the IMF slopes we derived for the two S209 clusters are
$\Gamma_1 = -1.0$, which is slightly top-heavy compared to the Salpeter
slope.
The method we used to derive the IMF for S209 is basically the same as
that used by \citet{{Muench2002},{Muench2003}} to derive the IMF for the
Trapezium and for IC 348.
They confirmed that their IMF derivations were not significantly
different from other derivations, e.g., from detailed spectroscopy.
The differences we found here therefore should not be due to major
systematic differences in the methods used to derive the IMF.
However, Table~\ref{tab:bestfit_main_sub} shows that, while the result
for the S209 main cluster has a small scatter ($\Delta \Gamma_1 = 0.1$)
and is somewhat robust, there is a large scatter for the sub-cluster
($\Delta \Gamma_1 = 0.2$--0.3) so this result is not conclusive.

Next, we compare the first break mass ($m_1$), which is the mass at
which the IMF slope first transitions from the high-mass side to the
lower-mass side.
This mass is often discussed together with the characteristic mass
($M_C$), which is the mass at the peak of the IMF.
In a power-law IMF, $M_C$ is defined as the midpoint of the plateau
between the two break masses (defined as $m_1$ and $m_2$ in this paper).
From the compilation by \citet{Elmegreen2008} for the solar
neighborhood, the break mass ($m_1$; the upper mass of the IMF plateau)
is about 0.5--1 $M_\odot$ and $M_C$ is about 0.3 $M_\odot$ [$\log (M_C/
M_\odot) \sim -0.5 \pm 0.5$] (see also Figure~3 of
\citealt{Bastian2010}).
For comparison, we estimate the value of $m_1$ for the S209 clusters to
be $\simeq$0.1 $M_\odot$.
Because the break mass ($m_1$) is an upper limit for $M_C$, it is
 suggested that $M_C$ in the S209 clusters are smaller than in the solar
 neighborhood,
even considering the scatter in the values of $m_1 $ in S209 for both
the main and sub-clusters ($m_1$ is at most 0.2 $M_\odot$ for both S209
clusters, considering the 1$\sigma$ uncertainties). 
However, in order to determine the actual mass range of the IMF
plateau and the characteristic mass, it is necessary to extend detection
to stars in the mass range where the IMF begins to fall. In the future,
deeper observations using next-generation telescopes (e.g., JWST) will
thus be important.

\subsection{Comparison with Previous IMF Derivations for Low Metallicity}
\label{sec:metal_imf}

As introduced in Section~\ref{sec:intro}, for the LMC/SMC, which are the
most typical examples used for IMF studies in low-metallicity
environments, detailed observations based on spectroscopic
classification have shown that the IMFs are generally comparable to the
Salpeter IMF for massive stars ($M \gtrsim 10$ $M_\odot$).
Although some studies have suggested that the IMFs may differ from the
Salpeter IMF, this may be due to the fact that only size scales much
larger than the scales of individual clusters (generally $\sim$10--100
pc) can be resolved as star-forming regions.
The star-formation history is consequently more complex in those regions
due to the presence of multiple generations of stars.
One recent result pointed out indications of a top-heavy IMF in the mass
range 15--200 $M_\odot$, taking into account the star-formation history
in the 30 Doradus starburst star-forming region in the LMC
\citep {Schneider2018}.
This is consistent with our results presented here for S209, although
the mass ranges of the targets are different.
However, a review by \citet{Kroupa2013} pointed out that biases due to
crowding, resolution, and mass estimates from photometric data make
studies of such regions very challenging.
Also, because the detection of low-mass stars in the LMC/SMC has been
difficult due to limited observational sensitivity, there has been no
observational discussion of the characteristic mass in such regions.

We have been carrying out IMF derivations for star-forming clusters in
low-metallicity environments in the Galaxy \citep{{Yasui2006},
{Yasui2008ASPC}, {Yasui2008}, {Yasui2016a}, {Yasui2016b}, {Yasui2017},
{Yasui2021}}.
Although there are similar difficulties for such studies as for LMC/SMC
studies, at least the spatial resolution and sensitivity are
significantly higher than for the LMC/SMC due to the much smaller
distances to sources within the Galaxy.
In particular, for this study of S209 the distance is 20 times smaller
(50/2.5), and the spatial resolution is about five times better than
that of an LMC/SMC study using HST ($\Delta \theta \sim 0\farcs 1$;
cf. $\Delta \theta \sim 0\farcs 4$).
Indeed, we achieved a separation of 1000 AU at the S209 distance of 2.5
kpc with seeing of $0\farcs4$, and our mass-detection limits extend down
to $\simeq$0.1 $M_\odot$.
Furthermore, because the clusters in S209 allow us to derive data on the
scale of an individual cluster ($\sim$1 pc), the star-formation history
is not complex, so the IMF derivation becomes simpler and with fewer
assumptions.
As a result, the high-mass slope and characteristic mass of the IMF we
obtained are similar to those of the Salpeter IMF obtained for clusters
in previous studies, which appears to contradict the results for the
S209 clusters.
This is probably due to the fact that the number of cluster members was
not very large ($N_* \sim 100$; see Figure~12 in \citealt{Yasui2021})
for the previous samples, and thus it was not possible to derive the IMF
with high accuracy and determine the presence of differences.
In addition, in those previous samples, cluster members were identified
only from limited $A_V$ ranges (e.g., Figure~5 in \citealt{Yasui2021}),
without considering Mass-$A_V$--limited samples, as in this paper.
We used the masses of stars with limiting magnitudes at the minimum of
the $A_V$ range as the mass-detection limit.
For objects with large $A_V$ values---even though they have only been
detected up to stars of higher masses than the {adopted} mass-detection
limit---this effect was not strictly taken into account in our previous
models.
This may have led to a lower estimate of the number of low-mass stars,
resulting in an estimate of the characteristic mass of $\sim$0.5
$M_\odot$, which is similar to the values seen in the solar
neighborhood.

For globular clusters in the Galaxy, the metallicity dependence of the
IMF has been discussed in combination with the density dependence.
Metal-poor globular clusters, which are subject to weaker tidal fields,
are denser than slightly metal-rich globular clusters \citep{Marks2010},
and higher-density clusters are thought to have top-heavy IMFs because
they provide sufficient feedback energy to blow off the residual gas
\citep{Marks2012}.
At the density of a globular cluster, the slope of the Salpeter IMF is
constant until the metallicity is ${\rm [Fe/H]} \simeq -0.5$, and it
increases gradually when the metallicity is ${\rm [Fe/H]} < -0.5$,
gradually becoming a top-heavy IMF \citep{Marks2012}.
This trend appears to be consistent with the slightly top-heavy IMF seen
in S209, although the difference in the IMF slope by as much as {0.35}
at $-$0.5 dex seems to be a more pronounced change than the results for
globular clusters.
In addition, because the star-forming region in S209 has a low density
of star-forming clusters---comparable to that of young open clusters in
the solar neighborhood (although the initial density may be slightly
different due to the lower metallicity)---it seems unlikely that a
theory appropriate for globular clusters can explain the rather
top-heavy IMF in S209.
On the other hand, because globular clusters are generally old
($\sim$$10^{10}$ yr), and their relatively massive stars have already
completed their lives, the IMFs are obtained for the very small mass
range $\lesssim$1 $M_\odot$. 
In contrast, for S209, we were able to derive IMFs up to a mass $\sim$20
$M_\odot$, which is one of the significant features of this study.

For globular clusters, \citet{Paresce2000} showed that the
characteristic mass does not change significantly from that of the solar
metallicity ($\simeq$0.3 $M_\odot$) in the metallicity range
$-$0.7--$-$2.2 dex.
From this result, they concluded that the global mass functions of
globular clusters should not have changed significantly due to
evaporation or tidal interactions within the Galaxy and should thus
reflect the initial distribution of stellar masses.
Nevertheless, it has been pointed out that, for globular clusters,
considering their old age, a significant number of low-mass stars may
have been scattered outside the cluster due to dynamical evolution.
Therefore, when studying the mass function of a very dense stellar
system, it is necessary to take account of the dynamical state of the
cluster \citep{Portegies Zwart2010}.
In contrast, because it is considered to be largely unaffected by these
processes in the case of younger clusters,
it is reasonable to assume that the PDMF is similar to the IMF
\citep[e.g.,][]{Lee2020}.
If we assume that the low characteristic masses suggested for the S209
clusters are indicative of a metallicity dependence of the IMF, {this
may have been true for the globular clusters in the early stages.
If this is the case, then the observational results by
\citet{Paresce2000} may simply indicate PDMFs with a characteristic mass
similar to that of the IMF in the solar neighborhood as a result of the
dispersion of very low-mass stars during their evolutionary phase, even
though the globular clusters also had a lower characteristic mass than
the IMF in the solar neighborhood at the beginning.}

Theoretical studies also suggest that a metallicity-dependent difference
may appear in the high-mass slope of the IMF due to dust cooling,
resulting in a more top-heavy IMF in a low-metallicity environment.
This trend in the metallicity-dependence of the high-mass slope of the
IMF appears to be similar to the results we obtained here.
However, this difference is expected to occur in environments with much
lower metallicities than the metallicity of interest here; the critical
metal content is expected to be $Z \sim 10^{-6}$--$10^{-5}$
\citep[e.g.,][]{Omukai2005}.
Therefore, although the trend of the metallicity dependence of the
high-mass slope of the IMF is similar to our results, these theoretical
studies cannot be applied directly.

Theoretical investigations have identified two main explanations for the
metallicity dependence of the characteristic mass:
\begin{enumerate}
 \item[i)] The Jeans mass ($M_J \propto \rho^{-1/2} T^{3/2}$)
	 is expected to be higher for a gas with lower density and
	higher temperature, which results in a higher average stellar
	mass.
	The temperature and density of the gas depend on the
	metallicity, and because the cooling efficiency decreases in
	low-metallicity environments, more massive stars should form
	 from low-metallicity gas.

 \item[ii)] A forming star self-regulates its mass by radiative feedback
	 \citep{Adams1996}, which is also metallicity dependent. Because
	 this effect works more effectively in a low-metallicity
	 environment, more high-mass stars should be produced in such
	 environments.
	  
\end{enumerate}

Comparison with the results for S209 clusters obtained here shows that
the above two effects are at least consistent with the top-heavy IMF
suggested for the clusters, but the characteristic mass of the clusters
seem to contradict both of these effects.
However, in the future, it will be important to treat a larger number of
objects, discuss them statistically, and use deeper observations to
detect even smaller-mass stars, where the IMF starts to decrease.
If this can be confirmed, it will be necessary to develop a theory to
explain this.
It is also important to note that lower-mass stars (thus fainter
sources) have generally the greater effect of residual field stars.


\section{Conclusions} \label{sec:conclusion}

The low-metallicity environments in the Galaxy are the closest---and so
far the only---suitable sites for population studies of resolved stars
(1000 AU separation in this study) on the same basis as for the solar
neighborhood.
However, the relatively small number of members ($\sim$100) in
previously studied regions has prevented us from deriving the IMFs with
high precision.
In the present study, we have focused on the star-forming region Sh
2-209, which contains two young clusters (one of which has $\sim$1500
members), with the goal of obtaining the IMFs with high precision for
the first time.
These clusters have been identified as young star-forming clusters
($\lesssim$5 Myr) that exist in a low-metallicity environment
(${\rm [O/H]} \lesssim -0.5$ dex)
at the very close distance of 2.5 kpc, as summarized later in this
section.
By observing individual star-forming clusters (on scales $\sim$1 pc),
which are the smallest units where many stars are born almost
simultaneously, we can get closer to the genuine IMF because the effects
of complex star-formation histories are eliminated.
The derivation of IMFs in young star-forming regions has several
advantages, such as the ability to derive IMFs over a wide range of
masses ($\sim$0.1--20 $M_\odot$ in the case of S209).
Also, the PDMFs are likely to be close to the IMFs because such young
clusters have been subject to little $N$-body relaxation and mass
separation.
Therefore, we consider S209 to be a prototype star-forming region for
deriving IMFs in low-metallicity environments.

Our main results can be summarized as follows:

\begin{enumerate}

\item The \ion{H}{2} region S209 has been identified previously as a
      star-forming region due to the presence of H$\alpha$, MIR, and
      radio-continuum emission. Although previous studies considered the
      distance to this nebulosity to be $\simeq$10 kpc, based on its
      photometric/kinematic distance, we pointed out---based on recent
      results from Gaia EDR3---that the distance may actually be as
      close as 2.5 kpc. From optical spectroscopic observations of this
      \ion{H}{2} region, the estimated oxygen abundance of S209 is
      $\simeq$$-$0.5 dex.
      Also, its electron temperature ($T_e$)---which is a very sensitive
      indicator of abundance---is one of the highest temperatures among
      the H II regions in the Galaxy, which is consistent with this
      low-metallicity estimate.

 \item We obtained deep {\it JHK}$_S$-band images using Subaru/MOIRCS
       with a $4\arcmin \times 7\arcmin$ field of view, which covers the
       central part of the S209 \ion{H}{2} region. The 10$\sigma$
       limiting magnitudes, based on pixel-to-pixel noise for the S209
       frames, reach up to $J \simeq 22.0$ mag, $H \simeq 21.0$ mag, and
       $K_S \simeq 20.6$ mag.

\item From the spatial distribution of stellar densities detected in the
      Subaru/MOIRCS JHK imaging, we identified two clusters, one in the
      north and one in the south. Their radii are 82$\arcsec$ and
      41$\arcsec$, respectively, and we named them the S209 main and
      sub-clusters. Using the JHK color--color diagram for the detected
      sources, we derived the distributions of reddening, extinction
      ($A_V$), and K-band disk excess ($K_{\rm excess}$). Using the
      resulting distribution of $A_V$, we extracted Mass-$A_V$--limited
      samples using the color--magnitude diagram and derived the cluster
      KLFs.

 \item We compared the cluster scale to the mass of the most massive
      star in the S209 main cluster, which we estimated from the NIR
      imaging results assuming two possible distances for S209 (10 kpc
       or 2.5 kpc).
       The results indicate that the 2.5 kpc distance is the more
      plausible. In fact, the NIR images show that the one of the Gaia
      sources lies exactly on the bright rim in the shock plane of the
      molecular cloud and the \ion{H}{2} region and is not a foreground
      or background star. We therefore consider the 2.5 kpc distance to
      be the most plausible distance to S209. In addition, the CO data
      confirm that the velocities of the molecular clouds associated
      with the S209 main and sub-clusters are the same, confirming that
      the two regions probably exist in the same environment. For the
      2.5 kpc distance, the radii are 1.0 pc for the S209 main cluster
      and 0.5 pc for the sub-cluster. These radii and the resulting
      densities of these clusters are similar to those of nearby young
      clusters. In other words, we found S209 to be an optimal
      environment for deriving IMFs at the scale of individual clusters
      on the same basis as for clusters in the solar neighborhood.

 \item In general, the derivation of IMFs requires independent
      information about the age of the region and of individual
       stars.
      However, because no age information exists so far for S209, in
      this paper we derived the IMF and the cluster age together as a
      set.
      Using the characteristic that the KLF is highly dependent on both
      the IMF and the age, we estimated the most reliable age and IMF by
      comparing various models with the KLF of the S209 cluster obtained
      from observations. Although our procedure is based on previous
      studies \citep{{Muench2000},{Muench2002}}, a unique point of this
      paper is that we added age as a fitting parameter. To construct
      the model KLFs, we assumed a three-power-law IMF, used reddening
      properties derived from observations, employed standard model
      $M$--$L$ relations, and considered Mass-$A_V$--limited samples.

\item For the 2.5 kpc distance, the estimated ages of the S209 main and
      sub-clusters are {5 Myr and 3 Myr}, respectively, both roughly
      consistent with the ages suggested by the morphology of the CO
      molecular cloud.
      For both clusters, we found no second break mass above the
      mass-detection limit, suggesting that the underlying IMF actually
      has a two-power-law form.

      Although the IMF parameters we obtained were very similar for the
      two clusters ($\Gamma_1 \simeq -1.0$ and $\log m_1 \simeq -1.0$),
      the parameters for the S209 main cluster, with $N_{\rm
      stars}\simeq 1500$, were determined with relatively high accuracy
      {($\Gamma_1 = -1.0 \pm 0.1$, $\log m_1 = -0.9^{+0.1}_{-0.2}$)},
      while those for the sub-cluster, with $N_{\rm stars}\simeq 350$,
      have large scatter {($\Gamma_1 = -1.0^{+0.3}_{-0.2}$, $\log m_1 =
      -0.9^{+0.1}_{-0.4}$).}

\item Compared to the Salpeter IMF ($\Gamma = 1.35$), which is commonly
      found in the solar neighborhood, the IMFs we obtained for two S209
      clusters are slightly top-heavy.
      However, note that while the results for the S209 main cluster are
      relatively robust, the results for the sub-cluster show large
      scatter.
      In addition, the break mass in the solar neighborhood is 0.5--1.0
      $M_\odot$, while the values obtained for the S209 clusters are
      smaller. However, because there is relatively large scatter in the
      estimated value of $m_1$ for both S209 clusters, this difference
      is not significant.
      In order to determine the actual mass range of the IMF plateau and
      the characteristic mass, it will be necessary to extend detections
      to stars in the mass range where the IMF begins to fall. In the
      future, deeper observations using next-generation telescopes
      (e.g., JWST) will therefore be important.

 \item Previous IMF studies for low-metallicity environments
      ($\sim$$-$1 dex)---both theoretical and observational---have shown
      little clear difference in the high-mass slope and characteristic
      mass from the IMF obtained for the solar neighborhood, at least
      for clusters of similar scale to the S209 clusters.
      This seems to contradict the results obtained here, where we found
      slightly top-heavy high-mass slopes and slightly low
      characteristic masses for the S209 clusters.
      However, this is the first derivation of the IMF for very young
      open clusters---and on the scale of individual star-forming
      clusters---and the IMF we obtained is literally close to the
      initial and genuine IMF.
      In the future, statistical discussions based on observations of
      more objects will be necessary to confirm whether the IMF we
      obtained for S209 reflects the IMF produced at low metallicity.

\end{enumerate}

\acknowledgments

We thank the anonymous referee for a detailed reading and for a
constructive report which helped to improve this manuscript. 
Dr. Ichi Tanaka, a Subaru/MOIRCS support astronomer, helped us a lot for
the observation, discussed how to reduce photometric uncertainties, and
provided s to sky flats for reducing observed data.
%
C.Y. is supported by KAKENHI (18H05441) Grant-in-Aid for Scientific
Research on Innovative Areas.

{\it Facillities:} Subaru Telescope, Gaia.

{\it Software:} IRAF \citep{Tody1993}, MATPLOTLIB, a PYTHON library for
publication quality graphics \citep{Hunter2007}, and NUMPY \citep{van
der Walt2011}.


\begin{table*}[h]
\caption{Properties of S209.} \label{tab:targets}
\begin{center}
\begin{tabular}{llcccccccc}
\hline
\hline
Name & Sh 2-209 \\
Galactic longitude (deg) &   151.6062 (1) \\
Galactic latitude (deg) &  $-$0.2400 (1) \\

R.A. (J2000.0) & 04 11 06.7 (1) \\
Decl. (J2000.0) & $+$51 09 44 (1) \\
Photometric heliocentric distance (kpc) & 10.9 (2)\\
Kinematic heliocentric distance (kpc) & 10.6 (3) \\

Photometric/kinematic Galactocentric distance$^{\rm a}$ (kpc) &  $\simeq$18\\ 

Gaia astrometric distance (kpc) & $\simeq$2.5 (4) \\
Gaia astrometric Galactocentric distance$^{\rm a}$ (kpc) &  $\simeq$10.3\\ 
Oxygen abundance $12 + \log {\rm (O/H)}$  
& $8.15^{+0.16}_{-0.26}$ (5, 6), $8.44^{+0.15}_{-0.22}$ (6, 7) \\
Metallicity [O/H] (dex)$^{\rm b}$ & $\simeq$$-$0.5 \\
\hline
\end{tabular}
\end{center}
{{\small {\bf Notes.} References are shown in the parentheses. \\
$^{\rm a}$Assumed solar Galactocentric distance $R_\odot =
 8.0$\,kpc. \\
$^{\rm b}$Assumed solar abundance $12+ \log {\rm (O/H)} = 8.73$
 \citep{Asplund2009}. \\ 
{\small \bf References. }
 (1) SIMBAD \citep{Wenger2000}, (2) \citet{Chini1984}, (3)
 \citet{Foster2015}, (4) \citet{Gaia2021}, 
 (5) \citet{Vilchez1996}, (6) \citet{Rudolph2006},
 and (7) \citet{Caplan2000}.}} 
\end{table*}

\begin{table*}[h]
 \caption{Properties from Gaia EDR3 for the probable dominant exciting
 sources of S209 suggested by \citet{Chini1984}---CW1, CW2, and
 CW3---and sources in the cluster regions for which parallaxes are
 derived with parallax-over-error $>$5.}  \label{tab:gaia}
\begin{center}
\begin{tabular}{llllllllll}
\hline
\hline
Gaia source ID & $P$ (mas) & $\sigma_P$ (mas) & ZP corr 
& $P_{\rm final}$ (mas)  & D (kpc) & Member? & Notes \\
(1) & (2) & (3) & (4) & (5) & (6) & (7) & (8)\\
\hline
\multicolumn{5}{l}{\it S209 main cluster} \\
271701112917796096 & 0.35 & 0.02 & -0.04
& 0.39 & $2.6^{+0.2}_{-0.1}$ & Y & CW1 \\

271701009838634752 & 0.29 & 0.06 & -0.04 & 0.34
& $3.0^{+0.6}_{-0.4}$ & Y & CW2 \\

271701112917794176 & 0.29 & 0.08 & -0.03
& 0.32 & $3.1^{+1.0}_{-0.6}$ & Y & CW3 \\

271701009838583168 & 0.91 & 0.06 & -0.03 & 0.94 &
$1.1\pm0.1$ & N & \\

271701108615685760 & 0.39 & 0.02 & -0.04 & 0.43
& $2.3 \pm 0.1$ & Y & \\
\hline
\multicolumn{5}{l}{\it S209 sub cluster} \\
271697951821873792 & 0.69 & 0.03 & -0.04
& 0.73 & $1.38 \pm 0.05$ & N &  \\
271701009838585856 & 1.80 & 0.03 & -0.04
& 1.84 & $0.544 \pm 0.008$ & N & \\
\hline
\end{tabular}
\end{center}
 {{\small {\bf Notes.} References are shown in the parentheses. \\
Col. (2): Absolute stellar parallax.
Col. (3): Standard error of absolute stellar parallax.
Col. (4): Parallax zero-point correction.  Col. (5): Final stellar
parallax considering parallax zero-point correction.  Col. (6): Distance
obtained from the parallax in Column 5 ($P_{\rm final}$).  Col. (7):
 Possibility of whether each source is a member of S209.
Y and N represent yes and no, respectively.}}
\end{table*}

\begin{table*}[!h]
\caption{Log of observations.} \label{tab:LOG}
\begin{center}
\begin{tabular}{lcccccccc}
\hline
\hline
Mode & Date & Band & $t_{\rm total}$ & $t$ & Coadd & $N_{\rm total}$
 & Seeing & Sky condition\\ 
& & & (1) & (2) & (3) & (4) & & (5)\\
\hline
$J$-long & Sep 3, 2006 & $J$ & 480 (360) & 120 & 1 & 4 (3) & $0\farcs4$&P \\
$H$-long & Sep 3, 2006 & $H$ & 480 (360) & 20 & 6 & 4 (3) & $0\farcs4$&P \\
$K_S$-long & Sep 3, 2006 & $K_S$ & 720 (540) & 30 & 3 & 8 (6) & $0\farcs4$ &P\\
$J$-short & Nov 8, 2006 & $J$ & 52 (39) & 13 & 1 & 4 (3) & $1\farcs5$ & H\\
$H$-short & Nov 8, 2006 & $H$  & 52 (39) & 13 & 1 & 4 (3) & $1\farcs8$ & H\\
$K_S$-short & Nov 8, 2006 & $K_S$ & 52 (39) & 13 & 1 & 4 (3) & $1\farcs2$ & H\\
\hline
\end{tabular}
\end{center}
{\small N{\scriptsize OTES.---}
Col.(1): total exposure time (s). {The values for the sky
 frames are shown in parentheses.} 
Col.(2): single-exposure time (s). 
Col.(3): number of coadds.
Col.(4): total number of frames. {The values for the sky
 frames are shown in parentheses.}
Col.(5): P: photometric and H: high humidity.} 
\end{table*}

\begin{table*}[!h]
\caption{Limiting Magnitudes (10$\sigma$) of Long-exposure Images for MOIRCS
Observations.}\label{tab:limit}
\begin{center}
\begin{tabular}{lcccl}
\hline
\hline
Target & $J$ (ch1/ch2) & $H$  (ch1/ch2)
& $K_S$ (ch1/ch2) \\
\hline
Cluster & 21.9/22.0 & 20.9/21.0 & 20.5/20.7 \\
Sky & 22.3/22.4 & 21.2/21.2 & 20.8/21.0 & \\
\hline
\end{tabular}
\end{center}
\end{table*}

\begin{table*}[!h]
\caption{Upper lines show the parameters of the best-fit IMF assuming a
distance of 2.5 kpc for the S209 main cluster for each age.
The best parameter set for each age is obtained when the reduced
$\chi^2$ value is closest to 1.
The obtained reduced $\chi^2$ values and corresponding probabilities shown 
for each age are the smallest and largest values, respectively. 
For parameters within the 90\% confidence level, the ranges are shown in
brackets.
The bottom line shows the best-fit IMF parameters for all ages combined,
with the ranges of the parameters within the 90\% confidence level shown
in parentheses.}
\label{tab:fit_main_D2.5}
\begin{center}
\begin{tabular}{lllllllll}
\hline
\hline
Age (Myr) &
$\log m_1$ & $\log m_2$  & $\Gamma_1$ & $\Gamma_2$ & $\Gamma_3$ 
& $\chi^2_{\rm red, min}$ & $P_{\rm max}$ \\
\hline
\multicolumn{5}{l}{\bf S209 main cluster ($D=2.5$ kpc)} \\
0.5 &$-$1.4 & $-$1.5 & $-$1.3 & 0.3 & 1.8 & 5.11 & $<$0.001 \\
1 &$-$1.4 & $-$1.5 & $-$1.3 & $-$0.1 & 0.3 & 11.67 & $<$0.001 \\
2 & $-$1.4 & $-$1.5 & $-$1.1 & 0.1 & 1.2 & 11.24 & $<$0.001 \\
3 & $-$1.5 [$-$1.5, $-$1.4] & $-$1.6 [$-$1.6, $-$1.5] & $-$0.9 [$-$0.9]
   &  0.1 [$-$0.4, 0.4] & 1.7 [$-$0.4, 2.0] & 1.60 & 0.08 \\
5 & $-$0.9 [$-$1.2, $-$0.7] & $-$1.2 [$-$1.6, $-$1.0] & $-$1.0 [$-$1.2, $-$0.9]
& $-$0.4 [$-$0.4, 0.4] & 1.0 [$-$0.4, 2.0] & 0.96 & 0.50 \\
7 & $-$0.7 [$-$0.7] & $-$1.2 [$-$1.2] & $-$1.1 [$-$1.1]
  & $-$0.4 [$-$0.4] & $-$0.4 [$-$0.4] & 1.66 & 0.07 \\
10 & $-$0.4 & $-$1.0 & $-$2.0 & 0.3 & $-$0.2 & 128.1 & $<$0.001 \\
\hline
 5 [3,7] & $-$0.9 [$-$1.5, $-$0.7] & $-$1.2 [$-$1.6, $-$1.0] & $-$1.0 [$-$1.2, $-$0.9] & $-$0.4 [$-$0.4, 0.4] &  1.0 [$-$0.4, 2.0] & & \\
\hline
\end{tabular}
\end{center}
\end{table*}

\begin{table*}[!h]
\caption{Same as Table~\ref{tab:fit_main_D2.5} but for the S209
sub-cluster, assuming a distance of 2.5 kpc.}
\label{tab:fit_sub_D2.5}
\begin{center}
\begin{tabular}{lllllllll}
\hline
\hline
Age (Myr) &
$\log m_1$ & $\log m_2$  & $\Gamma_1$ & $\Gamma_2$ & $\Gamma_3$ 
& $\chi^2_{\rm red, min}$ & $P_{\rm max}$ \\
\hline
\multicolumn{5}{l}{\bf S209 sub cluster ($D=2.5$ kpc)} \\
0.5 & $-$0.8 & $-$1.2 & $-$1.8 & $-$0.4 & 1.0 & 52.58 & $<$0.001\\
1 & $-$1.2 & $-$1.5 & $-$1.6 & 0.3 & $-$0.1 & 3.67 & 0.0005 \\
2 & $-$1.3 & $-$1.4 & $-$1.1 & 0.3 & $-$0.4 & 2.35 & 0.02 \\
3 & $-$1.0 [$-$1.5, $-$0.6] & $-$1.2 [$-$1.6, $-$0.9] & $-$0.9 [$-$1.4, $-$0.7] & 0.2 [$-$0.4, 0.4] & $-$0.3 [$-$0.4, 2.0] & 1.01 &  0.72 \\
5 & $-$0.7 [$-$0.9, $-$0.6] & $-$1.0 [$-$1.6, $-$0.8] & $-$1.0 [$-$1.3, $-$0.9]
& $-$0.4 [$-$0.4, 0.4] & 1.8 [$-$0.4, 2.0] & 1.35 & 0.22 \\
7 & $-$0.6 [$-$0.7, $-$0.5] & $-$0.9 [$-$1.2, $-$0.7] & $-$1.2 [$-$1.3, $-$1.0]
& $-$0.1 [$-$0.4, 0.4] & 0.7 [$-$0.3, 2.0] & 1.59 & 0.13 \\
10 & $-$0.3 & $-$1.0 & $-$1.5 & $-$0.4 & 1.2 & 2.14 & 0.04 \\
\hline
3 [3, 7] & $-$1.0 [$-$1.5, $-$0.5] & $-$1.2 [$-$1.6, $-$0.7]
& $-$0.9 [$-$1.4, $-$0.7] & 0.2 [$-$0.4, 0.4] &  $-$0.3 [$-$0.4, 2.0] & & \\
\hline
\end{tabular}
\end{center}
\end{table*}

\begin{table*}[!h]
\caption{Same as Table~\ref{tab:fit_main_D2.5} but assuming a distance
of 10 kpc.}  \label{tab:fit_main_D10}
\begin{center}
\begin{tabular}{lllllllll}
\hline
\hline
Age (Myr) &
$\log m_1$ & $\log m_2$  & $\Gamma_1$ & $\Gamma_2$ & $\Gamma_3$ 
& $\chi^2_{\rm red}$ & $P_{\rm max}$ \\
\hline
\multicolumn{5}{l}{\bf S209 main cluster ($D=10$ kpc)} \\
0.5 & $-$0.8 & $-$0.9 & $-$1.0 & $-$0.3 & 1.0 & 2.07 & 0.02\\
1 & $-$0.2 & $-$0.8 & $-$1.2 & 0.1 & $-$0.2 & 2.72 & 0.001 \\
2 & $-$0.3 & $-$0.7 & $-$1.2 & $-$0.4 & 0.8 & 4.06 & $<$0.001\\
3 & 0 & $-$0.3 & $-$1.4 & $-$0.4 & $-$0.3 & 6.39 & $<$0.001\\
5 & $-$0.2 & $-$0.3 & $-$1.4 & 0.1 & 0.8 & 12.33 & $<$0.001\\
7 & $-$0.1 & $-$0.3 & $-$1.7 & 0.0 & 1.4 & 17.66 & $<$0.001\\
10 & 0.0 & $-$0.1 & $-$2.0 & $-$0.1 & 1.0 & 31.36 & $<$0.001\\
 \hline
\end{tabular}
\end{center}
\end{table*}

\begin{table*}[!h]
\caption{Same as Table~\ref{tab:fit_main_D2.5} but for the S209
sub-cluster, assuming a distance of 10 kpc.}
\label{tab:fit_sub_D10}
\begin{center}
\begin{tabular}{lllllllll}
\hline
\hline
Age (Myr) & $\log m_1$ & $\log m_2$  & $\Gamma_1$ & $\Gamma_2$ & $\Gamma_3$ 
& $\chi^2_{\rm red, min}$ & P$_{\rm max}$ \\
\hline
\multicolumn{5}{l}{\bf S209 sub cluster ($D=10$ kpc)} \\
0.5 & $-$0.7 & $-$0.8  & $-$0.9 & 0.2 & 1.4 & 2.41 & 0.02 \\
1 & $-$0.1 [$-$0.2, 0.1] & $-$0.7 [$-$1.0, 0.0] & $-$1.5 [$-$2.0, $-$1.1] 
& 0.4 [$-$0.4, 0.4] & 0.6 [$-$0.4, 2.0] & 1.00 & 0.70 \\
2 & 0.1 [0.0, 0.1] & $-$0.1 [$-$0.7, 0.0] & $-$1.8 [$-$2.0, $-$1.5] 
& 0.1 [$-$0.4, 0.4] & $-$0.3 [$-$0.4, 2.0] & 1.33 & 0.23 \\
3 & 0.1 & $-$0.5 & $-$1.9 & $-$0.2 & $-$0.3 & 2.82 & $<$0.001 \\
5 & 0.1 & $-$0.4 & $-$1.7 & $-$0.1 & 0.6 & 5.29 & $<$0.001 \\
7 & 0.1 & $-$0.3 & $-$1.9 & $-$0.2 & 1.6 & 5.67 & $<$0.001 \\
10 & 0.0 & $-$0.1 & $-$2.0 & $-$0.2 & 1.8 & 7.00 & $<$0.001 \\
\hline
1 [1, 2] & $-$0.1 [$-$0.2, 0.1] & $-$0.7 [$-$1.0, 0.0] & $-$1.5 [$-$2.0, $-$1.1] 
& 0.4 [$-$0.4, 0.4] & 0.6 [$-$0.4, 2.0] & & \\
\hline
\end{tabular}
\end{center}
\end{table*}

\begin{table*}[!h]
 \caption{Parameters of the best-fit IMF for the S209 main and
sub-clusters. The 1$\sigma$ uncertainties are shown.}
\label{tab:bestfit_main_sub}
\begin{center}
\begin{tabular}{lllllllll}
\hline
\hline
Cluster & $\Gamma_1$ & $\log m_1$ & Age \\
 & & ($M_\odot$) & (Myr)  \\
\hline
Main & $-1.0 \pm 0.1$ & $-0.9^{+0.1}_{-0.2}$ & 5 \\
Sub (again) & $-1.0_{-0.2}^{+0.3}$ & $-0.9_{-0.4}^{+0.1}$ & 3 \\
\hline
\end{tabular}
\end{center}
\end{table*}


\clearpage
\appendix

\section{Subaru/MOIRCS Point-source Catalog for S209}

\startlongtable
\begin{deluxetable*}{lllrrrl}
 \tablecaption{Subaru/MOIRCS S209 Catalog.} 
\label{tab:S209_moircs}
\tablewidth{700pt}
\tabletypesize{\scriptsize}
\tablehead{
\colhead{ID} & \colhead{R.A.} & 
\colhead{Decl.} &
\colhead{$J$} & 
\colhead{$H$} &
\colhead{$K_S$} & 
\colhead{Notes}  \\
\colhead{} &
\colhead{(J2000.0)} & 
\colhead{(J2000.0)} & 
\colhead{(mag)} & 
\colhead{(mag)} & 
\colhead{(mag)} & 
\colhead{}  \\
 \colhead{(1)} & \colhead{(2)} & \colhead{(3)} & \colhead{(4)} & 
 \colhead{(5)} & \colhead{(6)} & \colhead{(7)}
} 
\startdata
1 & 62.683425 & 51.156903 & 18.96 ( 0.03 ) & 19.40 ( 0.04 ) & 20.23 ( 0.07 ) & ...\\
2 & 62.684710 & 51.156051 & 19.93 ( 0.06 ) & 20.79 ( 0.07 ) & ... (...)  & ...\\
3 & 62.684945 & 51.152882 & 21.10 ( 0.13 ) & 20.99 ( 0.10 ) & 21.45 ( 0.18 ) & ...\\
4 & 62.685047 & 51.158731 & 19.64 ( 0.04 ) & 20.33 ( 0.07 ) & 19.82 ( 0.16 ) & ...\\
5 & 62.685150 & 51.157613 & 18.76 ( 0.03 ) & 19.33 ( 0.03 ) & 21.40 ( 0.12 ) & ...\\
6 & 62.685374 & 51.152773 & 19.17 ( 0.04 ) & 19.80 ( 0.08 ) & 21.19 ( 0.11 ) & ...\\
7 & 62.685571 & 51.153886 & 17.38 ( 0.02 ) & 17.78 ( 0.02 ) & 18.92 ( 0.05 ) & ...\\
8 & 62.685782 & 51.159704 & 20.78 ( 0.13 ) & 21.05 ( 0.17 ) & 19.85 ( 0.18 ) & ...\\
9 & 62.686496 & 51.157573 & 19.04 ( 0.02 ) & 19.75 ( 0.04 ) & 21.62 ( 0.09 ) & ...\\
10 & 62.686930 & 51.156448 & 18.05 ( 0.01 ) & 18.41 ( 0.02 ) & 19.25 ( 0.07 ) & ...\\
\enddata
\tablecomments{
Table~\ref{tab:S209_moircs} is published in its entirety in the electric
edition of the Astrophysical Journal. A portion is shown here for
guidance regarding its form and content.
Columns (2) and (3): Right ascension and declination in degrees.
Columns (4)--(7): Subaru/MOIRCS magnitudes.
$J$ band in Column (4),
$H$ band in Column (5),
and $K_S$ band magnitudes in Column (6). 
Magnitude errors are shown in parentheses.
Column (7): Sources in the S209 main and sub-cluster regions are labeled
 ``Main'' and ``Sub,'' respectively.}
 
\end{deluxetable*}






\begin{thebibliography}{}

\bibitem[Adams \& Fatuzzo(1996)]{Adams1996}
Adams, F.~C. \& Fatuzzo, M.\ 1996, \apj, 464, 256.

\bibitem[Adams et al.(2006)]{Adams2006}
Adams, F.~C., Proszkow, E.~M., Fatuzzo, M., \& Myers, P.~C.\ 2006, \apj, 641, 504 

\bibitem[Asplund et al.(2009)]{Asplund2009}
Asplund, M., Grevesse, N., Sauval, A.~J., \& Scott, P.\ 2009, \araa, 47,
481

\bibitem[Balser et al.(2011)]{Balser2011}
Balser, D.~S., Rood, R.~T., Bania, T.~M., \& Anderson, L.~D.\ 2011,
\apj, 738, 27

\bibitem[Baraffe et al.(1998)]{Baraffe1998}
Baraffe, I., Chabrier, G., Allard, F., et al.\ 1998, \aap, 337, 403

\bibitem[Baraffe et al.(1997)]{Baraffe1997}
Baraffe, I., Chabrier, G., Allard, F., et al.\ 1997, \aap, 327, 1054

 \bibitem[Bastian et al.(2010)]{Bastian2010}
Bastian, N., Covey, K.~R., \& Meyer, M.~R.\ 2010, \araa, 48, 339. 

\bibitem[Beichman et al.(1988)]{Beichman1988}
Beichman, C.~A., Neugebauer, G., Habing, H.~J., Clegg, P.~E., \&
Chester, T.~J.\ 1988, Infrared astronomical satellite (IRAS) catalogs
and atlases.~Volume 1: Explanatory supplement, 1,

\bibitem[Bica et al.(2003)]{Bica2003}
Bica, E., Dutra, C.~M., Soares, J., \& Barbuy, B.\ 2003, \aap, 404, 223 

\bibitem[Blitz et al.(1982)]{Blitz1982}
Blitz, L., Fich, M., \& Stark, A.~A.\ 1982, \apjs, 49, 183

\bibitem[Caplan et al.(2000)]{Caplan2000}
Caplan, J., Deharveng, L., Pe{\~n}a, M., Costero, R., \& Blondel, C.\
2000, \mnras, 311, 317

\bibitem[Chini \& Wink(1984)]{Chini1984}
Chini, R., \& Wink, J.~E.\ 1984, \aap, 139, L5 

\bibitem[Condon et al.(1998)]{Condon1998}
Condon, J.~J., Cotton, W.~D., Greisen, E.~W., et al.\ 1998, \aj, 115,
1693

\bibitem[D'Antona \& Mazzitelli(1997)]{D'Antona1997}
D'Antona, F., \& Mazzitelli, I.\ 1997, Memorie della Societa Astronomica
Italiana, 68, 807

\bibitem[D'Antona \& Mazzitelli(1998)]{D'Antona1998}
D'Antona, F., \& Mazzitelli, I.\ 1998, ASP Conf.~Ser.~134: Brown Dwarfs
and Extrasolar Planets, 134, 442

\bibitem[Deharveng et al.(2000)]{Deharveng2000}
Deharveng, L., Pe{\~n}a, M., Caplan, J., \& Costero, R.\ 2000, \mnras,
311, 329

\bibitem[De Marchi et al.(2017)]{De Marchi2017}
De Marchi, G., Panagia, N., \& Beccari, G.\ 2017, \apj, 846, 110.

\bibitem[Drew et al.(2005)]{Drew2005}
Drew, J.~E., Greimel, R., Irwin, M.~J., et al.\ 2005, \mnras, 362, 753

\bibitem[Drilling \& Landolt(2000)]{Drilling2000}
Drilling, J.~S. \& Landolt, A.~U.\ 2000, Allen's Astrophysical
Quantities, 381

\bibitem[Elmegreen \& Lada(1977)]{Elmegreen1977}
Elmegreen, B.~G. \& Lada, C.~J.\ 1977, \apj, 214, 725.

\bibitem[Elmegreen et al.(2008)]{Elmegreen2008}
Elmegreen, B.~G., Klessen, R.~S., \& Wilson, C.~D.\ 2008, \apj, 681, 365 

\bibitem[Fern{\'a}ndez-Mart{\'\i}n et al.(2017)]{Fernandez-Martin2017}
Fern{\'a}ndez-Mart{\'\i}n, A., P{\'e}rez-Montero, E., V{\'\i}lchez,
J.~M., et al.\ 2017, \aap, 597, A84. 
		
\bibitem[Fich et al.(1990)]{Fich1990}
Fich, M., Dahl, G.~P., \& Treffers, R.~R.\ 1990, \aj, 99, 622

\bibitem[Foster \& Brunt(2015)]{Foster2015}
Foster, T., \& Brunt, C.~M.\ 2015, \aj, 150, 147 


\bibitem[Frerking et al.(1982)]{Frerking1982}
Frerking, M.~A., Langer, W.~D., \& Wilson, R.~W.\ 1982, \apj, 262, 590.

\bibitem[Gaia Collaboration et al.(2021)]{Gaia2021}
{Gaia Collaboration, Brown, A.~G.~A., Vallenari, A., et al.\ 2021, \aap, 649, A1}

\bibitem[Hillenbrand(1997)]{Hillenbrand1997}
Hillenbrand, L.~A.\ 1997, \aj, 113, 1733. 

\bibitem[Hunter(2007)]{Hunter2007}
Hunter, J.~D.\ 2007, Computing in Science and Engineering, 9, 90.

\bibitem[Ichikawa et al.(2006)]{Ichikawa2006}
Ichikawa, T., et al. 2006, \procspie, 6269

\bibitem[Kerber \& Santiago(2006)]{Kerber2006}
Kerber, L.~O. \& Santiago, B.~X.\ 2006, \aap, 452, 155.

\bibitem[Klein et al.(2005)]{Klein2005}
Klein, R., Posselt, B., Schreyer, K., Forbrich, J., \& Henning, T.\
2005, \apjs, 161, 361

\bibitem[Tody(1993)]{Tody1993}
Tody, D.\ 1993, Astronomical Data Analysis Software and Systems II, 52, 173

\bibitem[Koenig et al.(2012)]{Koenig2012}
Koenig, X.~P., Leisawitz, D.~T., Benford, D.~J., et al.\ 2012, \apj, 744, 130

\bibitem[Kroupa(2002)]{Kroupa2002}
Kroupa, P. 2002, Science  295, 82
		
\bibitem[Kroupa et al.(2013)]{Kroupa2013}
Kroupa, P., Weidner, C., Pflamm-Altenburg, J., et al.\ 2013, Planets,
Stars and Stellar Systems. Volume 5: Galactic Structure and Stellar
Populations, 115. 
		
\bibitem[Lada(1999)]{Lada1999}
Lada, E. A. 1999, in The Origin of Stars and Planetary Systems,
ed. C. J. Lada \& N. D. Kylafis (Dordrecht: Kluwer), 441

\bibitem[Lada \& Adams(1992)]{Lada1992} 
Lada, C.~J., \& Adams, F.~C.\ 1992, ApJ, 393, 278 

\bibitem[Lada \& Lada(2003)]{LadaLada2003}
Lada, C.~J., \& Lada, E.~A.\ 2003, \araa, 41, 57 

\bibitem[Lada \& Lada(1995)]{Lada1995}
Lada, E.~A. \& Lada, C.~J.\ 1995, \aj, 109, 1682. 

\bibitem[Lee et al.(2020)]{Lee2020}
Lee, Y.-N., Offner, S.~S.~R., Hennebelle, P., et al.\ 2020, \ssr, 216,
70. 

\bibitem[Leggett et al.(2006)]{Leggett2006}
Leggett, S.~K., et al.\ 2006, \mnras, 373, 781 

\bibitem[Lejeune \& Schaerer(2001)]{Lejeune2001}
Lejeune, T., \& Schaerer, D.\ 2001, \aap, 366, 538

\bibitem[Lindegren et al.(2018)]{Lindegren2018} 
Lindegren, L., Hern{\'a}ndez, J., Bombrun, A., et al.\ 2018, \aap, 616, A2

\bibitem[Lindegren et al.(2021)]{Lindegren2021}
Lindegren, L., Klioner, S.~A., Hern{\'a}ndez, J., et al.\ 2021, \aap, 649, A2.

\bibitem[Luhman et al.(2000)]{Luhman2000}
Luhman, K.~L., Rieke, G.~H., Young, E.~T., et al.\ 2000, \apj, 540,
1016.

\bibitem[Marks \& Kroupa(2010)]{Marks2010}
Marks, M. \& Kroupa, P.\ 2010, \mnras, 406, 2000. 

\bibitem[Marks et al.(2012)]{Marks2012}
Marks, M., Kroupa, P., Dabringhausen, J., et al.\ 2012, \mnras, 422, 2246. 

\bibitem[Massey(2003)]{Massey2003}
Massey, P.\ 2003, \araa, 41, 15. 

\bibitem[Mayor \& Queloz(1995)]{Mayor1995}
Mayor, M., \& Queloz, D.\ 1995, Nature, 378, 355 

\bibitem[Meyer et al.(1997)]{Meyer1997}
Meyer, M. R., Calvet, N., \& Hillenbrand, L. A. 1997, AJ, 114, 288

\bibitem[Miller \& Scalo(1979)]{Miller1979}
Miller, G.~E., \& Scalo, J.~M. 1979, ApJS  41, 513

\bibitem[Muench et al.(2000)]{Muench2000}
Muench, A. A., Lada, E. A., \& Lada, C. J. 2000, ApJ, 533, 358

\bibitem[Muench et al.(2002)]{Muench2002}
Muench, A.~A., Lada, E.~A., Lada, C.~J., \& Alves, J.\ 2002, \apj, 573, 366

\bibitem[Muench et al.(2003)]{Muench2003}
Muench, A.~A., Lada, E.~A., Lada, C.~J., et al.\ 2003, \aj, 125, 2029 

\bibitem[Omar et al.(2002)]{Omar2002}
Omar, A., Chengalur, J.~N., \& Anish Roshi, D.\ 2002, \aap, 395, 227 

\bibitem[Omukai et al.(2005)]{Omukai2005}
Omukai, K., Tsuribe, T., Schneider, R., et al.\ 2005, \apj, 626, 627.

\bibitem[Paresce \& De Marchi(2000)]{Paresce2000}
Paresce, F. \& De Marchi, G.\ 2000, \apj, 534, 870. 

\bibitem[Pei \& Fall(1995)]{Pei1995}
Pei, Y.~C. \& Fall, S.~M.\ 1995, \apj, 454, 69.

\bibitem[Portegies Zwart et al.(2010)]{Portegies Zwart2010}
Portegies Zwart, S.~F., McMillan, S.~L.~W., \& Gieles, M.\ 2010, \araa,
48, 431. 

\bibitem[Quireza et al.(2006)]{Quireza2006ApJ}
Quireza, C., Rood, R.~T., Bania, T.~M., Balser, D.~S., \& Maciel, W.~J.\
2006, \apj, 653, 1226

\bibitem[Richards et al.(2012)]{Richards2012} Richards, E.~E.,
Lang, C.~C., Trombley, C., \& Figer, D.~F.\ 2012, \aj,
144, 89  

		
\bibitem[Rieke \& Lebofsky(1985)]{Rieke1985}
Rieke, G.~H., \& Lebofsky, M. J. 1985, \apj, 288, 618

\bibitem[Rudolph et al.(2006)]{Rudolph2006}
Rudolph, A.~L., Fich, M., Bell, G.~R., Norsen, T., Simpson, J.~P., Haas,
M.~R., \& Erickson, E.~F.\ 2006, \apjs, 162, 346

\bibitem[Scalo(1998)]{Scalo1998}
Scalo, J. 1998, in ASP Conf. Ser. 142, The IMF Revisited: A Case for
Variations, ed. G. Gilmore \& D. Howell (San Francisco, CA: ASP), 201

\bibitem[Salpeter(1955)]{Salpeter1955}
Salpeter, E.~E.\ 1955, \apj, 121, 161. doi:10.1086/145971

\bibitem[Schneider et al.(2018)]{Schneider2018}
Schneider, F.~R.~N., Sana, H., Evans, C.~J., et al.\ 2018, Science, 359,
69. 

\bibitem[Sharpless(1959)]{Sharpless1959}
Sharpless, S.\ 1959, \apjs, 4, 257 

\bibitem[Shaver et al.(1983)]{Shaver1983}
Shaver, P.~A., McGee, R.~X., Newton, L.~M., Danks, A.~C., \& Pottasch,
S.~R.\ 1983, \mnras, 204, 53

\bibitem[Siess et al.(2000)]{Siess2000}
Siess, L., Dufour, E., \& Forestini, M.\ 2000, \aap, 358, 593

\bibitem[Simons \& Tokunaga(2002)]{Simons2002}
Simons, D.~A., \& Tokunaga, A.\ 2002, PASP, 114, 169

\bibitem[Sirianni et al.(2000)]{Sirianni2000} 
Sirianni, M., Nota, A., Leitherer, C., De Marchi, G., \& Clampin, M.\
2000, \apj, 533, 203 

\bibitem[Skrutskie et al.(2006)]{Skrutskie2006}
Skrutskie, M.~F., Cutri, R.~M., Stiening, R., et al.\ 2006, \aj, 131,
1163

\bibitem[Soderblom et al.(2014)]{Soderblom2014}
Soderblom, D.~R., Hillenbrand, L.~A., Jeffries, R.~D., et al.\ 2014,
Protostars and Planets VI, 219.

\bibitem[Suzuki et al.(2008)]{Suzuki2008}
Suzuki, R., et al.\ 2008, \pasj, 60, 1347

\bibitem[Tokunaga et al.(2002)]{Tokunaga2002}
Tokunaga, A. T., Simons, D. A., \& Vacca, W. D. 2002, PASP, 114, 180

\bibitem[Vall{\'e}e(2005)]{Vallee2005}
Vall{\'e}e, J.~P.\ 2005, \aj, 130, 569

\bibitem[van der Walt et al.(2011)]{van der Walt2011}
van der Walt, S., Colbert, S.~C., \& Varoquaux, G.\ 2011, Computing in
Science and Engineering, 13, 22.

\bibitem[Vilchez \& Esteban(1996)]{Vilchez1996}
Vilchez, J.~M., \& Esteban, C.\ 1996, \mnras, 280, 720 

\bibitem[Weidner \& Kroupa(2006)]{Weidner2006}
Weidner, C. \& Kroupa, P.\ 2006, \mnras, 365, 1333.

\bibitem[Wenger et al.(2000)]{Wenger2000}
Wenger, M., Ochsenbein, F., Egret, D., et al.\ 2000, \aaps, 143, 9 

\bibitem[Wright et al.(2010)]{Wright2010}
Wright, E.~L., Eisenhardt, P.~R.~M., Mainzer, A.~K., et al.\ 2010, \aj,
140, 1868

\bibitem[Yasui et al.(2006)]{Yasui2006} 
Yasui, C., Kobayashi, N., Tokunaga, A.~T., et al.\ 2006, \apj, 649, 753.

\bibitem[Yasui et al.(2008a)]{Yasui2008ASPC} 
Yasui, C., Kobayashi, N., Tokunaga, A.~T., Saito, M., \& Tokoku, C.\
2008a, Formation and Evolution of Galaxy Disks, 396, 225

\bibitem[Yasui et al.(2008b)]{Yasui2008}
Yasui, C., Kobayashi, N., Tokunaga, A.~T., Terada, H., \& Saito, M.\
2008b, \apj, 675, 443

\bibitem[Yasui et al.(2009)]{Yasui2009} 
Yasui, C., Kobayashi, N., Tokunaga, A.~T., Saito, M., \& Tokoku, C.\
2009, \apj, 705, 54 

\bibitem[Yasui et al.(2010)]{Yasui2010} 
Yasui, C., Kobayashi, N., Tokunaga, A.~T., Saito, M., \& Tokoku, C.\
2010, \apjl, 723, L113

\bibitem[Yasui et al.(2016a)]{Yasui2016a}
Yasui, C., Kobayashi, N., Tokunaga, A.~T., et al.\ 2016a, \aj, 151, 50. 

\bibitem[Yasui et al.(2016b)]{Yasui2016b} 
Yasui, C., Kobayashi, N., Saito, M., et al.\ 2016b, \aj, 151, 115. 

\bibitem[Yasui et al.(2017)]{Yasui2017}
Yasui, C., Izumi, N., Saito, M., et al.\ 2017, Formation and Evolution
of Galaxy Outskirts, 321, 34. 

\bibitem[Yasui et al.(2021)]{Yasui2021} 
Yasui, C., Kobayashi, N., Saito, M., et al.\ 2021, \aj, 161, 139.

\end{thebibliography}
\end{document}